\documentclass[]{aa}  

\usepackage{graphicx}
\usepackage{txfonts}
\usepackage{subfig}
\usepackage{textgreek}
\usepackage{multirow}
\usepackage{mathtools}

\bibpunct{(}{)}{;}{a}{}{,}

\newcommand{\FIG}[1]{}

\def\mso{\,{\rm M}_\odot}

\def\lso{\,{\rm L}_\odot}

\def\kms{\, {\rm km}\, {\rm s}^{-1}}

\begin{document}


   \title{Simplified models of circumstellar morphologies for interpreting high-resolution data}

   \subtitle{Analytical approach to the equatorial density enhancement}

   \author{Ward Homan
          \inst{1}
          \and
          Jels Boulangier
          \inst{1}
          \and
          Leen Decin
          \inst{1}
          \and
          Alex de Koter
          \inst{1,2}
          }

   \offprints{W. Homan}          
          
   \institute{$^1$ Institute of Astronomy, KU Leuven, Celestijnenlaan 200D B2401, 3001 Leuven, Belgium \\
             \email{Ward.Homan@ster.kuleuven.be} \\
             \email{Leen.Decin@ster.kuleuven.be} \\
             $^2$ Sterrenkundig Instituut `Anton Pannekoek', Science Park 904, 1098 XH Amsterdam, The Netherlands\\
             }

   \date{Received <date> / Accepted <date>}
 
   \abstract  
   {Equatorial density enhancements (EDEs) are a very common astronomical phenomenon. Studies of the circumstellar environments (CSE) of young stellar objects and of evolved stars have shown that these objects often possess these features. These are believed to originate from different mechanisms, ranging from binary interactions to the gravitational collapse of interstellar material. Quantifying the effect of the presence of this type of EDE on the observables is essential for a correct interpretation of high-resolution data.}    
   {We seek to investigate the manifestation in the observables of a circumstellar EDE, to assess which properties can be constrained, and to provide an intuitive bedrock on which to compare and interpret upcoming high-resolution data (e.g. \emph{ALMA} data) using 3D models.}    
   {We develop a simplified analytical parametrised description of a 3D EDE, with a possible substructure such as warps, gaps, and spiral instabilities. In addition, different velocity fields (Keplerian, radial, super-Keplerian, sub-Keplerian and rigid rotation) are considered. The effect of a bipolar outflow is also investigated. The geometrical models are fed into the 3D radiative transfer code {\tt LIME}, that produces 3D intensity maps throughout velocity space. We investigate the spectral signature of the $J$=3$-$2 up to $J$=7$-$6 rotational transitions of CO in the models, as well as the spatial aspect of this emission by means of channel maps, wide-slit position-velocity (PV) diagrams, stereograms, and spectral lines. Additionally, we discuss methods of constraining the geometry of the EDE, the inclination, the mass-contrast between the EDE and the bipolar outflow, and the global velocity field. Finally, we simulated \emph{ALMA} observations to explore the effects of interferometric noise and artefacts on the emission signatures.}    
   {The effects of the different velocity fields are most evident in the PV diagrams. These diagrams also enable us to constrain the EDE height and inclination. A level of degeneracy may occur in the shapes of individual PV diagrams for different global velocity fields. The orthogonal PV diagrams may completely eliminate this ambiguity. Information on the EDE substructure is evident in the channel maps, but cannot be recovered from the PV diagrams, nor from the spectral lines. However, stereograms enable the detection of warping. For most inclinations the spectral lines are relatively broad, making it difficult to distinguish from an eventual superposed bipolar outflow component. Only under low inclination angles can one distinguish between these structures. Simulations of synthetic ALMA observations show how emission is affected when the largest angular scale of an antenna configuration is exceeded. For a rotating EDE, the emission around zero velocity will first fade because of destructive interference.}
   {}
   

   \maketitle


\section{Introduction}

The recent development of high-resolution observation instruments (e.g. \emph{ALMA}) has provoked a serious expansion of the astrophysical and cosmological frontiers owing to their data of unprecedented detail and quality. Amongst these, the field of circumstellar environments (CSEs) has advanced significally, since these instruments now enable us to perform detailed studies of the hydro- and thermodynamical, as well as the chemical properties of these systems. The forward modelling of circumstellar systems (including but not confined to hydrodynamical modelling, chemical networks or radiative transfer) is still proving to be extremely challenging in CPU-time consumption, even for present-day computational facilities. Therefore, to analyse observations and retrieve information on the active physical and chemical laws, the systems must be simplified (e.g. by assuming analytical prescriptions). In this paper, we make these kinds of simplifying assumptions in order to evaluate the impact of equatorial density enhancements (EDEs) on both the spatially and spectrally resolved emission.

EDEs are a very common astronomical phenomenon, caused by either directional loss of material by the central object, or by the interplay between conservation of angular momentum and the viscosity of the material orbiting around a strong central source of gravity. They are observed in a broad astronomical range: from planets on the smallest spatial scales (Saturn for example) to young stellar objects \citep[e.g.][]{Rebollido2015} and interacting binaries \citep[e.g.][]{Terquem2015}. This latter phenomenon is also likely the cause of EDEs around post-asymptotic giant branch (post-AGB) stars. These are believed to be formed as a result of close-binary interactions (1-2 AU), with a possible lower mass main-sequence star as a companion \citep{VanWinckel2003,VanWinckel2006}. On the widest spatial scales, EDEs manifest as spiral galaxies \citep{Bekiaris2016}, believed to be rotating around supermassive black holes \citep[e.g.][]{Schodel2002}.

We shall compare the intrinsic emission signatures of different analytical EDE toy models to provide the reader with an insight on how these objects manifest in the observables to aid the interpretation and analysis of future data. In particular, we shall focus on the effect of the velocity structure in the EDE, the combined emission signal of a Keplerian rotating disk with a bipolar outflow, and a density substructure within the EDE.

This paper is organised as follows: In Sect. 2 we provide a detailed description of the mathematical expressions generating the desired morphologies, followed by the physical and numerical setup of our models. Section 3 presents the results of the radiative transfer calculations in the form of channel maps, PV diagrams, and integrated spectra. In Sect. 4, we link the calculated intensities to intrinsic geometrical properties and discuss the use of stereograms. Section 5 shows the effect of the \emph{ALMA} 'eye' on the characteristics of the intrinsic emission. Finally, a summary of the findings is given in Sect. 6.  

\section{Morphological models and scientific motivation}

\subsection{General assumptions of the EDE}
Spherical coordinates are defined as
\begin{equation}
 r_{xyz} = \sqrt{x^2 + y^2 + z^2},
\end{equation}
\begin{equation}
 \theta = {\rm arccos} \left(\frac{z}{r_{\rm xyz}}\right),
\end{equation}
\begin{equation}
 \phi = {\rm arctan}\left(\frac{y}{x}\right),
\end{equation}
representing radial distance, and azimuthal and equatorial angle respectively. Cylindrical coordinates are defined as
\begin{equation}
 r_{xy} = \sqrt{x^2 + y^2},
\end{equation}
\begin{equation}
 \phi = {\rm arctan}\left(\frac{y}{x}\right),
\end{equation}
\begin{equation}
 z = z,
\end{equation}
representing radial distance, and equatorial angle, and vertical height (with respect to the equatorial plane) respectively.

The shape of an analytical EDE is, in principle, comprised of three major spatial dependencies. One radial component ($R$), one component describing any possible equatorial variations ($\Phi$), and a final constituent describing its height ($Z$). Each of these can be dependent on any of the cylindrical coordinates. In dimensionless units we express this as
\begin{equation}
 D(r_{xy},\phi,z) = R(r_{xy},\phi,z)\ \Phi(r_{xy},\phi,z)\ Z(r_{xy},\phi,z),
\end{equation}
where $D$ is the physical property that is being described.

To describe the physical properties of the EDE, one needs to specify temperature, density and velocity. We assume a power-law temperature dependence for the gas constituting the EDE,
\begin{equation}
 T_{EDE}(r_{xy}) = T_*\left(\frac{r_{xy}}{R_*}\right)^{-\epsilon},
\end{equation}
with $T_*$ the stellar temperature, $R_*$ the stellar radius, and $\epsilon$ the temperature power-law index. We note that this temperature profile is inconsistent with optically thick EDEs, since for a fixed radial distance these types of EDEs would have relatively hot surface layers and a relatively cool mid-plane zone. As these obscured mid-plane regions are mostly unobservable at the (infrared and submillimeter) line frequencies investigated here, they will only marginally affect our results. In addition, the assumed temperature profile is independent of the azimuthal height coordinate $z$. This is again a considerable simplification, since in reality, EDE temperatures increase as one transitions from the equatorial plane towards the surface \citep[e.g.][and references therein]{Pavlyuchenkov2007}.

A universal turbulent velocity $v_{\rm turb}$ is present in the entire spatial domain of the numerical model.

Since we aim to understand different density structures and velocity fields, different prescriptions will be analysed. We detail the physical assumptions and analytical formulations in Sect. 2.2 and 2.3, respectively. As EDEs are often accompanied by a bipolar outflow, we also study the effect of a bipolar outflow on the observables. A mathematical description is given in Sect. 2.4.

\subsection{Density structure}

\subsubsection{Flared EDE}
\begin{figure}[htp]
\centering
\includegraphics[width=0.45\textwidth]{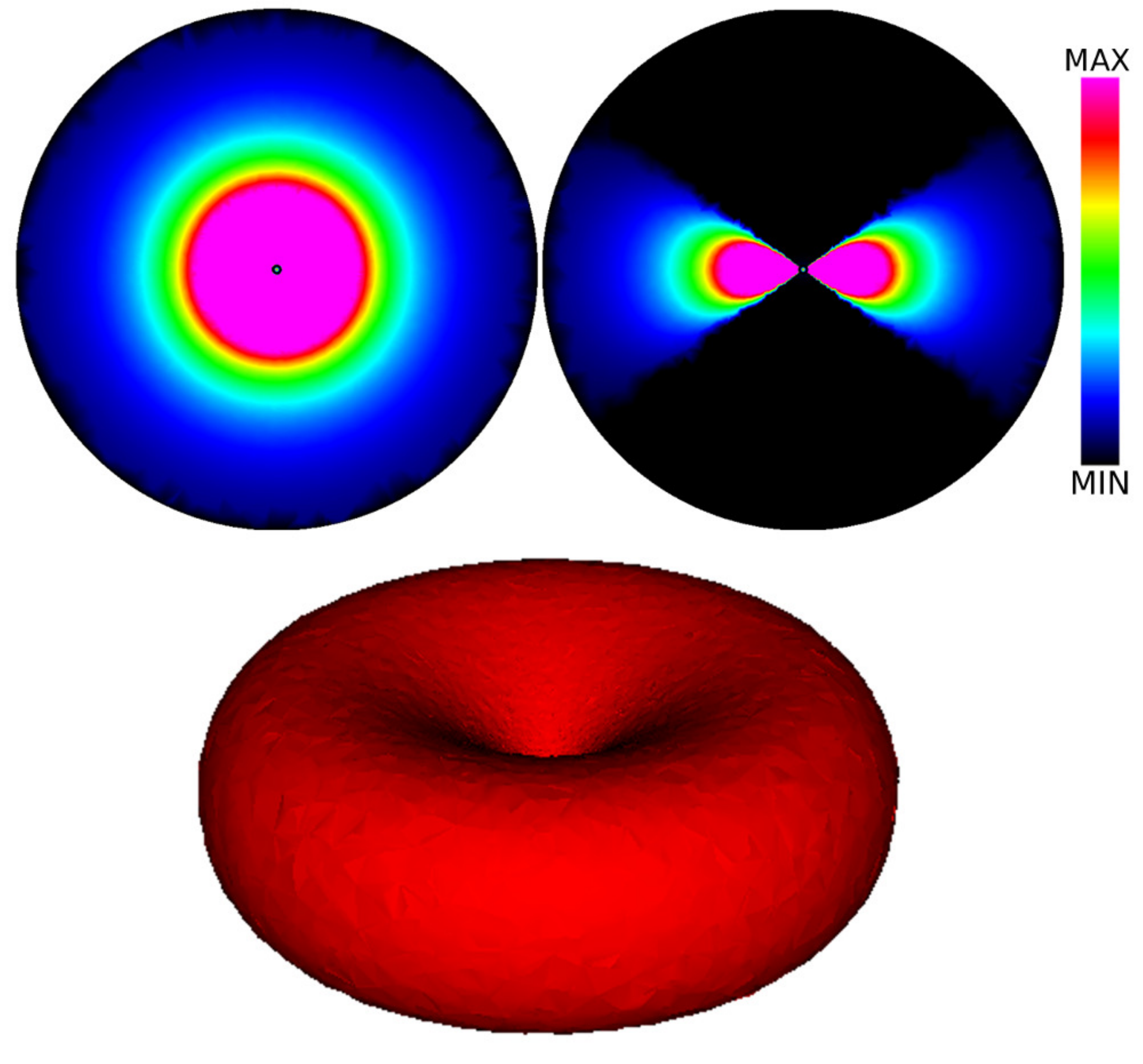}
\caption{Visualisation of the EDE density model. The top left image shows a cut through the equatorial plane, the top right image shows a cut through a meridional plane. MIN represents the minimum density in the models (in this case zero), and MAX has been chosen such as to enhance the visual appearance of the image. The bottom image shows a three-dimensional contour plot of the red regions in the top images of this figure. \label{dens_EDE}}
\end{figure}

We construct a simple EDE with no substructure by setting $R(r_{xy},\phi,z) = R(r_{xy})$, $\Phi(r_{xy},\phi,z) = 1$ and $Z(r_{xy},\phi,z) = Z(r_{xy},z)$, shown in Fig. \ref{dens_EDE}. The radial component of the EDE density is assumed to decay as a power law. Vertically, a simple Gaussian distribution is assumed. This is consistent with the previously made assumption of a vertically isothermal disk. In addition we define a critical radius $r_c$ within which the density of the EDE has been set to zero, to simulate its inner radius. Mathematically, this translates to an EDE density represented by 
\begin{equation}
 \rho_{EDE}(r_{xy},\phi,z) = \rho_0\left(\frac{r_{xy}}{r_c}\right)^{-p}\exp\left[\frac{-z^2}{2H(r_{xy})^2}\right],
\end{equation}
where $\rho_0$ represents the density at $r_{xy}=r_c$ and $z=0$, $p$ describes the rate at which the density decays radially, and $H(r_{xy})$ represents the one-sigma Gaussian height of the EDE. Adopting the solution of a disk in vertical hydrostatic equilibrium we express the Gaussian height $H(r_{xy})$ as
\begin{equation}
 H(r_{xy}) = H_c\left(\frac{r_{xy}}{r_c}\right)^h,
\end{equation}
where $H_c$ represents the initial one-sigma height at $r_{xy}=r_c$ and $z=0$, and the exponent $h$ represents the characteristical course with which the EDE height changes as a function of radius. By adopting $h>1$, which ensures the aspect ratio $H(r_{xy})/r_{xy}$ keeps increasing, we can refer to this property as the flaring of the EDE.

\begin{figure}[htp]
\centering
\includegraphics[width=0.45\textwidth]{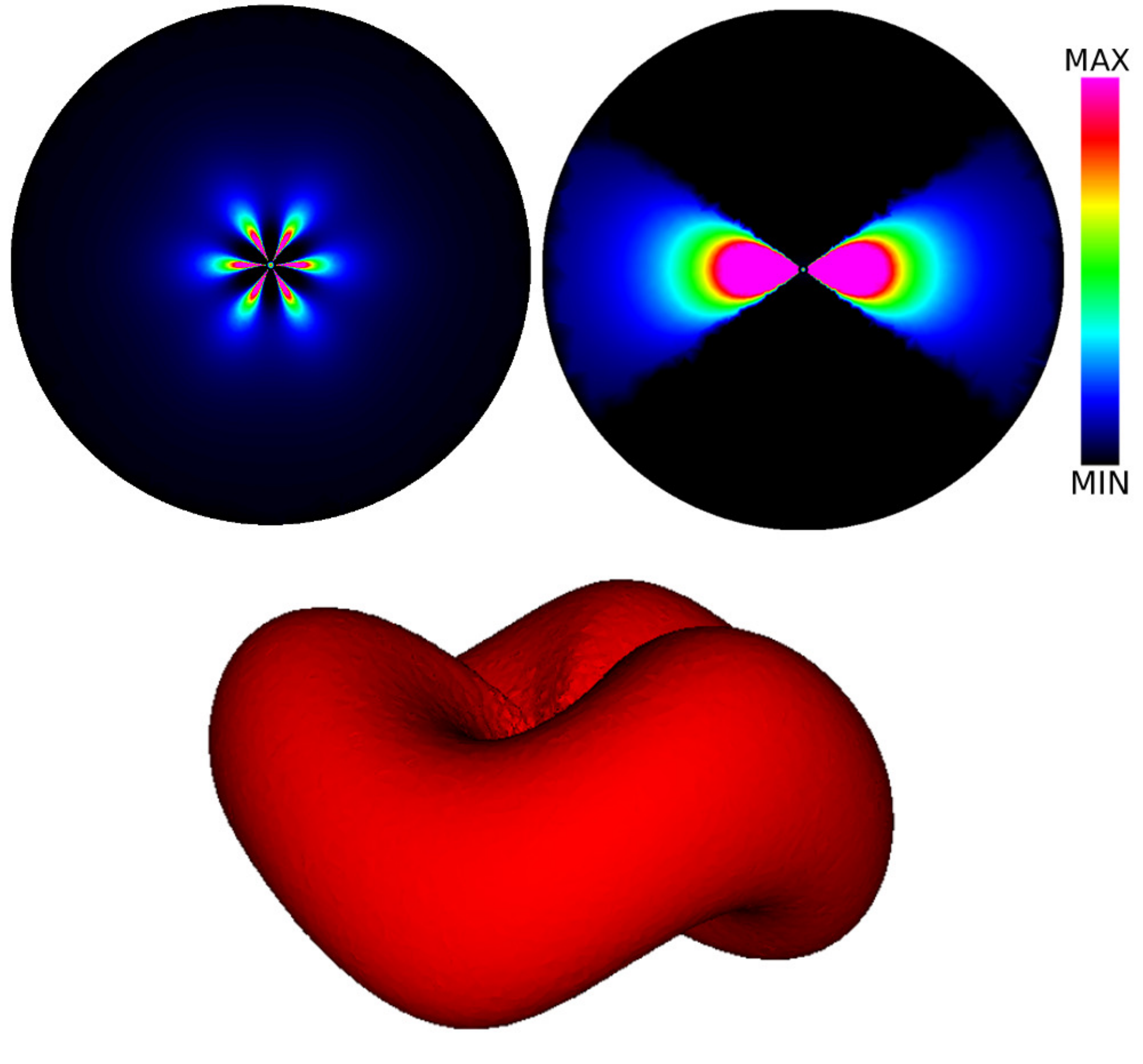}
\caption{Visualisation of the density model of the warped EDE with three undulations. The top left image shows a cut through the equatorial plane, the top right image shows a cut through a meridional plane. The meaning of the colours is clarified by the linear colourbar on the right. MIN represents the minimum density in the models (in this case zero), and MAX has been chosen such as to enhance the visual appearance of the image. The bottom image shows a three-dimensional contour plot of the model. \label{dens_warp}}
\end{figure}

\subsubsection{Warped EDE}
Warped EDEs have been observed \citep[eg.][]{Nixon2015}, and are believed to arise when a strong non-axisymmetrical force is acting on the EDE. It is theorised that compact, high-mass objects may bring about such a force \citep[][and references therein]{Nixon2015b,Montgomery2012,Ogilvie2013}. AGB stars and their descendants are often accompanied by a binary companion \citep{Raghavan2010,Duchene2013}. If the system acquires an EDE, and the companion is massive enough, the EDE may result in acquiring a warp. Though presently not well understood, warped EDEs have also been observed around solar-type (pre-) main sequence stars \citep[e.g.][]{Boccaletti2015}, and may somehow be related to planet formation \citep[e.g.][]{Casassus2015}.

The density of our warped EDE is represented by
\begin{equation}
 \rho_d(r_{xy},\phi,z) = \rho_0\left(\frac{r_{xy}}{r_c}\right)^{-p}\exp\left[\frac{-(z-A\sin(N\phi))^2}{2H(r_{xy})^2}\right],
\end{equation}
where $A$ is the amplitude of the warping, and $N$ the number of undulations. $H(r_{xy})$ is as described in Eq. 10. A visual representation of a warped EDE with three undulations can be found in Fig. \ref{dens_warp}.

\begin{figure}[htp]
\centering
\includegraphics[width=0.45\textwidth]{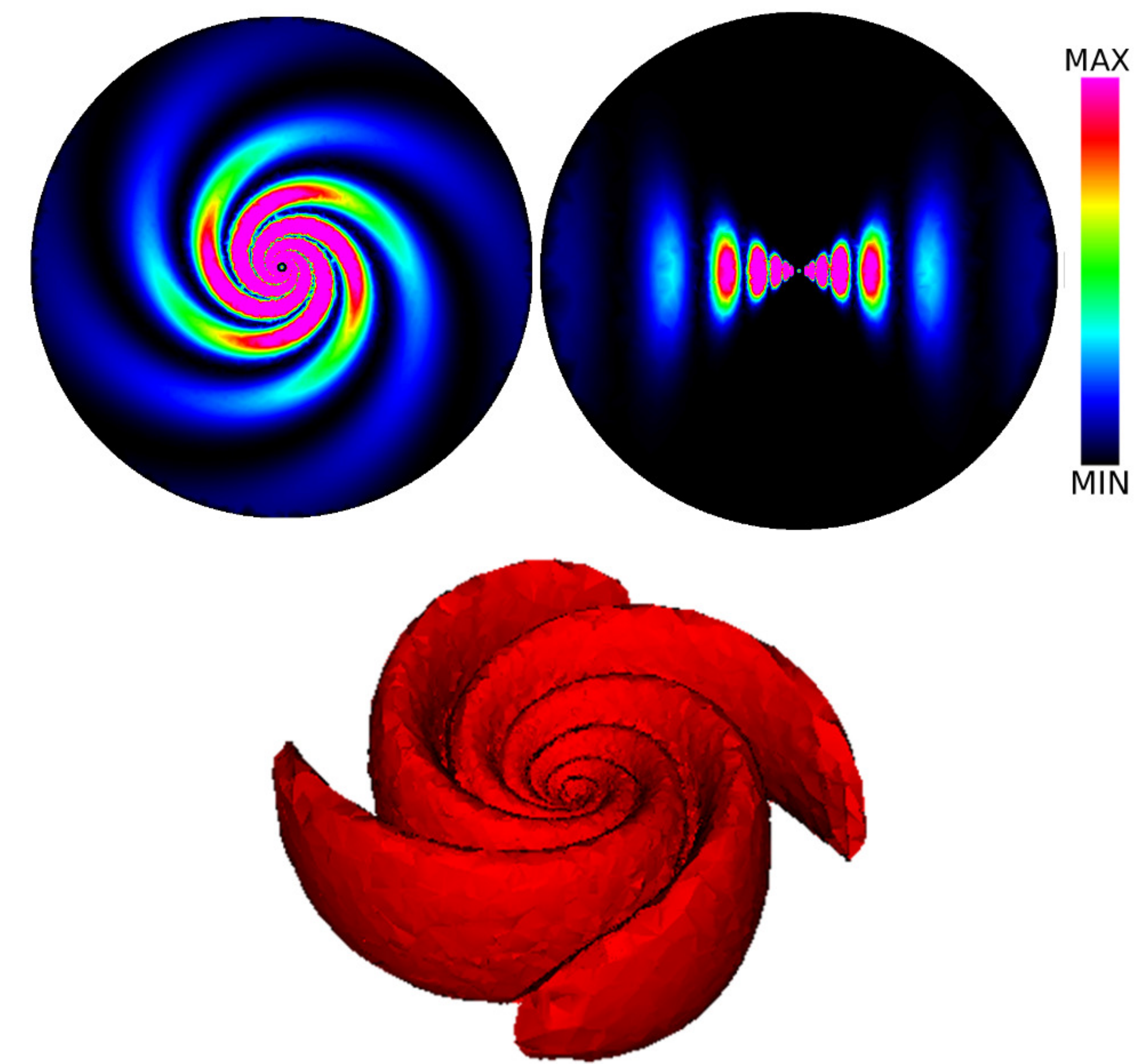}
\caption{Visualisation of the density model of the spiral-shaped density model. The panels have the same meaning as in Fig. \ref{dens_warp}. \label{dens_spir}}
\end{figure}

\subsubsection{Spiral-shaped density model}
Spiral-shaped instabilities arise when the attracting force of the primary source of gravity has a magnitude large enough to overcome both the internal pressure gradient of the gas (i.e. Jeans criterion) and the centrifugal forces, and also to halt the relative motion between different portions of the differentially rotating object \citep{Toomre1964}. Recent high-resolution observations of the CSE of young stars have clearly identified spiral-shaped instabilities \citep[][and reference therein]{Christiaens2014}, which have lead to an extensive study of their detectability (in continuum emission) via \emph{ALMA} simulations \citep{Dipierro2014}. Theoretical work on their formation and on the consequences of their presence is currently underway \citep[e.g.][]{Dipierro2015,Juhasz2015,Lin2015}. It is believed that massive planets inside protoplanetary EDEs may also cause such patterns \citep{Dong2015,Zhu2015}.

We model EDE containing spiral-shaped instabilities as
\begin{equation}
 \rho_d(r_{xy},\phi,z) = \rho_0 \sin(L\phi+\omega(r_{x,y}))\left(\frac{r_{xy}}{r_c}\right)^{-p}\exp\left[\frac{-z^2}{2H(r_{xy})^2}\right],
\end{equation}
where $L$ is the number of spiral arms and $\omega(r_{xy})$ is a radial dependent phase factor. We find that a phase factor which depends logarithmically on the radius $\omega(r_{xy}) \sim \log(r_{x,y})$ produces spiral shapes which, in visual appearance, closely resemble the instabilities generated by hydrodynamical simulations of young stellar objects and protostellar systems \citep{Lin2015,Vorobyov2015}. $H(r_{xy})$ is as described in Eq. 10. Fig. \ref{dens_spir} contains a visual representation of the density model.

\begin{figure}[htp]
\centering
\includegraphics[width=0.45\textwidth]{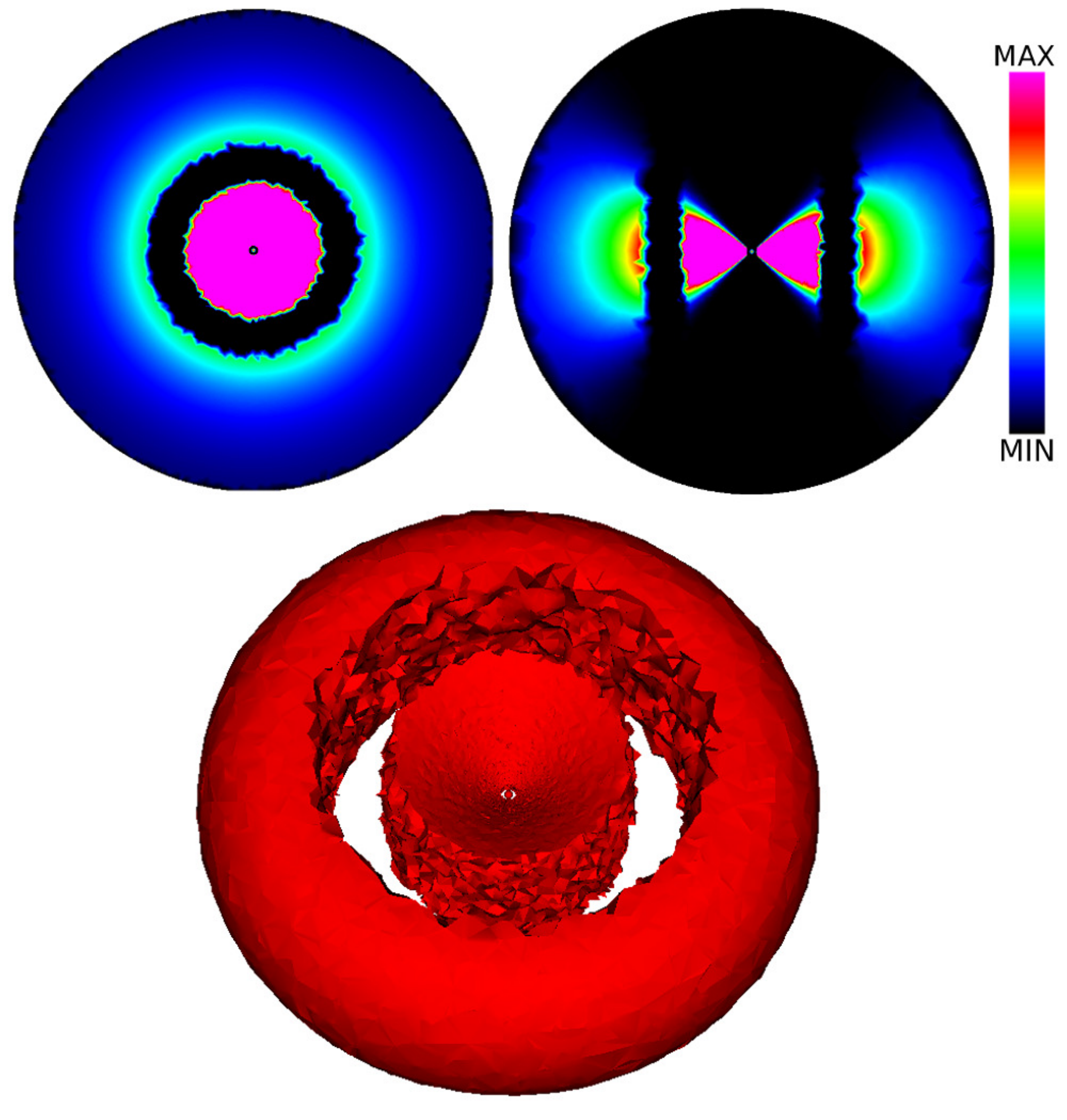}
\caption{Visualisation of the density model of the EDE containing an annular gap. The panels have the same meaning as in Fig. \ref{dens_warp}. The coarseness of the gap edges is purely a resolution effect. \label{dens_gap}}
\end{figure}

\subsubsection{EDE with annular gap}

The primary cause for the formation of an annular gap in an EDE would be the growth of a (proto) planet from the rotating EDE-material. Recent high-resolution imaging has proven able to resolve such gaps, which are the direct consequences on EDEs of planets in the making \citep[][and references therein]{Flock2015}. These discoveries have also lead to a theoretical development of the mechanics behind the formation of these gaps in protoplanetary systems \citep[eg.][]{Les2015}.

The annular gap in the EDE is achieved by setting a fixed radius-range in which the EDE density is substantially reduced. The images in Fig. \ref{dens_gap} show a three-dimensional contour plot of this geometry, and slices through such an EDE. 

\begin{table*}[htp!]
\centering
\caption{The different combinations of previously described morphological properties we shall explore in this paper. The last collumn refers to the section where the specific combination is discussed. \label{comb}}
\begin{tabular}{ l  l  l  l}
\hline
\hline

EDE Density & EDE Velocity & Extra feature & Section \\ \hline

Flared & Keplerian & & 4.1 \\ \hline

Flared & Radial & & 4.2.1 \\
Flared & Super-Keplerian & & 4.2.2 \\
Flared & Sub-Keplerian & & 4.2.3 \\
Flared & Rigid rotation & & 4.2.4 \\ \hline

Warped & Keplerian & & 4.3.1 \\
Annular gap & Keplerian & & 4.3.2 \\
Spiral instabilities & Keplerian & & 4.3.3 \\ \hline

       &           & Bipolar outflow & 4.4.1 \\
Flared & Keplerian & Bipolar outflow & 4.4.2 \\ \hline

\end{tabular}
\end{table*}

\subsection{Velocity field}
We also appraise five global velocity fields: Keplerian, super-Keplerian, sub-Keplerian, rigid rotation and radially outflowing.
\begin{itemize}
 \item \underline{\textbf{Keplerian field:}} An EDE can be assumed to rotate in a Keplerian way by simply assuming a zero net force on an orbiting test-particle. This is the most basic of the chosen velocity fields. This field has an exclusively tangential velocity component of the following magnitude
 \begin{equation}
  v_{EDE} = v_{EDE,\bot}(r_{xy}) = \sqrt{\frac{GM_*}{r_{xy}}},
 \end{equation}
 where $M_*$ is the mass of the central object to which the EDE is gravitationally bound.
 \item \underline{\textbf{Super-Keplerian field:}} Super-Keplerian EDEs, i.e. differentially rotating EDEs with a positive radial velocity component, are believed to exist around rapidly rotating stars, and are the result of material reaching the critical velocity for mass-loss from the equatorial regions \citep[][and references therein]{Lee2013}. Super-Keplerian EDEs have also been used to model EDEs surrounding the central regions of post-AGB stars \citep[e.g.][]{Bujarrabal2013b}. An increased rotation rate, or and augmented tangential velocity component of the field will result in the centripetal force on a test-particle to surpass the graviational attraction, resulting in the test-particle slowly travelling outward. It must be noted that in such a case the tangential velocity structure of the differential rotation field would be somewhat different from Keplerian. However, because we do not possess a clear analytical description of such a rotational field, we shall approximate it by modelling a Keplerian field superposed by an outward radial velocity field $v_{\rm out}$.
 \item \underline{\textbf{Sub-Keplerian field:}} Accretion of elliptically orbiting material onto EDEs, and the subsequent mixing with the EDE material, can exert a torque on the system which is great enough to slow the rotation down to sub-Keplerian regimes, pushing the material in towards the central star \citep[e.g.][]{Cassen1981,Visser2010}. We shall refer to this scenario as the sub-Keplerian case, where an inward velocity $v_{\rm in}$ has been superposed on top of a differentially rotating field. A particle travelling in this field will slowly spiral inward. Following the same argumentation as for the super-Keplerian field, we approximate the perturbed rotating field by a Keplerian velocity field.
 \item \underline{\textbf{Rigid rotation:}} The presence of a strong magnetic field can generate the required torque to force a rigid rotation. Simulations of protoplanetary EDEs with magneto-rotational instabilities can generate self-correcting (thus stable) rigid rotators \citep{Kato2009}. In addition, MHD (magneto-hydrodynamics) has shown the periodical formation of a rigidly rotating EDE in a line-driven wind of a rotating star with a properly aligned dipole surface field \citep[][and references therein]{UdDoula2008}. We model this rigid rotation (which is rotating with a constant angular velocity) by setting a maximum tangential velocity $v_{tan}$ at the boundary of the model.
 \item \underline{\textbf{Radial field:}} Radially outflowing EDEs have been frequently observed around post-AGB stars \citep[e.g.][]{Hirano2004,Chiu2006} and possibly also around AGB stars \citep[][]{Kervella2015}. We model our radially outflowing EDE by assuming a constant outflow velocity $v_{rad}$.
\end{itemize}

\subsection{Bipolar outflow}
Finally, we briefly discuss the emission characteristics of a bipolar outflow. We do this because we expect, due to the geometrical boundary conditions, EDEs and bipolar outflows to be commonly observed together in the evolved stars regime. If an EDE forms around an evolved star which is producing a wind, the EDE will obstruct the motion of the gas through its densest regions, and naturally collimate the outflow. Quantifying the appearance of bipolar outflow emission is therefore essential when studying cylindrically symmetrical morphologies of evolved star CSEs. \\\\

\textbf{Velocity $v_{\rm bo}$:} \newline
We assume the bipolar outflow to have an exclusively radial velocity field, expressed as
\begin{equation}
 v_{\rm bo}(r_{xyz}) = v_\infty\left(1-\frac{r_c}{r_{xyz}}\right)^\beta,
\end{equation}
where $v_\infty$ is the terminal wind velocity, and where $\beta$ is the parameter controlling the rate of acceleration of the flow. \\\\
\textbf{Density $\rho_{\rm bo}$:} \newline
The density of the bipolar outflow is exclusively radial, and given by
\begin{equation}\label{dens}
 \rho_{\rm bo}(r_{xyz}) = \frac{\dot{M}}{\Omega r_{xyz}^2 v_{\rm bo}(r_{xyz})},
\end{equation}
where $\dot{M}$ is the mass loss rate of the central star, $\Omega$ is the total solid angle taken up by the bipolar outflow, and $v_{\rm bo}(r_{xyz})$ is the radial velocity of the outflow. For an un-obscured wind, $\Omega=4\pi$. When the radial wind is obscured by an EDE, the solid angle of the bipolar outflow is limited by the boundary between the two structures, the description which is given in Sect. 4.4.2.\\\\
\textbf{Temperature $T_{\rm bo}$:} \newline
For the temperature structure we adopt the same power-law profile as for the EDE (Eq. 8), albeit using the spherical radial coordinate,
\begin{equation}
 T_{\rm bo}(r_{xyz}) = T_*\left(\frac{r_{xyz}}{R_*}\right)^{-\gamma},
\end{equation}
with $T_*$ the stellar temperature, $R_*$ the stellar radius, and $\gamma$ the temperature power-law index.

\section{Radiative transfer models and numerical methods}

\subsection{Parameter space}
In this section we describe the models we shall examine, as well as the model parameters of each specific combination of density and velocity field. Table \ref{comb} presents the combinations of specific morphological models we shall explore. We also present a listing of both the general parameters (Table \ref{Gen_par}), and of the specific physical and morphological parameters of the investigated geometries (Table \ref{EDE_par}).

\begin{table}[htp!]
\centering
\caption{General situational parameter values of every radiative transfer model. \label{Gen_par}}
\begin{tabular}{ l  l }
\hline
\hline
\multicolumn{2}{ c }{Stellar Parameters} \\
\hline
$T_*$ & $2500\ K$  \\ 
$L_*$ & $10000\ \lso$ \\
$M_*$ & $2\ \mso$ \\
Distance & $150 {\rm\ pc}$ \\
${\rm [CO/H]}_2$ & $5.0\times 10^{-4}$ \\ 
\hline

\multicolumn{2}{ c }{Dust Parameters} \\
\hline
Amorphous Carbon & 53\% \\ 
Silicon Carbide & 25\% \\
Magnesium Sulfide & 22\% \\
Gas/Dust by mass & 100 \\
\hline

\end{tabular}
\end{table}

\begin{table}[htp!]
\centering
\caption{Parameter values of the EDE  and bipolar outflow models. \label{EDE_par}}
\begin{tabular}{ l  l  l}
\hline
\hline

\multicolumn{2}{ c }{EDE density parameters} \\
\hline
$p$ & $2.25$ \\ 
$H_c$ & $2.0 {\rm\ AU}$ \\
$h$ & $1.25$ \\
$M_{\rm EDE}$ & $10^{-3} \mso$ \\

\hline

\multicolumn{2}{ c }{EDE velocity parameters} \\
\hline
$v_{\rm out}$ & $2.0 \kms$ & (super-Keplerian) \\
$v_{\rm in}$ & $-2.0 \kms$ & (sub-Keplerian) \\
$v_{\rm rad}$ & $10.0 \kms$ & (radial) \\
$v_{\rm tan}$ & $10.0 \kms$ & (rigid rotation) \\
$v_{\rm turb}$ & $1.0 \kms$ \\ 
\hline

\multicolumn{2}{ c }{EDE temperature parameters} \\
\hline
$\epsilon$ & $0.5$ \\ 
\hline

\multicolumn{2}{ c }{Bipolar outflow parameters} \\
\hline
$v_\infty$ & $10.0 \kms$ \\
$v_{\rm turb}$ & $1.0 \kms$ \\ 
$\beta$ & $0.5$ \\
$r_c$ & $10.0 {\rm\ AU}$ \\ 
$\gamma$ & $0.5$ \\
$M_{\rm bipolar}$ & $10^{-3} \mso$ \\
\hline

\end{tabular}
\end{table}

The choice for the stellar and dust parameters are based on estimations of the system properties of the carbon-rich AGB star CW Leo (See \citet{Homan2015} for details). 

The overview presented in Table \ref{EDE_par} is chosen such as to narrow-down the total explorable parameter space. The EDE parameters are based on values which are bound by theory, and whose numerical values have been confirmed by modelling of observations. Hydrostatic equilibrium for a non-self-gravitating Keplerian disk requires that $h=-\epsilon/2 + 3/2$. The parameter $\epsilon$ (see Eq. 8) adopts a value of 0.75 for the flat disk limit, and 0.5 for a more complex flaring disk  \citep{Kenyon1987}. \citet{Chiang1997,Hartmann1998} show the latter to be more consistent with reality, resulting in a flaring coefficient $h$ of 1.25. Expressing the disk midplane density as a function of its surface density (defined as the intregral of the mass volume density $\rho$ over the vertical disk height $z$) one comes to the relation $p=f+h$, where $f$ is the power of the rate with which the surface density drops with radius. Generally, $f$ is found to have a value between 0 and 1 \citep{Wilner2000,Kitamura2002,Andrews2007}. \citet{Andrews2009,Andrews2010} found a mean value of $<f> \sim 1$ for a sizable sample of disks. Adopting this value for the radial surface density decay rate and the previously adopted value for the flaring exponent, the value of $p=2.25$ is obtained. This value also happens to coincide with typical density decay rates found in EDEs around post-AGB stars and planetary nebulae \citep{Bujarrabal2013b, Bujarrabal2015, Bujarrabal2016}. The bipolar outflow parameters correspond to typical expected values of dust-driven winds generated by evolved stars.

At this point we wish to emphasise that the main aim of this work is to provide the reader with a basic intuition on how EDEs might manifest in the observables. The numerics involved in calculating the transport of radiation for such a vast set of cases is very CPU-intensive. Thus, in order to circumvent hydrodynamics calculations and thus mitigate calculation times no effort has been comitted to ensuring the adopted expressions are physically self-consistent. However, a very low degree of consistency has been implemented using analytical first-order approximations of disk morphologies. In addition, besides having made such reasonable assumptions, it must be noted that these assumptions are only valid for simple, non-self-gravitating, hydrostatically stable Keplerian disks. The assumption that the previously described physical distributions, and parameter values are universally valid for each case presented in Table \ref{comb} is therefore substantial. Nevertheless, we believe these simplifications to be valid first-order approximations, and shall therefore use them universally throughout the paper. The same can be said of the velocity fields, and especially of the sub- and super-Keplerian cases. These are both approximated by a regular Keplerian field, with a superposed radial component. While one would also expect the tangential velocity structure of such a steady-state to be different from the simple Keplerian field, the Keplerian field should still be a good first-order approximation of a differentially rotating field with its highest tangential velocities close to the central star.

\subsection{Radiative transfer}
To generate the 3D intensity channel maps of simulated emission of the above models, the Non-Local-Thermodynamical-Equilibrium (NLTE) full-3D, submillimeter and infrared (IR) radiative transfer code { \tt LIME} \citep{Brinch2010} was used. For a technical overview of the inner workings of the code, see \citet{Brinch2010}.

We focused mainly on the emission of the CO $J$=3$-$2 rotational transition (unless otherwise stated), for which the spectroscopic CO data of the LAMDA database \citep{Schoier2005} were used. The collisional rates were taken from \citet{Yang2010}. We have chosen this transition as the primary reference transition because the line formation zone encompasses most of the morphological structure, yet does not extend into the region where photodissociation effects must be taken into account. 

All models have been sampled with approximately $7 \times 10^5$ tetrahedral cells, whose size and position have been weighted by the density in the model.

\subsection{The making of the wide-slit position-velocity diagram}
\begin{figure}[t]
\centering
\includegraphics[width=0.4\textwidth]{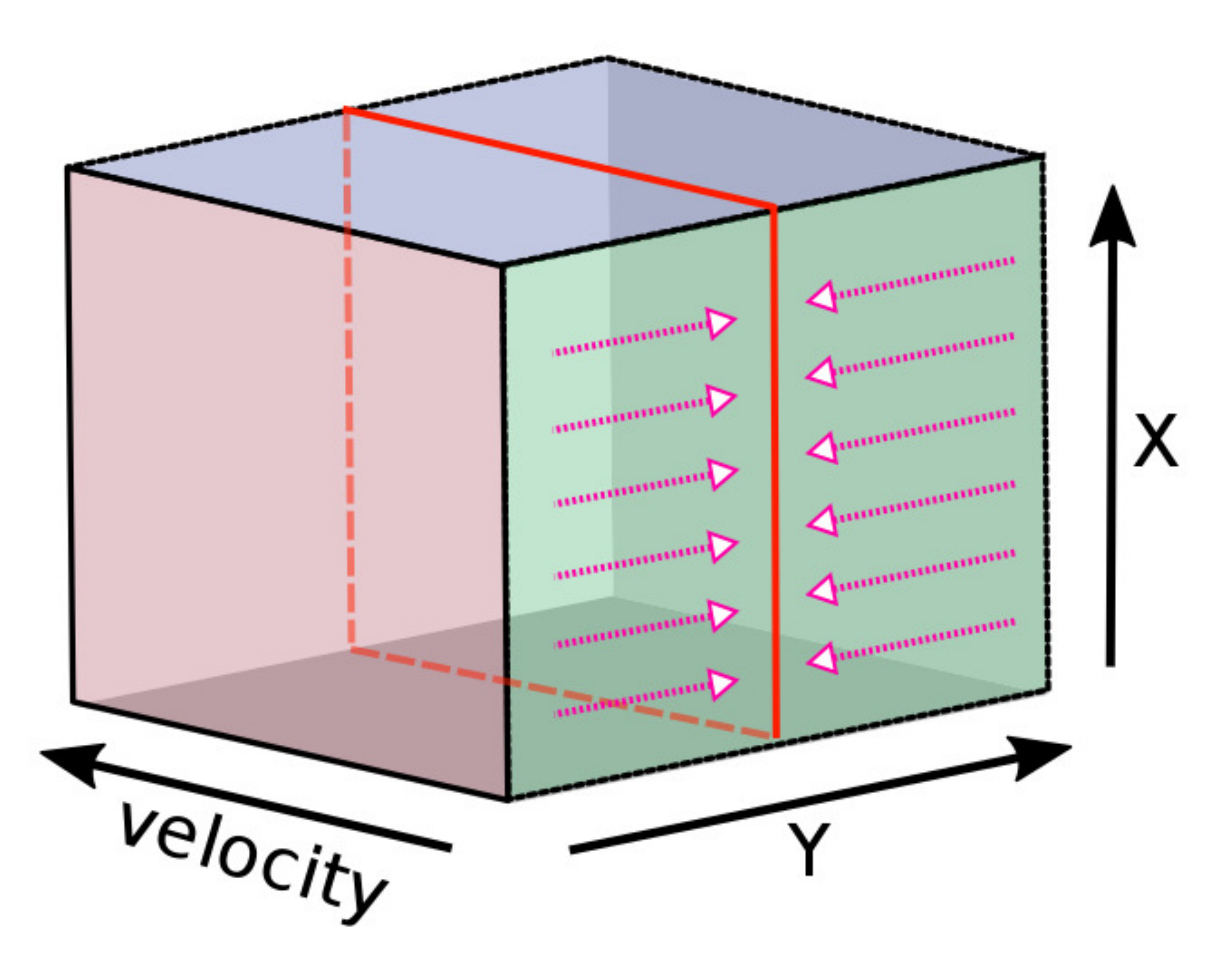}
\caption{Schematic representation of the production of a wide-slit PV diagram. The cube represents the 3D datacube, with two spatial axes (X,Y) and a velocity axis. A slit over the data is selected (red slice), provided the velocity axis remains intact. In this particular case the Y-axis is conserved. Subsequently, all the emission of the 3D data is collapsed onto the slit, by summing up the intensities with constant (X,v) coordinates. \label{PVCons}}
\end{figure}

In this paper we shall make abundant use of wide-slit position velocity diagrams. We discuss how they are produced here. 3D emission data consists of two linearly independent angular dimensions (representing the two angular coordinates in the plane of the sky), and one velocity dimension. The PV diagram is, in effect, a slice through the 3D data at an arbitrary angular axis in the plane of the sky (thus preserving the velocity axis). It is thus a 2D plot of the emission along this chosen axis, versus velocity. In principle any slit width can be chosen. If the slit has any width larger than one singular data pixel, then the emission is collapsed (onto each other) by summing up the emission with identical PV coordinates. To get a condensed overview of the overall wind morphology, which aids in the possible detection of potentially hidden structural characteristics, it is useful to make PV diagrams with a maximal slit width, namely the full size of the datacube. These will be referred to as wide-slit PV diagrams. The strengths of this visualisation technique have been verified by \citet{Decin2015}, where a spiral-shaped structure was detected in the inner wind of the AGB star CW Leo. We therefore consider it an invaluable tool for the analysis of 3D data, and shall use it throughout the discussions below. Fig. \ref{PVCons} is a schematic representation of this process. 

It is advantageous to construct two orthogonal wide-slit PV diagrams of the 3D data, choosing any set of linearly independent (thus perpendicular) angular dimensions oriented such as to maximally exploit the asymmetry of the data. The axes along which the asymmetry of the models appears strongest will be labeled as X and Y. X and Y are thus by no means fixed spatial coordinates, but are rather chosen entirely as a function of the geometrical properties of the observed system. In the case of disk-like structures, one slit should be oriented along (or parallel to) the equatorial axis, and the other along (or parallel to) the polar axis of the structure. We note that these orientations cease to have meaning when the structure is viewed face-on. All PV diagrams presented in this paper have been constructed following these instructions, with X being the polar axis, and Y being the equatorial axis (for non-face-on views). Much easier to interpret than the full 3D data, well-chosen wide-slit PV diagrams provide the user with clear correlated structural trends of the complete circumstellar environment. In all the following sections, the PV diagrams which maximally exhibit the asymmetry of the data (along X and Y) will be referred to as PV1 and PV2 respectively. Each PV diagram will be centred on $v=0 \kms$ and zero offset. 

\section{Radiative transfer results}

\begin{figure}[htp]
\centering
 \includegraphics[width=0.4\textwidth]{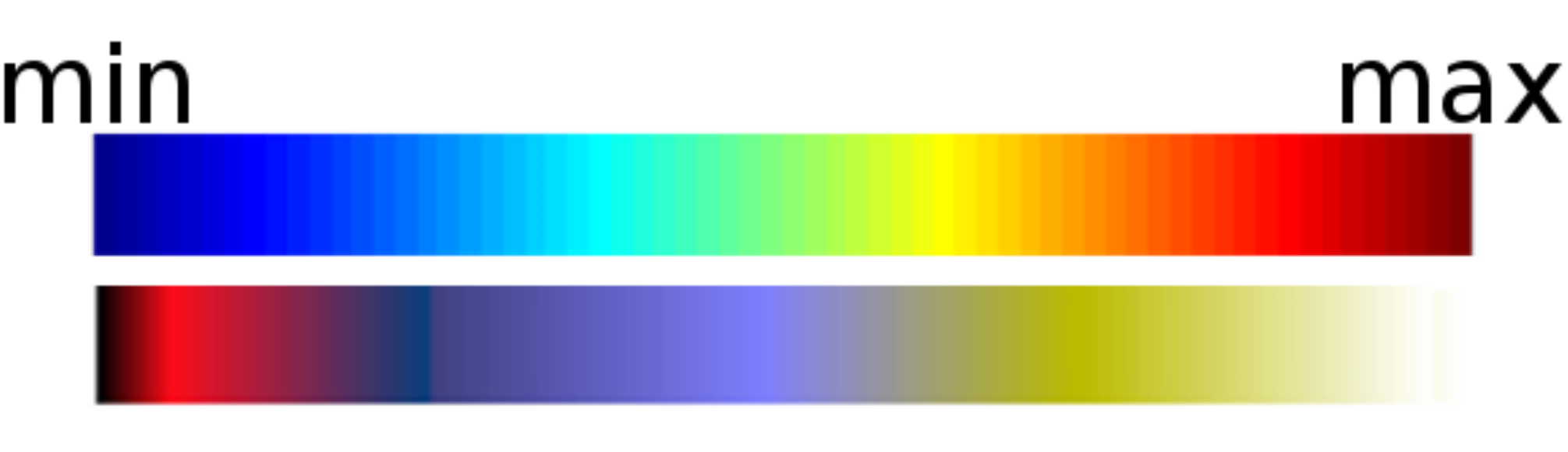}
\caption{The top linear colourmap is used in all images of PV diagrams below. The bottom linear colourmap is used in all images of channel maps. Both colourmaps represent the strength of the pixel intensities in the image. The colours are always scaled to the emission in the individual image, unless otherwise stated. \label{cmap}}
\end{figure}

In this section we describe the properties of the intrinsic emission in both the intensity channel maps and the spectral lines. The colour-coding used in all images below follows from the linear colourbars shown in Fig. \ref{cmap}. The colours are always scaled to the emission in the individual image, unless otherwise stated. We note that the quality of most images shown below is strongly enhanced when viewed on screen. The emission from abovementioned models will be viewed at six evenly spaced inclination angles between zero degrees (face-on) and ninety degrees (edge-on). The field of view consists of 500 pixels per dimension. The modelled disks have a radius of 500 AU, which encompasses the whole field of view. The spectral resolution is 0.5 $\kms$.

We have opted to omit the scales in all figures because due to the nature of the geometrical set-up, these systems are arbitrarily scalable in size and in absolute emission strength. Any absolute evaluation of correspondent angular or length scales is entirely model-dependent, and therefore meaningless. All results are continuum subtracted.

We shall begin with an in depth discussion of the emission features of the channel maps and the position-velocity diagrams of the flared Keplerian disk (Sect. 4.1). Subsequently, in Sect. 4.2, we focus on the effect of the different velocity fields, and compare them with the results from Sect. 4.1. In Sect. 4.3, we discuss how density substructure affects the emission patterns. We finish with a description of the bipolar outflow and its influence on the global emission features when combined with a flared disk (Sect. 4.4).

\subsection{The Keplerian disk}

\subsubsection{Looking through the channel maps}
The most straightforward way to investigate three-dimensional emission data is by looking at the channel maps, i.e. images of the Doppler-shifted emission at different velocities (The channel maps of all calculated models are presented in Appendix A). If the data is of high quality and of high enough spatial and spectral resolution, an exploratory investigation of the channel maps should reveal all the data has to offer.

\begin{figure}[htp]
\centering
\includegraphics[width=0.45\textwidth]{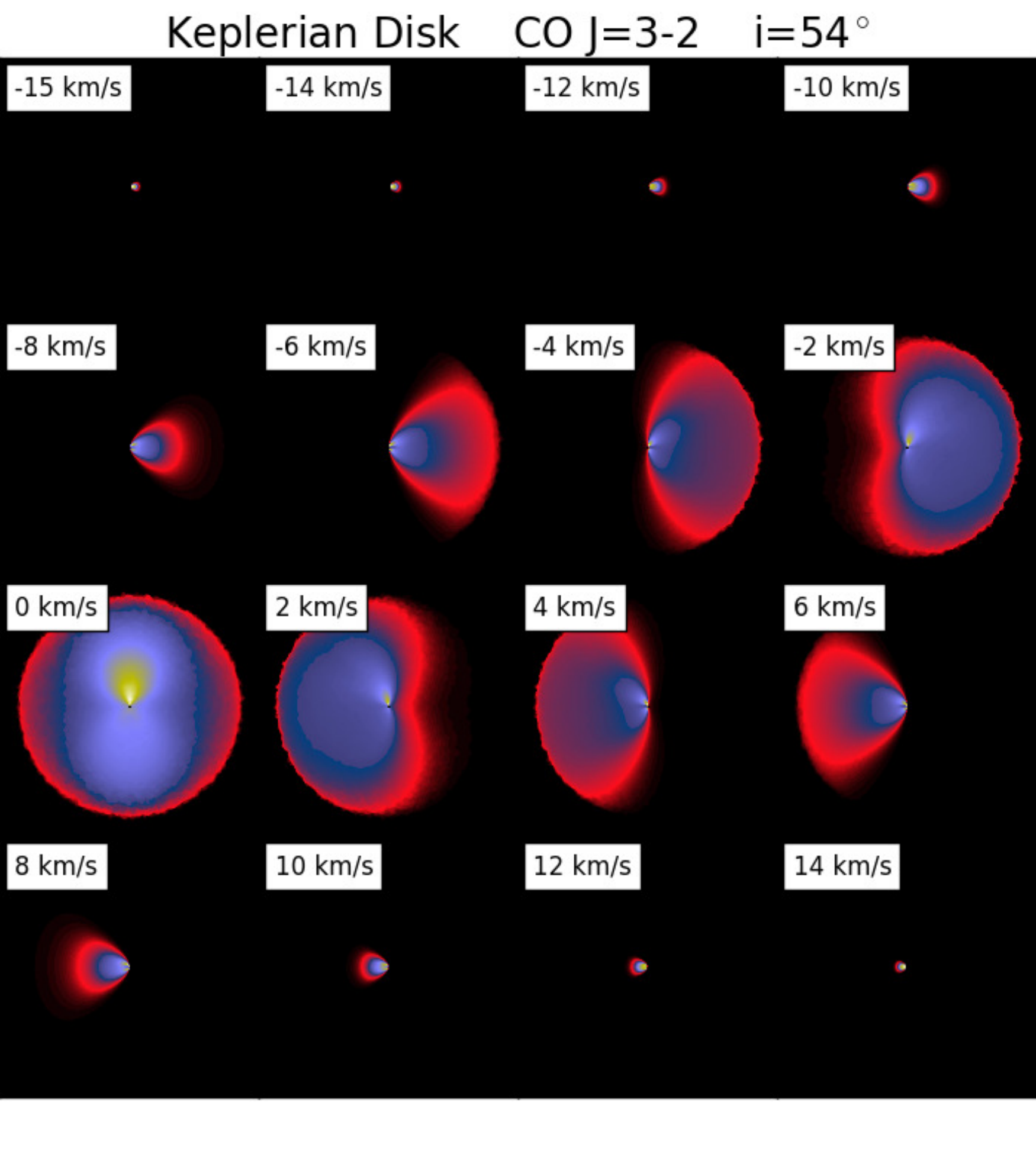}
\caption{Example of a matrix of channel maps of the Keplerian flared EDE model, seen under an inclination angle of $i=54^\circ$. Sect. C in the appendix contains an overview of the channel maps of all calculated models.\label{Kep_Cmap}}
\end{figure}

In order to illustrate this we shall discuss the properties of the channel maps of the Keplerian disk, seen under an inclination angle of $i=54^\circ$ (shown in Fig. \ref{Kep_Cmap}). We choose this particular inclination because one is most likely to observe objects with cylindrical symmetry under such or comparable inclination angles. Additionally, the main properties of the emission features are very similar for all other inclination angles which do not approach $i=0^\circ$ or $i=90^\circ$. The emission generated by the center of the EDE is found to be maximally Doppler-shifted, because it contains the highest speeds in the differentially rotating EDE. The right side of the EDE contains blue-shifted emission, and the left side redshifted emission, allowing for a direct deduction of the direction of rotation. As velocities approach the central velocity, the emission zone increases. This is due to the geometry of the EDE which, being flared, contains the material with lower projected velocities (along the line of sight) in a bigger volume. Around zero velocity a small asymmetry is seen in the intensities along the vertical offset axis. This is due to the geometrical implications of penetrating any inclined EDE with parallels to the line of sight. These parallels will probe different density regions for identical, but opposite, vertical offsets. The fact that the highest emission has a positive vertical offset indicates that this is the side of the EDE closest to the observer. The inclination orientation can thus be deduced from this difference in emission. 

\subsubsection{The Keplerian disk in the PV diagrams}
\begin{figure*}[htp]
\centering
\includegraphics[width=0.9\textwidth]{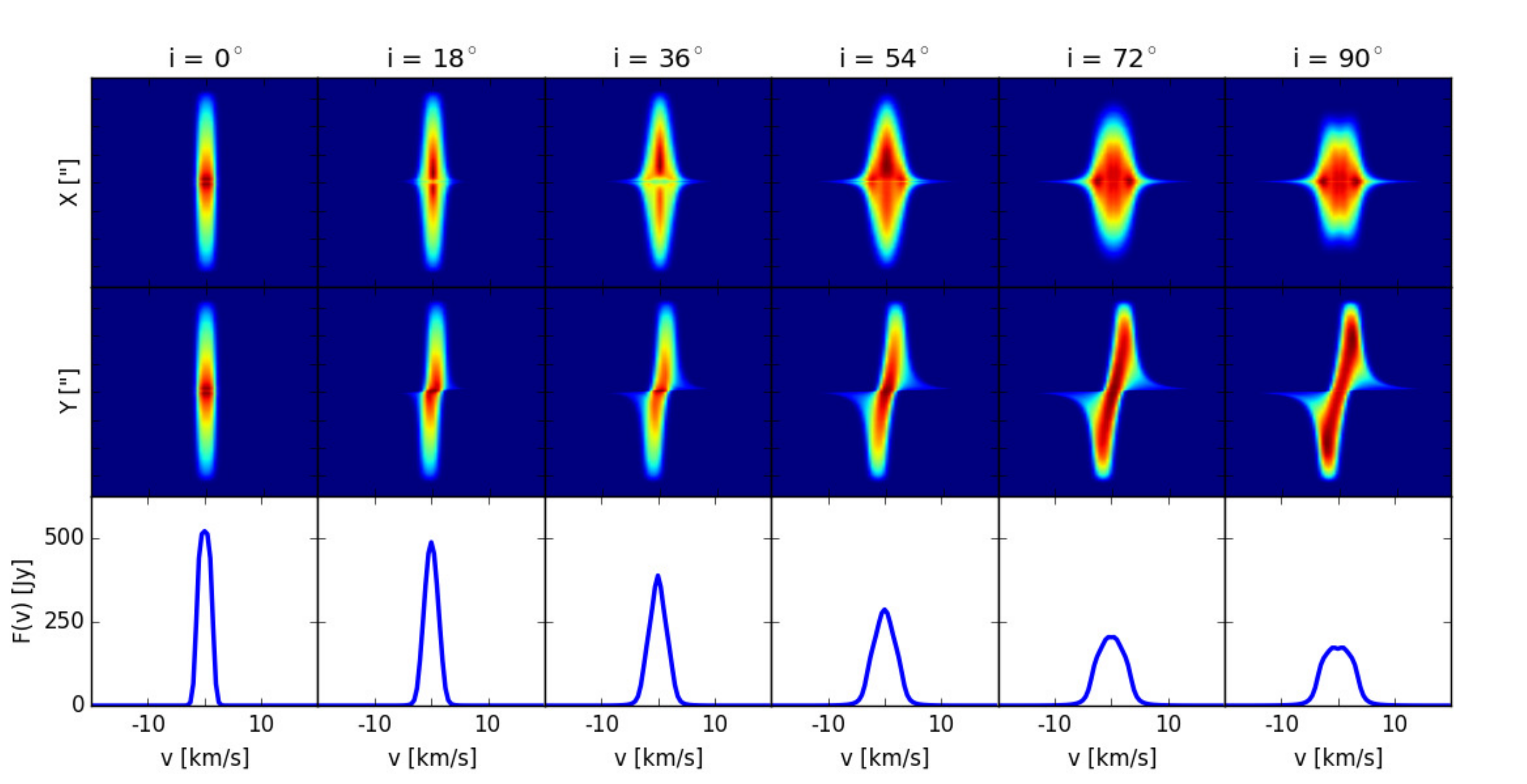}
\caption{Dependence on inclination of the PV diagrams (\emph{top two rows}) and the spectral lines (\emph{bottom row}) of the Keplerian EDE model. \label{Kep}}
\end{figure*}

When the data is noisy and/or when the embedded morphology is perturbed by local dynamics, position-velocity (PV) diagrams may be a better suited tool to analyse the three-dimensional data. Fig. \ref{Kep} shows the PV diagrams of the Keplerian EDE model as a function of inclination. Seen face-on ($i=0^\circ$), the EDE produces a narrow vertical emission band around the central velocity. This is due to the velocity field of the EDE. Being Keplerian, this means that there is no component of the velocity along the line of sight. The width of the signal is exclusively a manifestation of the turbulent velocity component of the velocity field. Because the face-on EDE model is perfectly symmetrical, both PV diagrams are identical. As the inclination increases towards an edge-on vantage point, the behaviour of the two PV diagrams diverges. The EDE emission in PV1 broadens, and its vertical size diminishes substantially. This can be explained by the elliptical aspect of a tilted EDE, projected on a plane perpendicular to the line of sight. If the object is not observed edge-on or face-on, the inclination direction can be deduced from the difference in emission between the positive and the negative offset side in PV1. The side with the higher emission is probing denser and hotter regions, meaning that this is the side which is oriented towards the observer. The inclination dependence of PV2 is different. The EDE contribution does not contract vertically, but slowly shifts in velocity space. The emission with positive offset shifts towards the red, while the emission with negative offset shifts to the blue. This shift originates in the rotation direction of the inclined EDE. Additionally, some EDE emission around zero offset gets stretched throughout the entire velocity space. This is created by the material with a very high tangential velocity, which resides very close to the center of the EDE. 

Fig. \ref{Kep} also displays the spectral lines corresponding to each inclination angle. These show a very narrow and bright peak around zero velocity for the face-on EDE. As the inclination increases, the strength of the line decreases, while its width gradually increases. The spectral lines provide very limited information on the specific geometric details of the observed object. This conclusion shall repeat itself throughout the paper.

\subsection{Effect of the velocity field on the emission}

In this section we will quantify the effect of modifying the global nature of the velocity field on the molecular emission patterns, addressing the cases of the radial field, the super-Keplerian field, the sub-Keplerian field and the rigid rotator. For the sake of simplicity, we use the same geometrical setup as described in Sect. 2.2.1 for each case. The effect of the different velocity fields on the emission redistribution is extremely pronounced. In fact, we shall confine the discussion only to the PV diagrams because they enable us to deduce all the information contained in the channel maps (which can be found in Appendix A).

\begin{figure*}[htp]
\centering
\includegraphics[width=0.9\textwidth]{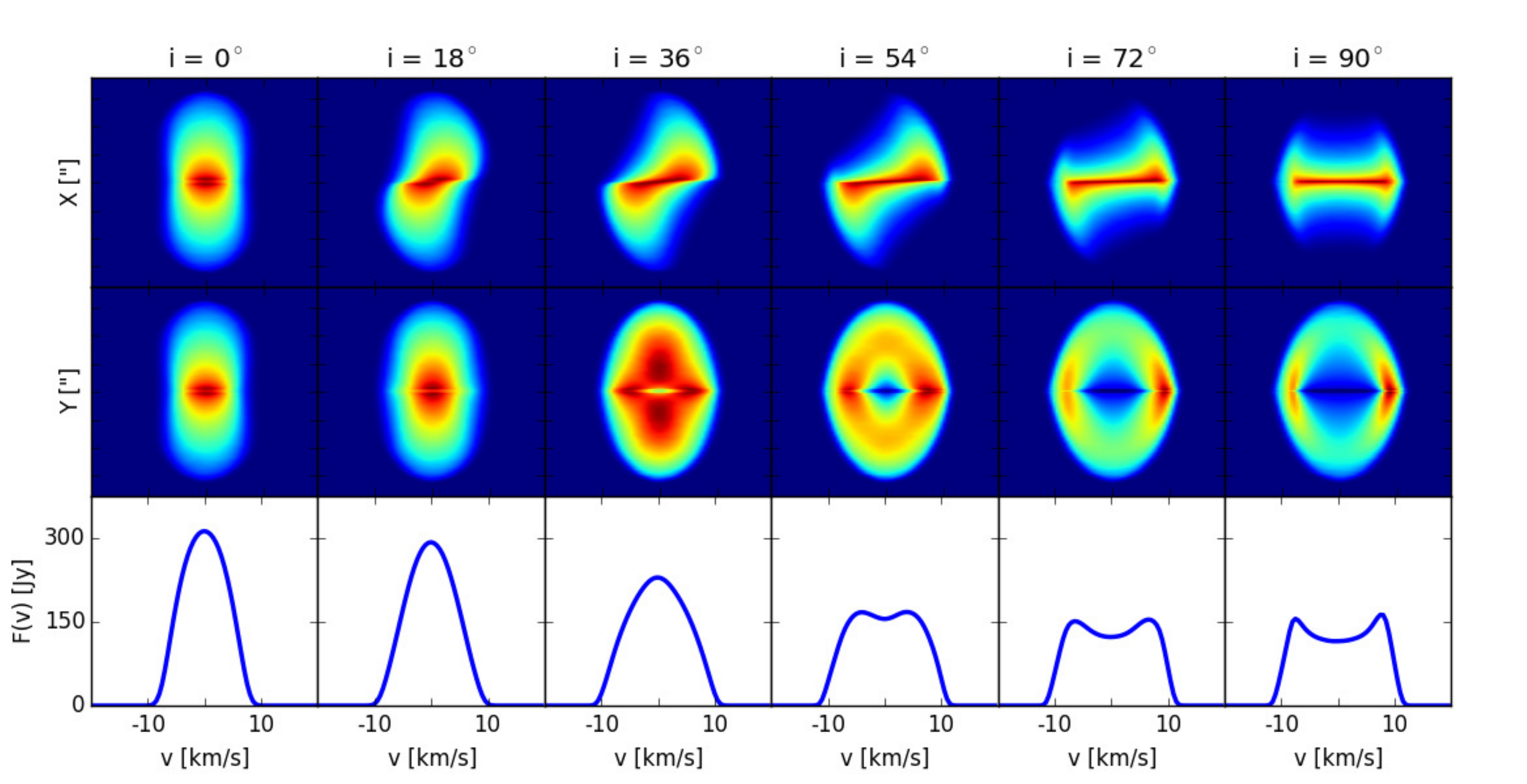}
\caption{Dependence on inclination of the PV diagrams (\emph{top two rows}) and the spectral lines (\emph{bottom row}) of the EDE model with a constant radial velocity. \label{RadVel}}
\end{figure*}

\begin{figure*}[htp]
\centering
\includegraphics[width=0.9\textwidth]{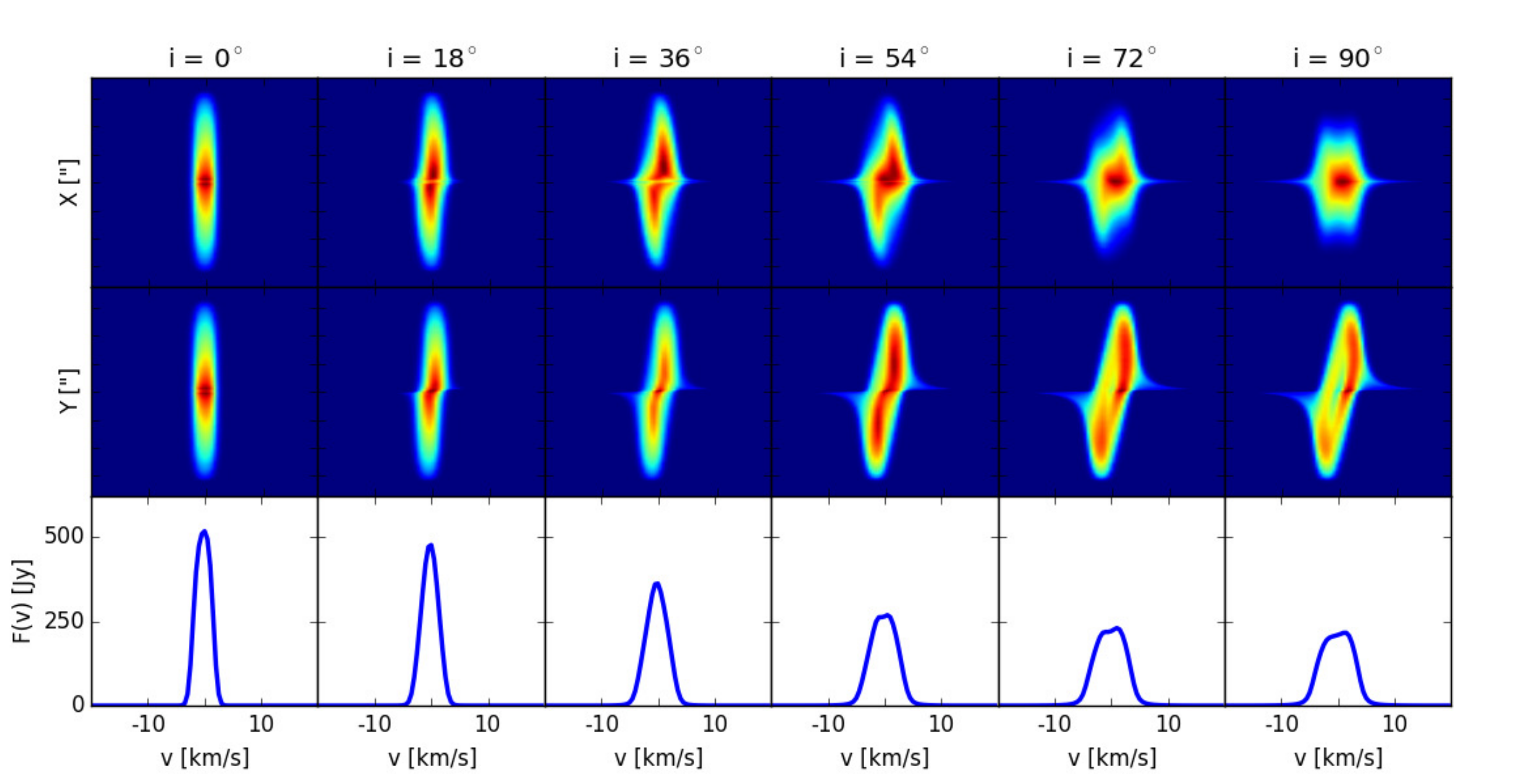}
\caption{Dependence on inclination of the PV diagrams (\emph{top two rows}) and the spectral lines (\emph{bottom row}) of the super-Keplerian EDE model. \label{SupKep}}
\end{figure*}

\begin{figure*}[htp]
\centering
\includegraphics[width=0.9\textwidth]{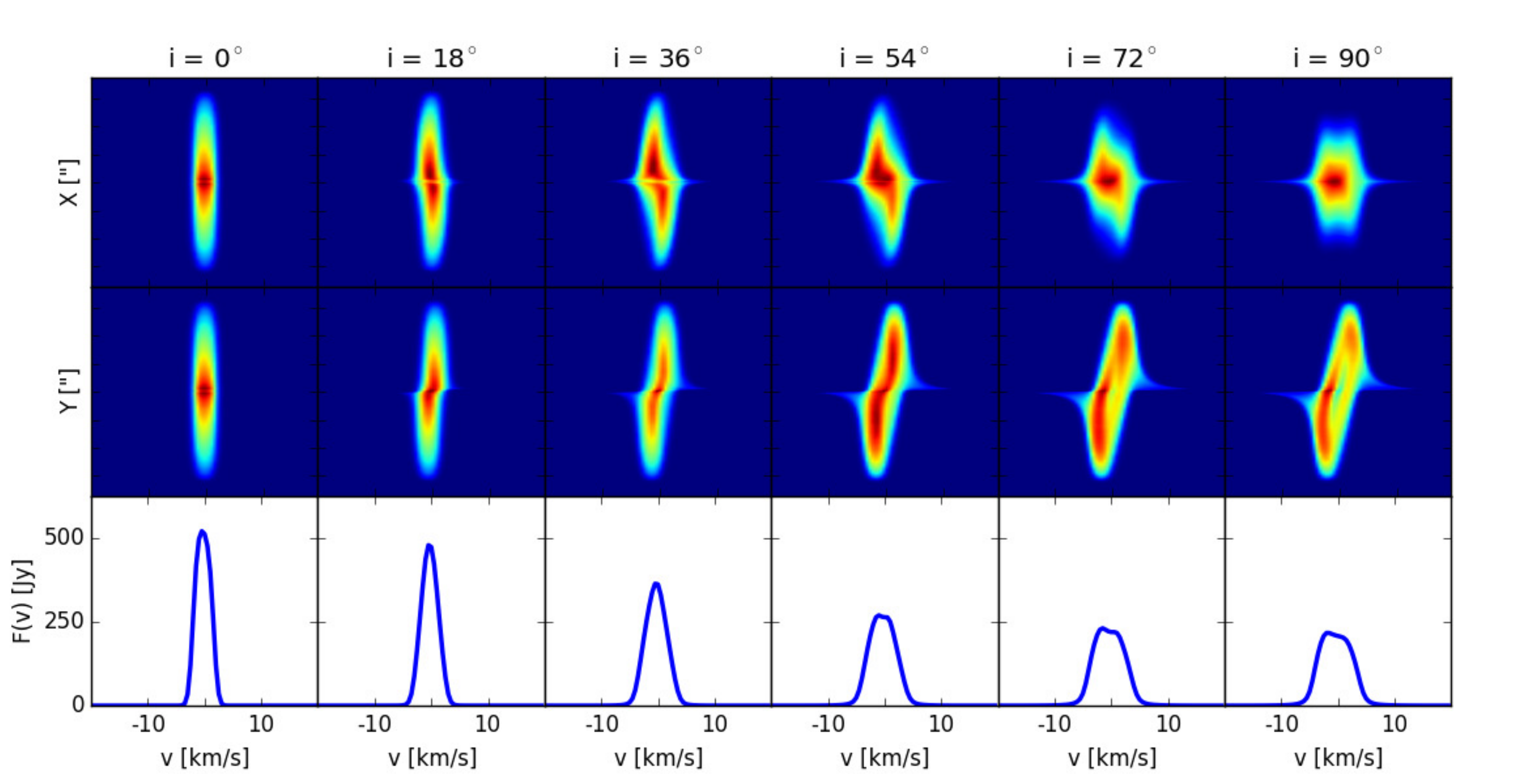}
\caption{Dependence on inclination of the PV diagrams (\emph{top two rows}) and the spectral lines (\emph{bottom row}) of the sub-Keplerian EDE model. \label{SubKep}}
\end{figure*}

\subsubsection{Radial EDE velocity}
We have simulated an expanding EDE, of which the PV diagrams and spectral lines are shown in Fig. \ref{RadVel}. Here, the expansion velocity of the EDE is 10 $\kms$. The overall morphology of the PV diagrams is very different from the Keplerian EDE. Seen face-on, the PV diagrams show a much broader central feature, which arises from the fact that the EDE emission region now contains a greater range of projected velocities along the line of sight. With increasing inclination PV1 shows a transition of the top half of the emission towards the red-shifted side, and the bottom side of the emission towards the blue-shifted side. This shift is caused by the orientation of the tilted EDE. Additionally, the emission is compressed vertically. Like for the Keplerian EDE, this is because the projected height of the cylindrical EDE diminishes as its orientation approaches a more edge-on position. Finally, in the edge-on view, the EDE appears as a strong and narrow emission feature centred around zero offset, extending over the whole of velocity space. 
In PV2 the emission rapidly stretches out towards terminal velocity, with an eye-shaped opening appearing in its center. This eye-shaped feature increases in size, whilst the highest emission zones migrate to the highest velocity regions. The resulting edge-on emission morphology in PV2 becomes progressively ring-like. This ring-like pattern is caused by a combination of the flaring of the EDE and the line-forming region, which both have the tendency to enhance the emission in specific regions of the EDE. Regarding the spectral lines, when viewed face-on, this expanding EDE generates a parabolic spectral feature, with a width which remains slightly more narrow than the 10 $\kms$ input expansion velocity. This feature flattens and broadens with increasing inclination angle turning increasingly double-peaked.

\begin{figure*}[htp]
\centering
\includegraphics[width=0.9\textwidth]{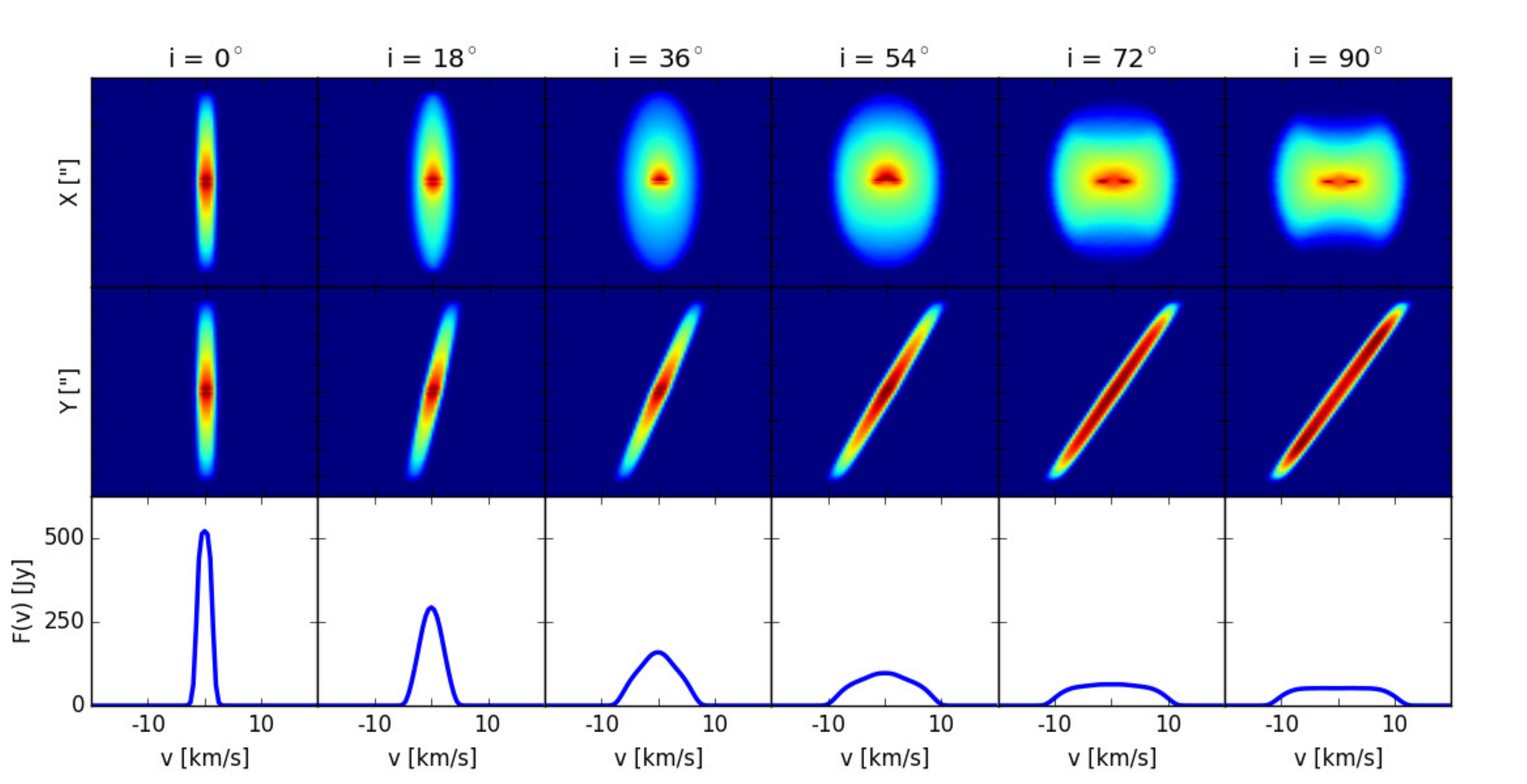}
\caption{Dependence on inclination of the PV diagrams (\emph{top two rows}) and the spectral lines (\emph{bottom row}) of the rigidly rotating EDE model. \label{RigRot}}
\end{figure*}

\subsubsection{Super-Keplerian rotation}
The morphology of the molecular emission for the super-Keplerian EDE (Fig. \ref{SupKep}) is similar to the emission signatures of the Keplerian disk. This is not surprising, because our velocity field is dominated by the differentially rotating component. However, the effect of the weaker radial component does not go unnoticed in the PV diagrams. The behaviour of PV1 as a function of inclination is comparable to the EDE with the Keplerian velocity field. However, due to the weaker velocity components of the radial field, the emission is slightly redistributed. The emission with a positive offset is moved into the red-shifted side of the diagram, and vice versa. The general morphology of PV2 is also similar to the morphology of PV2 in Fig. \ref{Kep}. In fact, because the emission redistribution in PV2 in Fig. \ref{RadVel} only becomes very characteristic for high inclination angles, the PV2 in Fig. \ref{SupKep} is indistinguihable from the PV2 in Fig. \ref{Kep} for low inclination angles. The highest inclination angles clearly show some 'internal substructure', with a central decreased emission, which is not present in the Keplerian PV diagrams. Due to the relatively low magnitude of the radial velocity component, the small amount of high-velocity emission, generated by the most inner regions of the EDE, where the Keplerian field dominates, is unchanged. The spectral lines are comparable to the spectral lines of the EDE with a Keplerian velocity field. This is because the rotating velocity field dominates over the radial field throughout most of velocity space. In the regions in velocity space where the emission is strongest, both velocity components are of comparable magnitude, with a combined effect.

\subsubsection{Sub-Keplerian rotation}
The emission signature of an accretion EDE is shown in Fig. \ref{SubKep}. Comparing these emission signatures with the super-Keplerian case some strong similarities are immediately noticable. In fact, the PV2 diagrams are identical point-reflected versions of each-other, and for the PV1 the shape of the emission is mirrored around zero velocity. This is because the projected components of a global negative radial velocity component are, in essence, visually completely equivalent to the super-Keplerian case with an inverted rotation direction. Essentially, the blue-shifted emission of the super-Keplerian case is transformed to red-shifted emission here, and vice-versa, by the introduction of this negative radial velocity component. The effect on the line profiles is negligible compared to the lines in Fig. \ref{SupKep}, for the same reasons as outlined for the super-Keplerian velocity field. This is to be expected, because the line profiles in the super-Keplerian case are symmetrical, therefore any mirroring around zero velocity leaves them unchanged.

\subsubsection{Rigid rotation}
We present the spectral results of a rigidly rotating EDE, rotating with a constant angular velocity, and a tangential velocity of 10 $\kms$ at the outer edge of the model in Fig. \ref{RigRot}. Face-on the PV diagrams do not differ from the Keplerian EDE. In PV1, an increasing inclination angle broadens the strong emission feature of the EDE and compresses it vertically. Again, this is a direct consequence of the flatter appearance of an inclined circular object projected onto a plane perpendicular to the line of sight. The effect of inclination on PV2 is strong, and results in an effective clockwise rotation of the narrow vertical emission feature (for the face-on EDE) to an equally narrow feature, rotated by a $45^\circ$ angle when viewed edge-on. This is essentially caused by the fact that the velocity increase with radius exactly cancels out the projection effects. The projected velocities of the outer regions of the EDE are identical to the projected velocities of all the material along the same line of sight. The effect on the spectral lines is pronounced, compared to the models above. Face-on, the spectral line does not differ from the Keplerian disk. However, as the inclination angle increases, the peak strength becomes substantially smaller, the line broadens and flattens, until (when edge-on) it becomes an outspoken flat-topped profile.

\subsection{Effect of EDE substructure}

In this section, we investigate the influence of additional density substructure on the emission patterns. In order to put reasonable limits on the available parameter space, we shall confine the discussion to the Keplerian EDE. We investigate the effect of EDE warping, annular gaps and spiral-shaped instabilities in the EDE on the molecular emission patterns. We shall discuss the channel maps here, but (in order to avoid the repetition of figures) refer the reader to Appendix A, where all channel maps are shown. A strong general characteristic of the channel maps of all three cases is the strong domination of the Keplerian velocity field. This is recognised by the strong overall morphological similarities between the channel maps of the EDEs with substructure and the Keplerian EDE without any. We shall discuss the effect of the substructure on the channel maps and on the PV diagrams per individual case. However, we have found that the resulting signatures in the PV diagrams are minimal, and are simply non-existent in the spectral lines. This is mainly due to the fact that the spectral lines and the wide-slit PV diagrams reprocess the spatial axes, which leads to a concealment of the substructure due to intensity-averaging. Only if the data is of high enough quality will the substructure be immediately apparant from the channel maps.

\begin{figure*}[htp]
\centering
\includegraphics[width=0.9\textwidth]{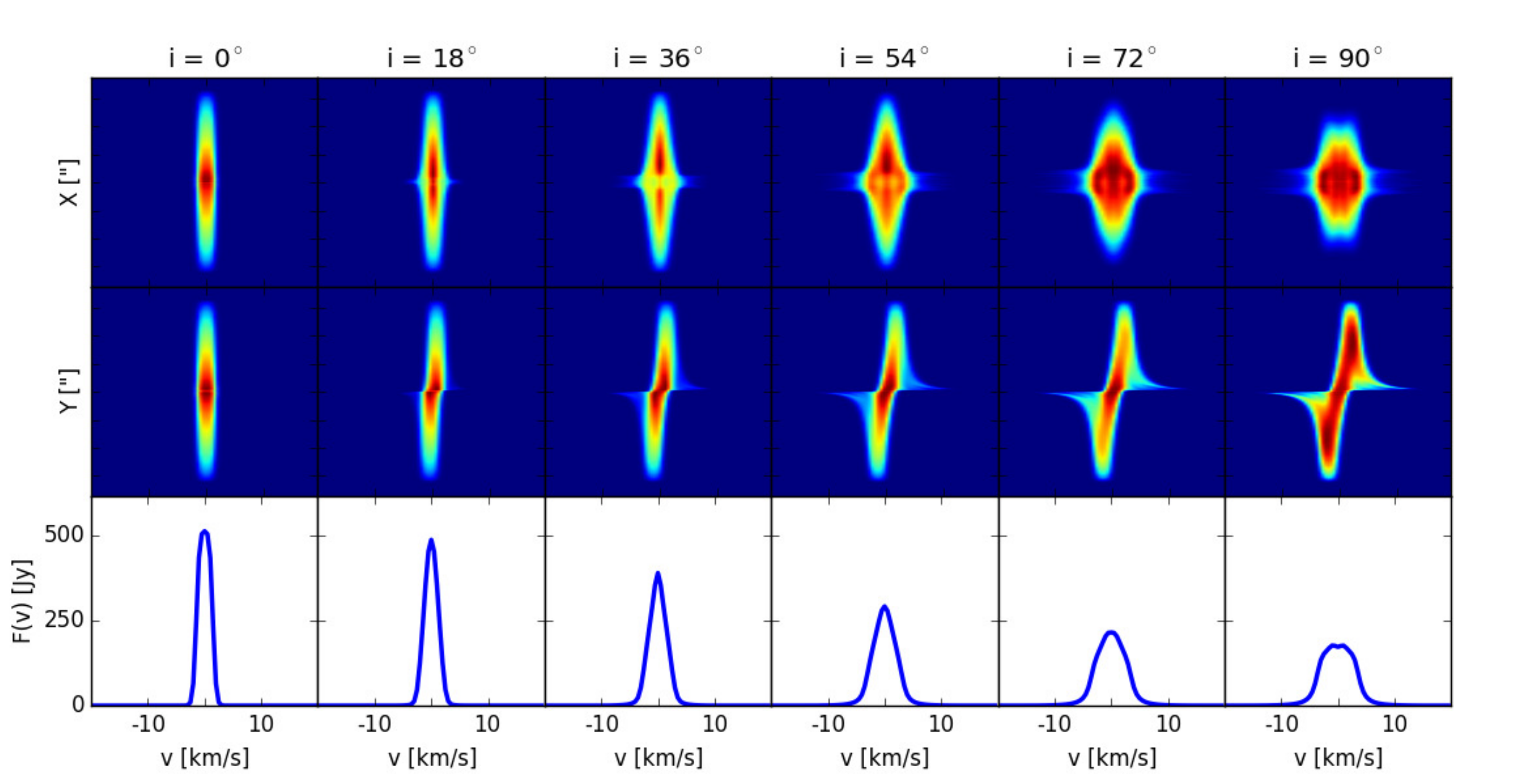}
\caption{Dependence on inclination of the PV diagrams (\emph{top two rows}) and the spectral lines (\emph{bottom row}) of the Keplerian EDE model containing a warp. \label{Warp}}
\end{figure*}

\begin{figure*}[htp]
\centering
\includegraphics[width=0.9\textwidth]{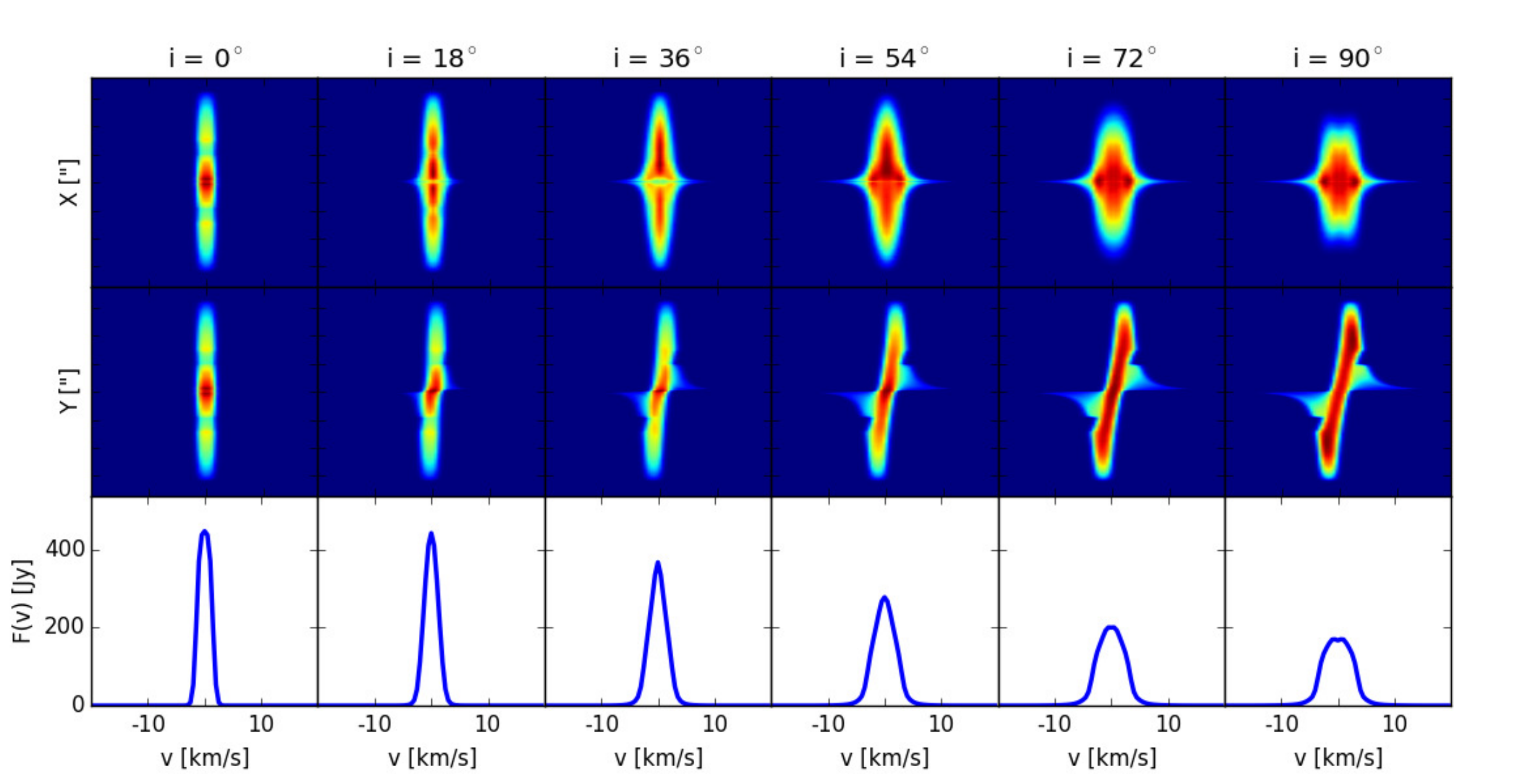}
\caption{Dependence on inclination of the PV diagrams (\emph{top two rows}) and the spectral lines (\emph{bottom row}) of the Keplerian EDE model containing an annular gap. \label{Gap}}
\end{figure*}

\begin{figure*}[htp]
\centering
\includegraphics[width=0.9\textwidth]{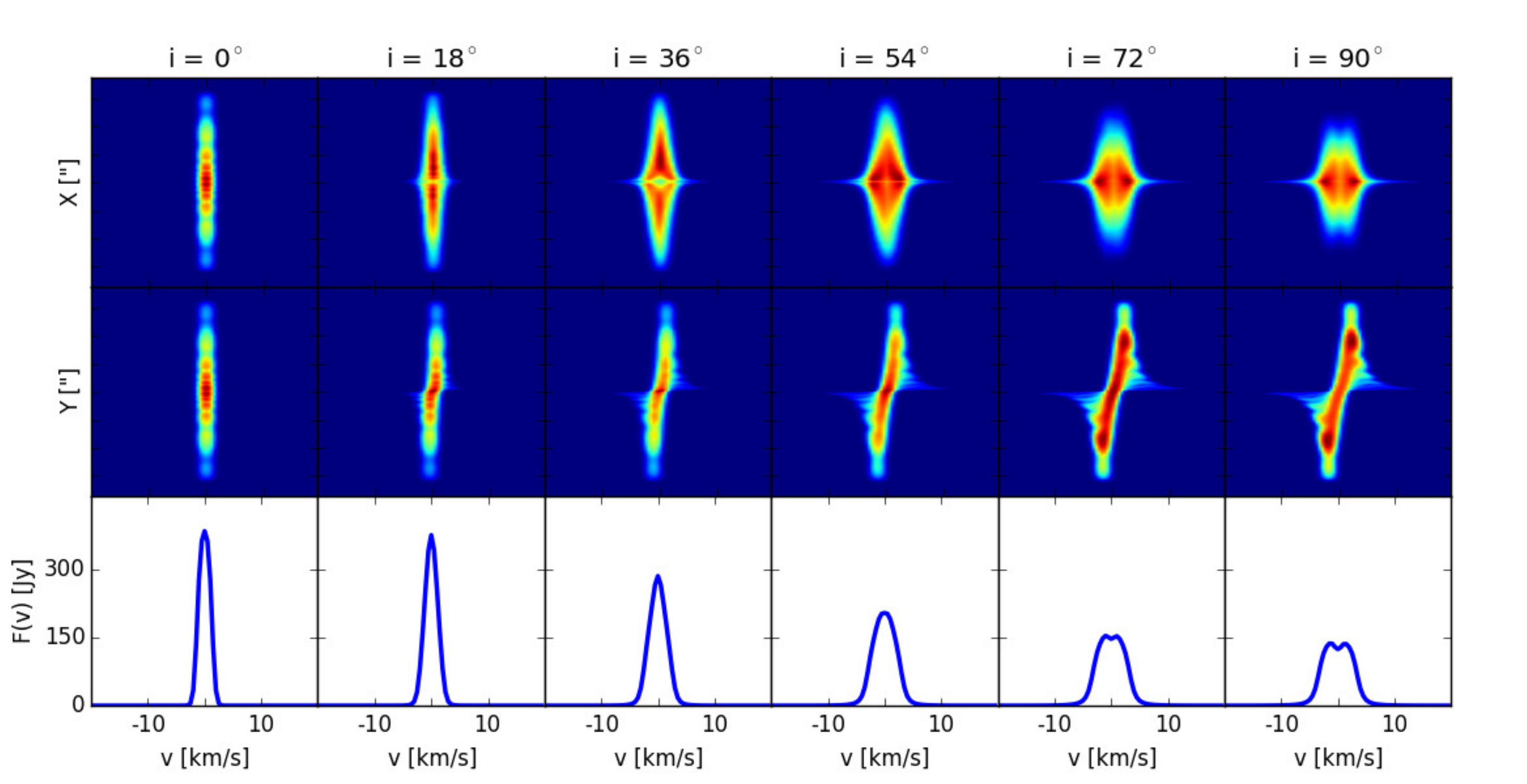}
\caption{Dependence on inclination of the PV diagrams (\emph{top two rows}) and the spectral lines (\emph{bottom row}) of the Keplerian EDE model with added spiral-shaped instabilities. \label{Spir}}
\end{figure*}

\subsubsection{Warped EDE}
Introducing a warping of the EDE does not radically alter the appearance of the channel maps. Here, we discuss an EDE with $L$=3 undulations, and with an amplitude A of one eighth of its radius. The lack of a clear warping signature in the channel maps (see Fig. \ref{warp_chan}) seems at first sight surprising. But this result can be easily explained by a combination of projection effects. The projected velocity of the EDE along the line of sight completely masks the warping for objects with an uneven number of undulations, when their inclination deviates from a face-on view. This is because the peaks of the warping of the blue-shifted side of the EDE have the same projected velocity as the valleys of the red-shifted side of the EDE, and vice versa. The warped EDE thus appears as a thicker unwarped EDE in the 3D datacube. EDEs with a warp consisting of an even number of undulations may not suffer this fate. However, this is only the case when the equatorial orientation of the EDE is such that the blue-shifted valleys coincide perfectly with red-shifted valleys (and the same for the peaks). In this case the 3D data will exhibit a wave-like pattern in the EDE emission. It is, however, much more likely to observe a warped EDE with an even number of undulations under an equatorial orientation which does not perfectly coincide with this condition. In that case (and thus in most cases) warped EDEs are thus more likely to manifest as a broader and unwarped in the emission.

When constructing a wide-slit PV diagram all the emission at the chosen position-velocity coordinate is summed, and any previously present pattern is lost, as shown in Fig. \ref{Warp}. In both the evenly and unevenly warped EDE the resulting emission feature in the PV diagrams is thus wider than the feature expected from an unwarped EDE. This can clearly be seen by comparing (the not face-on) PV1 of the warped EDE with the equivalent PV diagram of the Keplerian disk model (Fig. \ref{Kep}). Some small features can be found to differ between this morphological model and the Keplerian EDE. However, these differences are so small we expect them to be undetectable in real data. Again, like for the other morphologies, the presence of warping is not noticable in the spatially unresolved case. We refer to the Keplerian disk (Fig. \ref{Kep}) for a discussion of the spectral lines.

\subsubsection{EDE with an annular gap}
In Fig. \ref{gap_chan} we exhibit the effect of introducing an annular gap into the Keplerian disk model on the velocity channel maps. The gap has a width of 15\% of the EDE radius, and is located at a distance of 25\% of the EDE radius from the center. The effect of the gap is predictable. The gap generates an annular zone void of emission, clearly visible in the channel maps of the $i=0^\circ$ case. Focusing on the $v=0 \kms$ channel, we find that as the inclination increases, the projected gap size diminishes along the inclination axis, until the gap signal is completely hidden by the material which is located between the gap and the observer. However, the gap reappears at higher velocities (in this specific case around $v=\pm 6-8 \kms$) as it disappears at lower velocities. The emission gap is thus always present in the channel maps, and taking into account the inclination an estimation of its size can be readily performed.

The overall morphology of the PV diagrams is nearly unaltered by the introduction of a gap (Fig. \ref{Gap}). The PV diagrams of the face-on EDE nicely show a discontinuity in the emission at the offset at which the gap is located. As inclination increases, this discontinuity turns into a wedge of 'missing emission' in PV2. Higher inclination angles effectively erase any effect of the gap on the morphology of PV1. The annular gap has no noticable effect on the shape of the spectral lines. There is, however, a decrease in flux in the EDE contribution, as expected. If the probed rotational transition just so happens to be mainly produced in the region of the gap, this decrease may be substantial. However, such a spectral line would be morphologically indistinguishable from a gapless EDE, with an adapted mass.

\subsubsection{EDE with spiral-shaped instabilities}
Fig. \ref{spir_chan} shows the channel maps for the EDE with spiral-shaped density-enhancements, as a function of inclination. For low inclination angles the presence of the spiral is very clear. As inclination increases, the spiral in itself becomes less pronounced. In fact, for the highest inclination angles the spiral may very well be misintepreted as a series of narrow annular gaps. The only indication that it may be a logarithmic spiral is the rate at which the gaps appear as a function of offset. This conclusion propagates through to the PV diagrams.

In Fig. \ref{Spir} the PV diagrams are shown. PV2 shows a sequence of enhanced emission zones for low inclination angles, which turn into wedge-like cuts on the outside of the EDE emission for higher inclinations. In fact, the observed effect of introducing spiral-shaped density enhancements to the EDE is indistiguishable from introducing annular gaps at different radii, for any inclination angle. However, in the case of a logarithmic spiral, it can be seen that the angular size of the enhanced emission (in the low-inclination PV diagrams) or of the wedges (for the higher-inclination PV diagrams) decreases towards zero offset. Additionally, the rate at which these features appear increases towards zero offset. This can either be interpreted as a sequence of annular gaps, which decrease in frequency but increase in size with radius, or as logrithmic spiral-instabilities. The latter thus indicates the likely presence of spiral shapes. The effects on the spectral lines of introducing logarithmic spiral-instabilities in the EDE are comparable to these of annular gap.

\subsection{The emission of the bipolar outflow}

\subsubsection{The bipolar ouflow (without EDE)}
\begin{figure}[htp]
\centering
\includegraphics[width=0.45\textwidth]{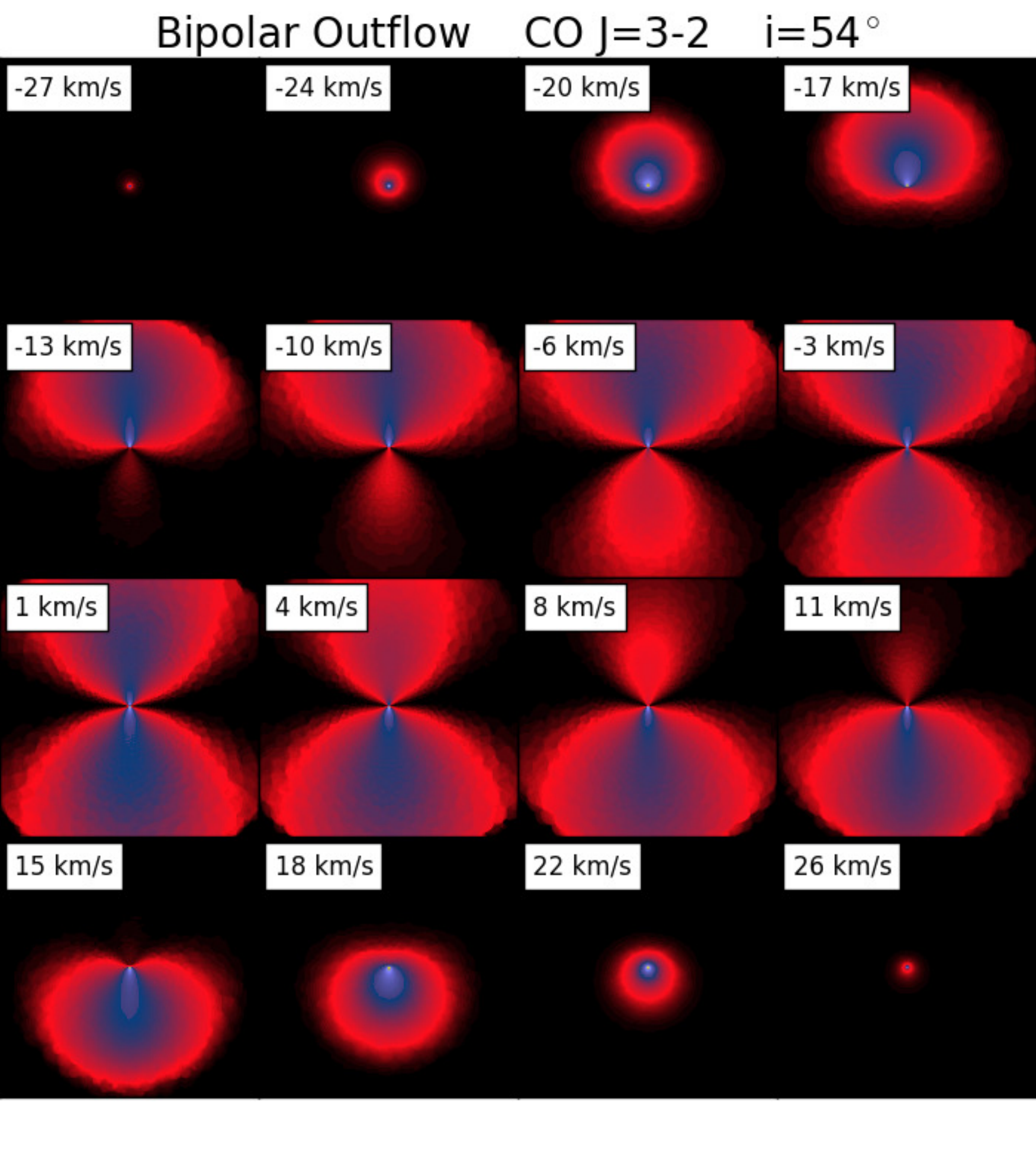}
\caption{Example of a matrix of channel maps of the bipolar outflow model, seen under an inclination angle of $i=54^\circ$. Sect. C in the appendix contains an overview of the channel maps of all calculated models. \label{BipOut_Cmap}}
\end{figure}

Fig. \ref{BipOut_Cmap} shows the channel maps of the intrinsic emission of a bipolar outflow, observed under an inclination angle of $i=54^\circ$. The reasoning behind this specific choice of angle follows the same argumentation as in Sect. 3.1.1. At high velocities the emission is strong, but the emission area is minimal. Only material with exclusively positive or negative offset is visible. As velocities approach zero, the channel maps present the portion of the bipolar outflow which is oriented towards the observer. In this case it is the portion of the outflow with a positive vertical offset which contains the blue-shifted emission, and the portion with the negative vertical offset which contains the red-shifted emission, clearly showing the orientation of the inclined hourglass. Around central velocity the bipolar outflow is symmetrical.

\begin{figure*}[htp]
\centering
\includegraphics[width=0.9\textwidth]{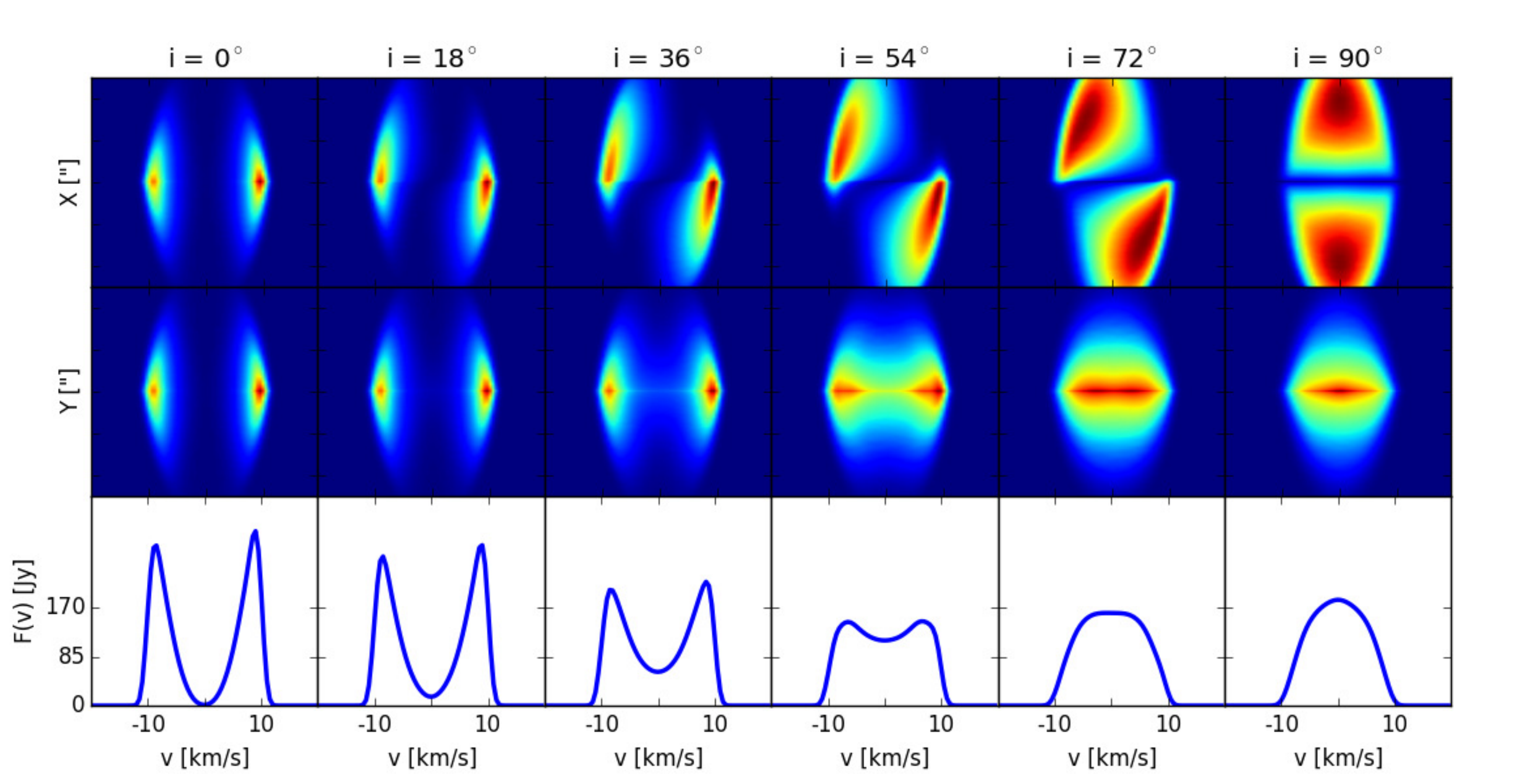}
\caption{Dependence on inclination of the PV diagrams (\emph{top two rows}) and the spectral lines (\emph{bottom row}) of the bipolar outflow model. \label{BipOut}}
\end{figure*}

Fig. \ref{BipOut} shows the PV diagrams of the bipolar outflow model as a function of inclination. Seen face-on ($i=0^\circ$), the bipolar outflow produces two emission features at high velocities. Along the velocity axis, a region void of emission resides between these two components, which is easily explained by the lack of material in the equatorial regions. Of course, because the face-on bipolar outflow is perfectly symmetrical, both PV diagrams are identical. As the inclination increases towards the $i=90^\circ$ vantage point, the behaviour of the two PV diagrams diverges. The emission in PV1 rotates around the center of the image, filling up the velocity-gap, and preserving point reflection symmetry. This is caused by the intrinsic correlation between doppler-shift, and orientation of the hourglass-shape. The inclination dependence of PV2 is different. The emission of the bipolar outflow in PV2 gets stretched-out horizontally towards zero, producing a large feature with a strongly enhanced central bar at $i=90^\circ$. The spectral lines of the outflow start-off as a pronounced double-peaked profile at $i=0^\circ$, after which its peaks slowly diminish in strength. As one approaches the $i=90^\circ$ vantage point, the relative emission contribution of the central velocity regions augments until these regions dominate the spectral lines, ultimately forming a parabolic line profile which is slightly narrower than the $i=0^\circ$ profile. This decrease in line width is explained by the decreased maximum-projected-velocity-component for the fully inclined outflow.

\subsubsection{Keplerian disk + bipolar ouflow}
\begin{figure}[htp]
\centering
\includegraphics[width=0.45\textwidth]{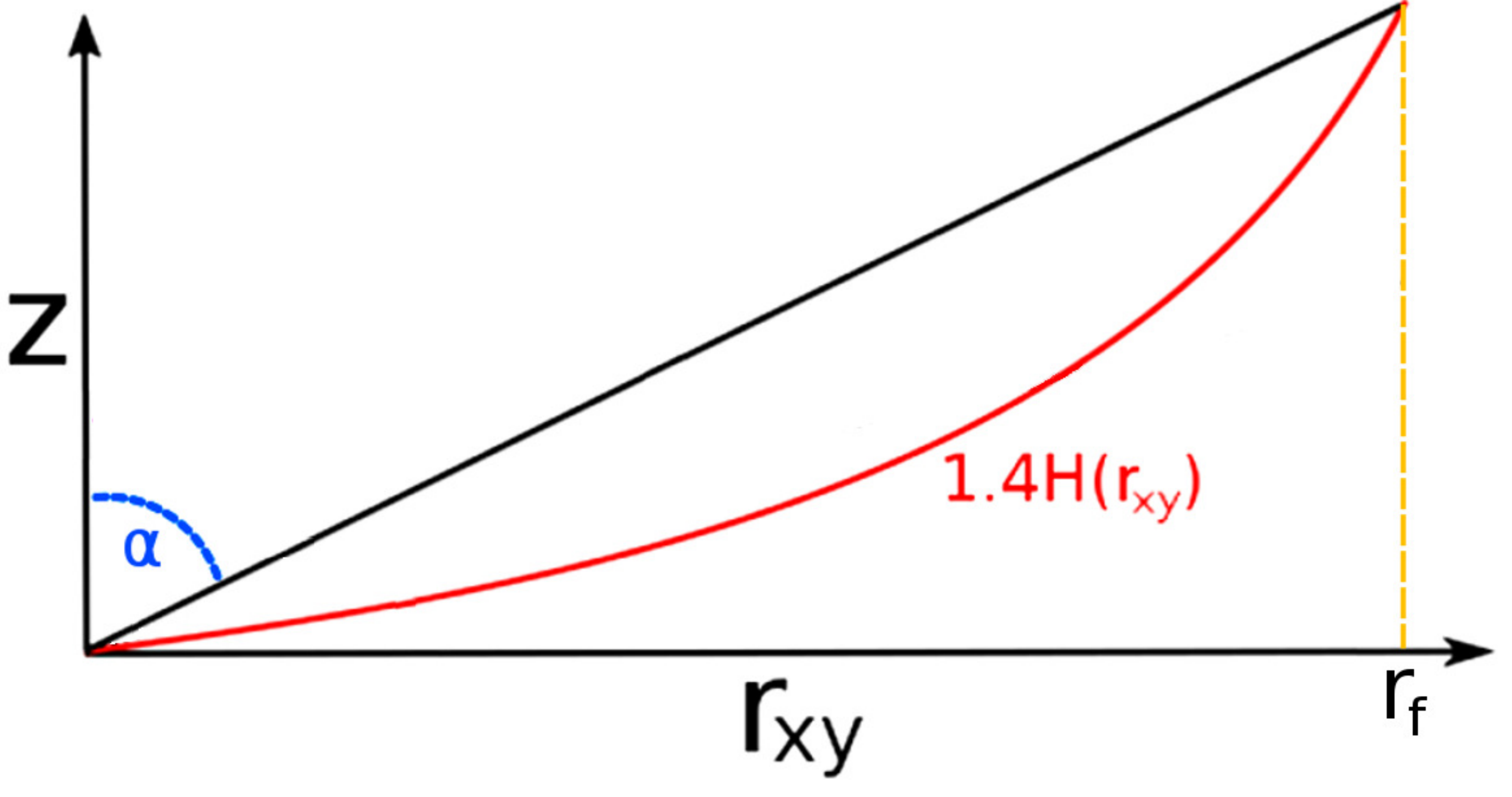}
\caption{Visual representation of the boundary conditions. \label{bound}}
\end{figure}

As mentioned previously, EDEs and bipolar outflows quite often go hand-in-hand in the evolved stars context. We therefore believe a discussion on their combined emission to be paramount for the completeness of this paper. When combining both the EDE and the bipolar outflow in a single radiative transfer model, we require a condition by which we can numerically differentiate between these components. One straightforward criterion is by defining the boundary as the spatial coordinates where the density of the EDE and the density of the bipolar outflow are identical. However, this introduces numerical anomalies for certain contrasts between outflow mass and EDE mass. We have therefore opted to separate both components by a geometrical criterion. We place the boundary at an angle 
\begin{equation}
 \alpha = \rm{arctan}\left(\frac{r_f}{1.4H(r_f)}\right),
\end{equation}
with $r_f$ the edge of the considered numerical domain, and where $H$ is given by Eq. 10. Cutting off the EDE by confining it within this wedge does not strongly affect its flaring because the higher density flare curve systematically lies below this line. This can be visualised in Fig. \ref{bound}. It does, however, discard the lowest density outer regions of the EDE, accounting for 5.7\% of the total EDE mass for the chosen $1.4$ prefactor of the denominator. This value has been empirically determined to best resemble the density criterion for the cases where no complications arise. We believe this not to be an issue because, for EDE's with a bipolar outflow, one may expect the outer layers of the rotating EDE to strongly interact with the radial outflow (generating thermodynamical instabilities at the boundary zone), and possibly even the outflow to blow away the outer low-density regions of the EDE. However, due to the complexities associated with defining and modelling such detail, we have opted not to include them.

\begin{figure*}[htp]
\centering
\includegraphics[width=0.9\textwidth]{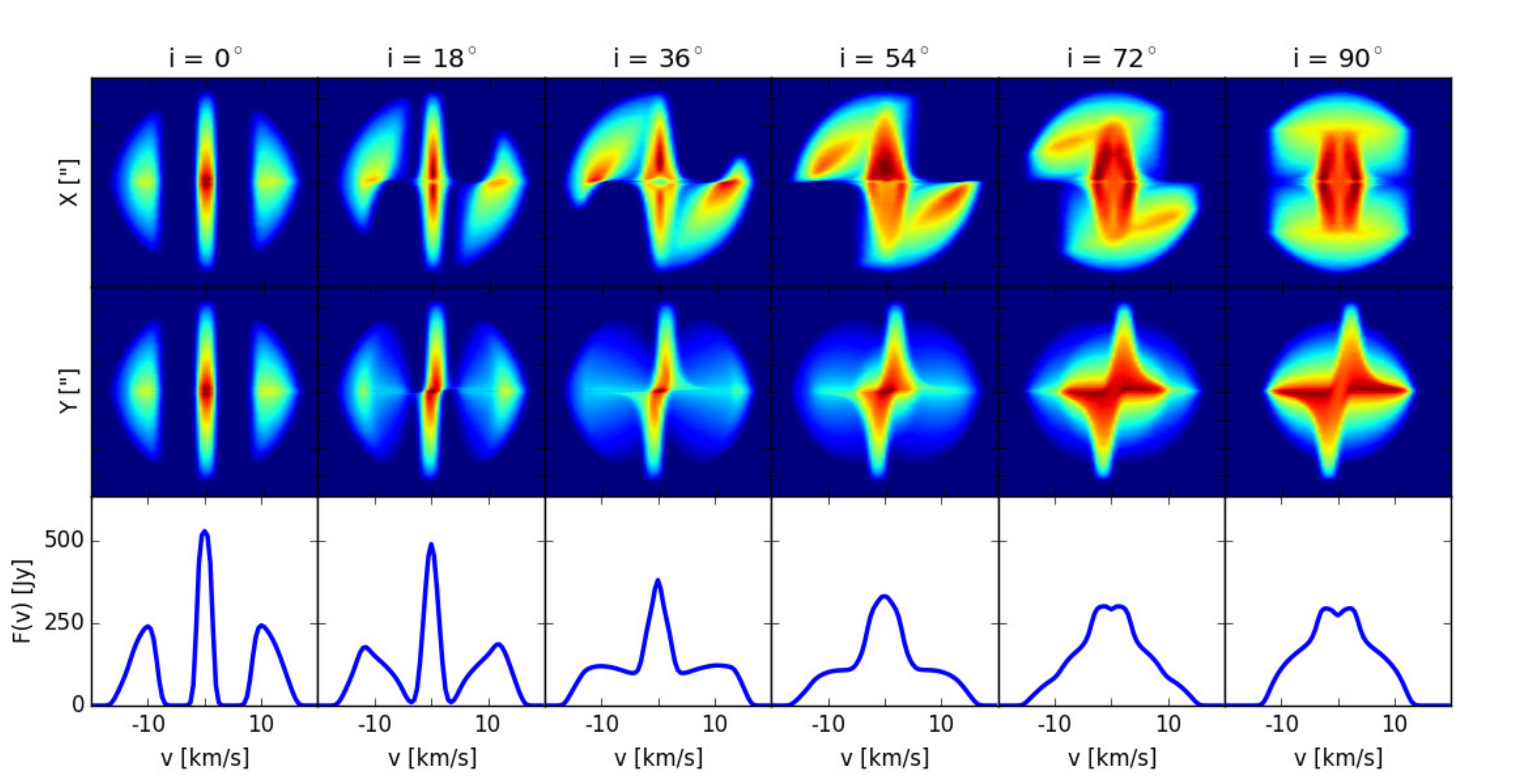}
\caption{Dependence on inclination of the PV diagrams (\emph{top two rows}) and the spectral lines (\emph{bottom row}) of the bipolar outflow model. \label{Comb}}
\end{figure*}

Fig. \ref{Comb} shows the PV diagrams of the combined Keplerian disk + bipolar ouflow model. Around $i=0^\circ$ the emission signature of the disk is clearly recognisable. This is because it is well separated (in velocity space) from the bipolar outflow emission features. This gap is caused by the previously described boundary condition criterion, which does not ensure continuity in velocity space. At the interface between the outflow and the disk, the velocity field abruptly transitions from radial to tangential. In reality we may expect this boundary to harbour some (thermo)dynamical instabilities, which could generate extra features which may fill-in this emission gap. Due to the separate behaviour of both the disk and the outflow the combined PV diagrams evolve in a very predictable way. However, as one approaches $i=90^\circ$, the signatures begin to overlap, creating emission patterns which may not be so easily disentangled. In PV2 the Keplerian disk signature is still recognisable, but now the addition of the bipolar outflow strongly enhances the regions around zero offset. This generates a much more horizontally elongated emission pattern. In PV1 the vertical size of the disk feature can hardly be recognised to diminish in size. In addition, the two emission features of the bipolar outflow now melt together with the central disk feature, forming a resultant emission distribution that may not be easily recognised as a direct combination of the disk and outflow features. The evolution of the spectral lines is generally more predictable. The lines can simply be recognised as a direct morphological combination of both the disk and the bipolar outflow lines. The dual nature of the combined object can be clearly recognised in all the lines.

\subsubsection{General findings on the combining of geometrically separate emission}
Because we assume the bipolar ouflow and any EDE to be geometrically separate, it is very easy to simulate their combined emission patterns to a good first order approximation (for any of the abovementioned models). If both the EDE and the bipolar wind are optically thin, then the resulting intensity pattern is simply the sum of both individual components. If the EDE is optically thick, and the wind is not, then the emission of the wind originating behind the EDE (along the line of sight) will be obscured, and therefore only the EDE emission should be considered in the overlap regions. {Conversely, when the EDE is optically thin, and the outflow optically thick, the emission originating in the  portions of the EDE oriented towards the observer can simply be added on top of the emission of the outflow. The emission from the portions of the EDE obscured by the outflow can simply be discarded. The only areas where some non-trivial optical depth effects may take place is right at the edges between the EDE emission and the outflow emission. However, these regions are very small compared to the global intensity map, and thus only minimally impact the global shapes present in the PV diagrams. We thus argue that non-trivial optical depth effects (in PV diagrams) are only of marginal importance when interpreting emission of combined EDE and bipolar outflow objects.


\section{Discussion}

\subsection{Model sensitivity to input parameters and distributions}

Because we wish to limit this study to a qualitative investigation of the overall morphological emission trends we have not researched the impact of each parameter on the emission. Nevertheless, we reserve this section for a short discussion on the impact of the model parameters.

The models are most sensitive to the parameters which directly impact the distribution of emission throughout the three dimensions (velocity, X and Y) which make up the datacube. This translates concretely to the global velocity field, and to the overall geometry of the morphology. Assuming a fixed velocity field and geometrical set-up, the general shape of the emission edges (as a function of velocity) will always remain unchanged, both in the channel maps and in the PV diagrams. This means that changes to the internal properties of the system only manisfest as a redistribution of the internal (confined by the edges) emission of the data. The most important of these internal properties are the density distribution, the temperature field and/or the molecular abundance (henceforth referred to as DTA). A very important point to address here is that DTA are fully degenerate when modelling CO emission. This is because the molecule has an extremely low electric dipole moment. The excitation of the CO energy levels is therefore mainly stimulated by collisions (meaning that it is therefore also insensitive to the radiation field of the central star and/or of the dust, and to the relative abundance and composition of the circumstellar dust). As any change in DTA will guarantee changes in the rates of collision, their effect on the resulting emission can impossibly be disentangled, or discussed separately. An in depth investigation of the available parameter space would therefore result in a virtually infinite discussion.

We can, however, briefly describe the `predictable' qualitative effect of modifying the rate with which any of the simplified radial power-laws describing any of the DTA fields changes the PV emission, by means of two examples. (1) An increased radial decay rate of one of the DTA distributions (whilst leaving the others unchanged) will result in an increased emission contrast between the material at zero offset compared to the material at greater offsets. This increased contrast can be undone by decreasing the rate at which any one of the other DTA distributions decreases radially, due to the abovementioned degeneracy issue. (2) More substantial morphological emission differences may arise when the DTA distributions are organised such that cool and dense material makes up the outer regions of the EDE. Such a scenario will cause absorption of the emission originating from hotter gas located further along the line-of-sight, which may result in concealed (or missing) portions of the expected intrinsic emission distribution at certain velocities.

Regarding the velocity parameters; $M_*$, $v_{rad}$ and $v_{tan}$ will simply stretch the PV signature throughout velocity-space. The effect of $v_{\rm in}$ and $v_{\rm out}$ is less trivial. Increasing their absolute value will open-up a bigger portion of the velocity-space to be influenced by the radial-velocity emission redistribution. As the values increase, the PV signatures will resemble the exclusively radial field more closely. However, the underlying Keplerian field will maintain its influnce in the relevant portion of velocity-space, most strongly manifested as a tilt of the `radial signature'.

\subsection{Considering extra transitions of CO}

In addition to the $J$=3$-$2 rotational transition in the lowest vibrational energy level of CO, we have considered the 4 subsequent rotational transitions for the Keplerian model, all the way up to the highest transition that can be observed with \emph{ALMA} ($J$=7$-$6). We have not modelled the two lowest rotational transitions to avoid having to make assumptions on the onset of photodissociation far from the central source. The results are shown in Fig. \ref{CO}. Higher $J$ transitions are formed in hotter regions, and thus probe the denser inner regions of the model. In this particular case the chosen temperature profile and EDE size are such that the higher energy levels of CO are more effectively populated. We therefore measure an increased emission for these higher $J$ transitions. When viewed face-on, the projected velocities all tend toward zero, leaving the turbulent velocity as the only broadening mechanism. This explains why the velocity range within which the lines of the face-on EDE are formed is unaffected. 

The formation of the double peak can be attributed to optical depth effects. For a Keplerian disk seen face-on there are no systematic velocities in the line of sight, leaving only the turbulent velocity as broadening mechanism. At line center, where the extinction is maximal, one reaches optical depth unity high up in the disk surface where for a given line-of-sight the temperature is lower than it is at the point where the beam crosses the mid-plane (see Eq. 11, which is a function of $r_{xyz}$ and not $r_{xy}$). In the line wings (which probe the wings of the Gaussian turbulent velocity distribution) the depth of the $\tau \sim 1$ layer is deeper in the disk, effectively increasing the volume of the emission zone. In addition, the increased optical depth probes higher temperatures (and thus an increased source function). These effects combined result in a slightly increased emission in the line wings, causing the double-peaked line shape. This double peak is not visible for the lower $J$ transitions because the main line-forming region is beyond the radial distance probed in these models. Hence optical depth differences between the line core and wings are not as pronounced as for the higher $J$ lines, consequently the difference between the source functions i much smaller, suppressing the double-peaked shape.

At high inclinations the higher transitions show a sharp peak around zero velocity. This peak is caused by the substantially larger volume of optically thinner emitting material around zero velocity, compared to the optically thick dense regions at high velocities. Also, compared to the lower transitions, this peak is more pronounced. This is due to the low excitation of the population of the lower levels, resulting in emission which is generally optically thin (throughout the whole of velocity space), leading to a more even emission profile.

The ratios between the integrated flux of each line compared to the integrated flux of the $J$=3$-$2 transition is independent of inclination, as shown in Table \ref{ratio}. This indicates that (at least for the rotational transitions of CO where \emph{ALMA} is sensitive, and if the temperature profile is known) the density profile can be properly probed, independent of its orientation with respect to the observer. 

\begin{figure*}[htp]
\centering
\includegraphics[width=0.9\textwidth]{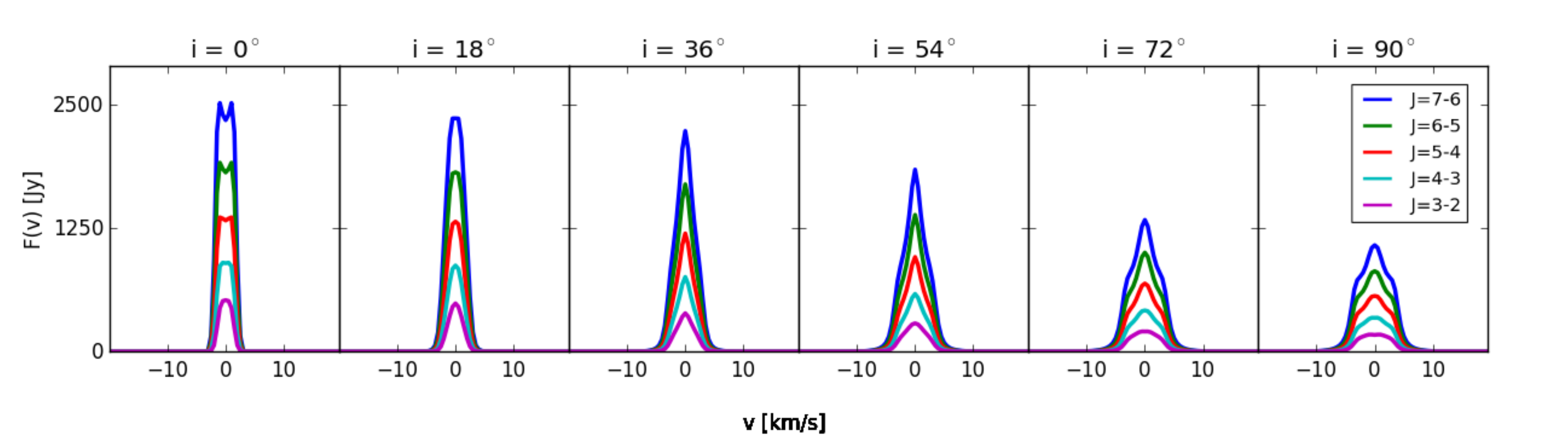}
\caption{CO transition lines of the Keplerian EDE model, as a function of inclination. \label{CO}}
\end{figure*}

As expected because of simple scaling arguments, no significant morphological effects are seen in the channel maps and PV diagrams of the higher $J$ transitions of CO. Therefore we refrain from showing them in this work.

\begin{table}[htp]
\centering
\begin{tabular}{ c | c c c c c c }

\hline
\hline
       & $i$=0$^\circ$ & $i$=18$^\circ$ & $i$=36$^\circ$ & $i$=54$^\circ$ & $i$=72$^\circ$ & $i$=90$^\circ$ \\ \hline
 $J$=3$-$2 & 1.00    & 1.00    & 1.00    & 1.00    & 1.00    & 1.00    \\
 $J$=4$-$3 & 2.61    & 2.61    & 2.62    & 2.62    & 2.62    & 2.60    \\
 $J$=5$-$4 & 5.27    & 5.27    & 5.29    & 5.30    & 5.30    & 5.26    \\
 $J$=6$-$5 & 9.16    & 9.15    & 9.19    & 9.20    & 9.19    & 9.09    \\
 $J$=7$-$6 & 14.36   & 14.33    & 14.38    & 14.37    & 14.30    & 14.10    \\
\hline

\end{tabular}
\caption{Ratios of the integrated flux of the higher rotational transitions of CO compared to the $J$=3$-$2 transition, per inclination angle. \label{ratio}}
\end{table}

\subsection{Constraining the geometry}

Approximating an EDE by a cylinder of diameter $\sqrt{x^2+y^2}=d$ and height $z=h$, it is trivial to calculate its projected dimensions as a function of inclination angle $i$. Obviously, the projected width $d'$ does not change, which is why the maximum offset of the EDE emission in PV2 does not change with inclination angle. Thus $d' = d$. The projected height $h'$ of the emission zone of the chosen spectral line changes with inclination angle as follows
\begin{equation}
 h' = \frac{h\ \tan i +d}{\sqrt{\tan^2 i +1}}.
\end{equation}
Having knowledge of either $h$ or $i$ can thus provide insight into the value of the other. Estimating a value for $i$, by comparison with the emission models above or with the channel maps shown in Sect. C of the Appendix, will permit the calculation of an estimated value for $h$.

\subsection{Constraining the velocity field}

Consider the following PV diagrams:
\begin{itemize}
 \item Fig. \ref{Kep}, PV2 at $i$=54$^\circ$-90$^\circ$.
 \item Fig. \ref{RadVel}, PV1 at $i$=36$^\circ$-54$^\circ$.
 \item Fig. \ref{SupKep}, all diagrams except $i$=0$^\circ$ and PV1 at $i$=90$^\circ$.
 \item Fig. \ref{SubKep}, all diagrams except $i$=0$^\circ$ and PV1 at $i$=90$^\circ$.
\end{itemize}
Clearly, there is possible ambiguity when comparing the abovementioned PV diagrams, as they all have a very similar morphology. The importance of constructing orthogonal wide-slit PV diagrams is yet again emphasised at this point. The morphology of the orthogonal PV diagram looks very different in the four cases, which would make it a vital diagnostic in determining the overall type of velocity field. As PV diagrams are becoming a very important method of analysing and evaluating the observed morphology \citep[e.g.][]{Hirano2004,Chiu2006}, the orthogonal PV diagram method can be used as an additional technique to acquire more constraints on the properties of the system.

\subsection{The use of stereograms as tool to identify substructure}

Next to wide-slit PV diagrams, it is in some cases also useful to plot integrated emission for a specific user-chosen velocity range. This means dividing the 3D datacube into portions along the velocity axis, and summing up the intensities with identical spatial coordinates for each portion. Essentially, it is simply a channel map with a wide velocity width. We shall refer to such diagrams as stereograms. They are especially useful to investigate relative velocities between emission zones.

We proceed with the stereograms shown in Fig. \ref{velside1}, where we have collapsed the exclusively red-shifted, blue-shifted, and central velocities of the 3D datacube of the edge-on Keplerian disk model with a bipolar outflow. It can immediately be seen that the bipolar outflow contains both approaching and receding emission. Also, the EDE can be recognised. The approaching and receding sides are mirrored, as expected from a rotating object. The central velocities, shown as the black contours, show the maximal height of the EDE.

\begin{figure}[htp]
\centering
 \resizebox{4.0cm}{!}{\includegraphics{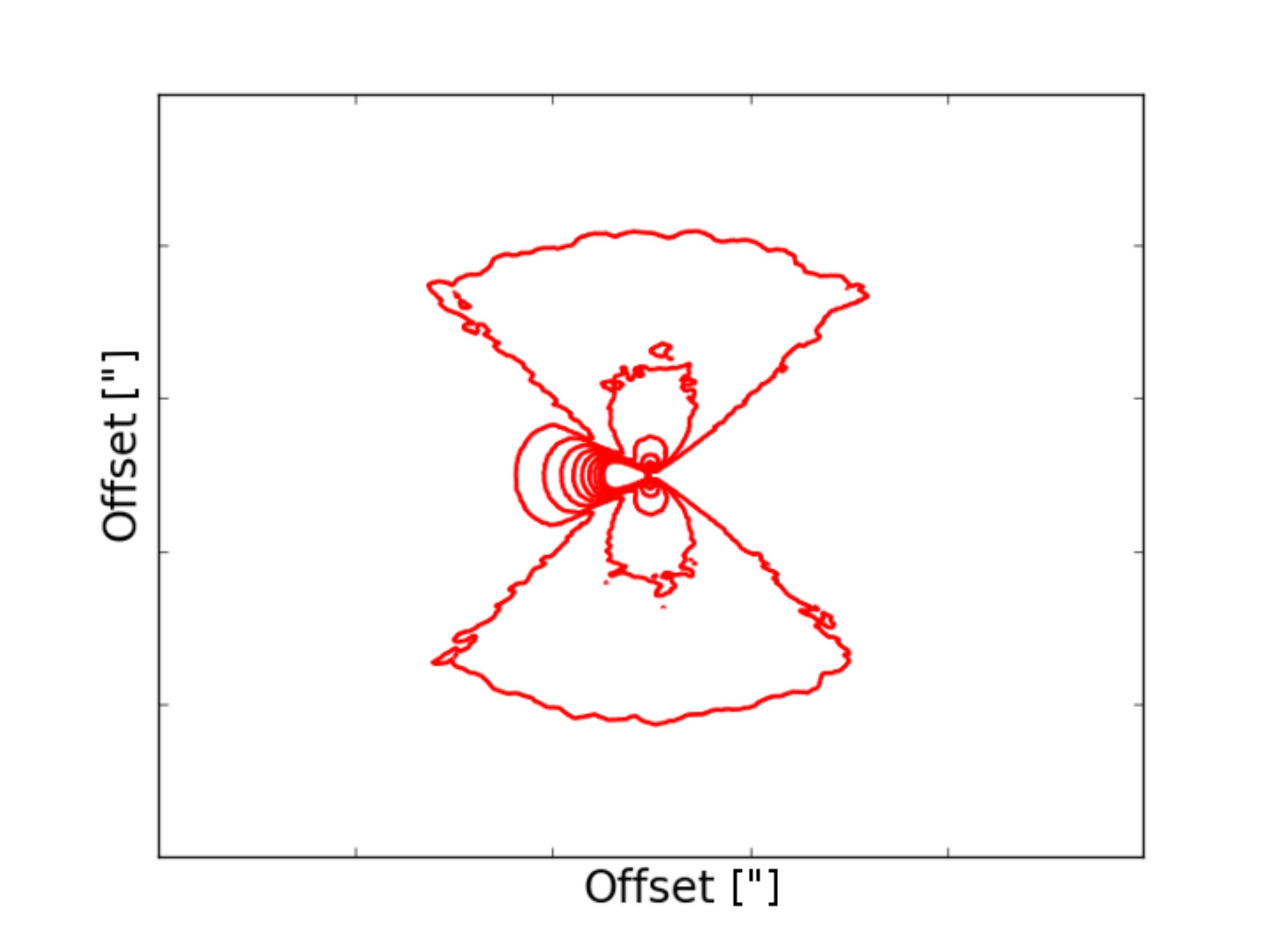}}
 \resizebox{4.0cm}{!}{\includegraphics{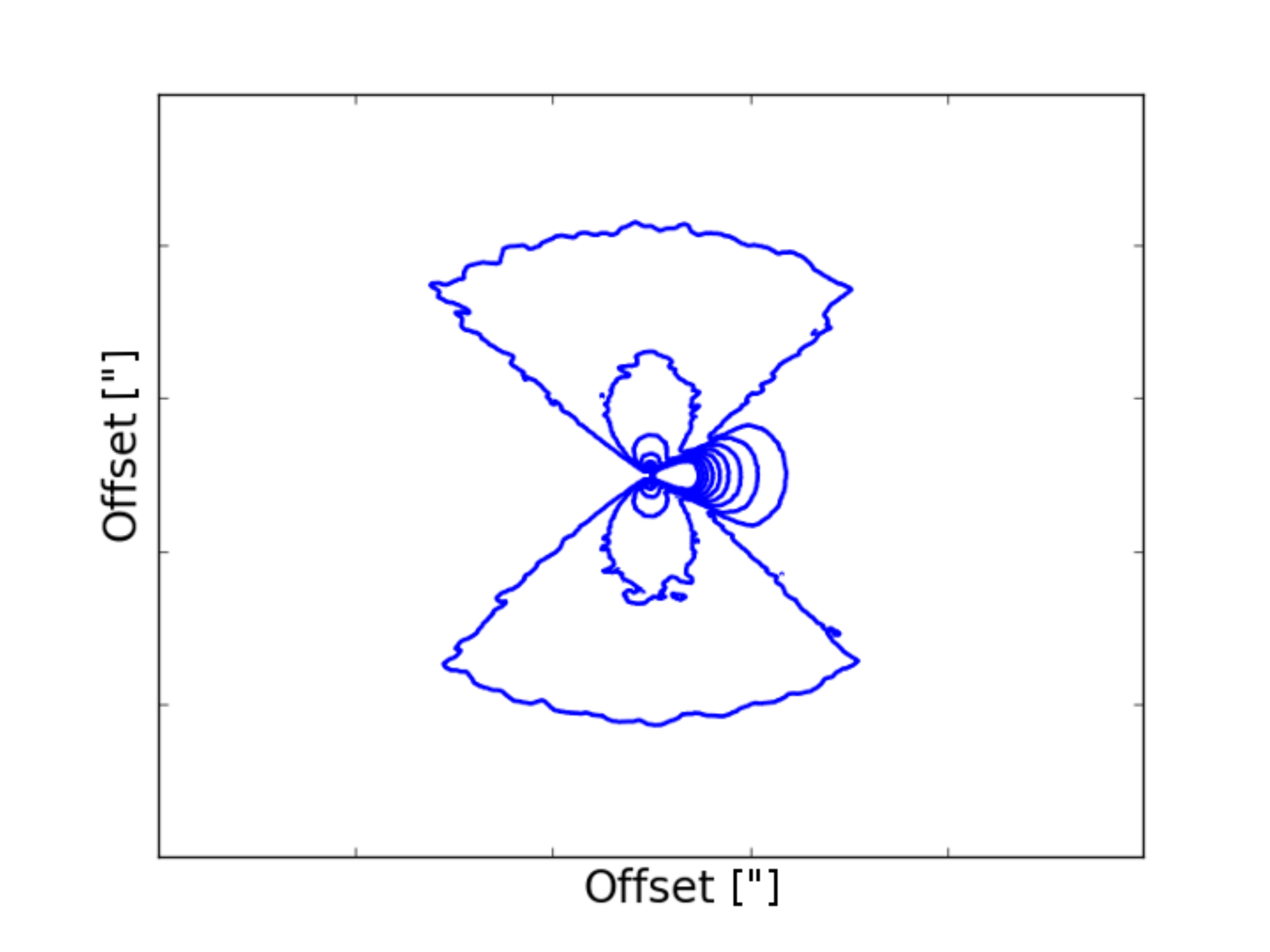}}
 \resizebox{4.0cm}{!}{\includegraphics{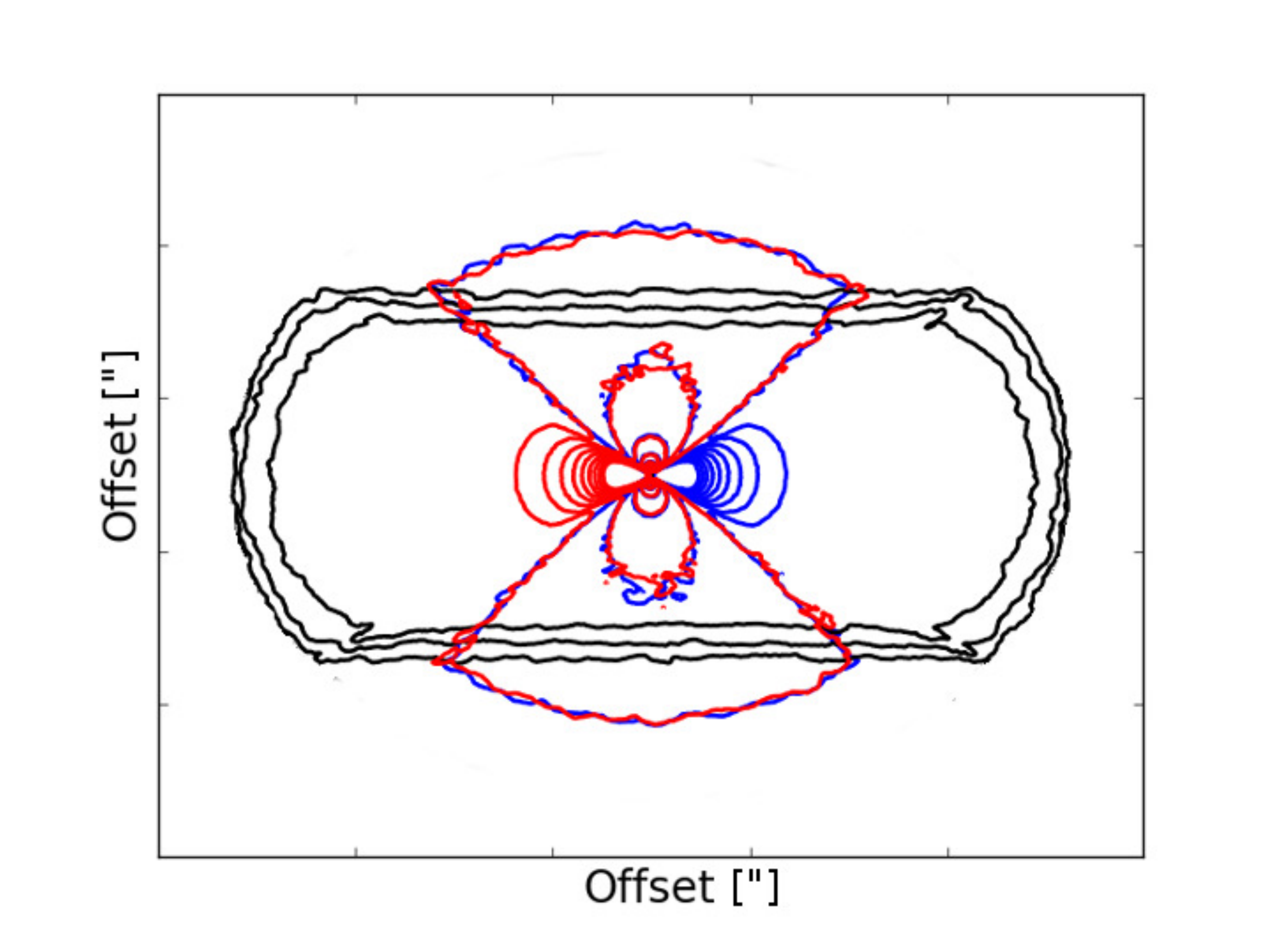}}
\caption{Stereograms of the Keplerian disk model with a bipolar outflow. The receding (red contours) and approaching (blue contours) sides have been constructed with the outermost 49\% of the velocity channels. The remaining two percent, the material with hardly any projected velocity along the line of sight, corresponds to the black contours. \label{velside1}}
\end{figure}

Because the stereogram does not reprocess the spatial axes of the datacube, it is a powerful tool to analyse emission from objects containing spatial asymmetries. For the considered models in this paper, its use is best demonstrated by focusing on the warped EDE, which was the only considered case where neither channel maps, PV diagrams nor spectral lines have proved able to detect any warping signatures. Fig. \ref{velside2} shows the velocity-bias of the emission of the warped EDE + bipolar outflow model, and in comparison the stereogram of the EDE with no substructure (Fig. \ref{velside1}). Some clear differences can be seen. The most striking difference arises from the portions of the bipolar outflow which are unobstructed by the warped EDE, leading to the appearance of petals in the higher-density contours of the stereogram. In this particular case three petals appear both in the red and in the blue emission. The petals in the red emission consist of one petal facing up, and two facing down. The opposite is found for the blue emission. The petals are created by the undulations, i.e. how many times the bipolar outflow penetrates through the orbital plane of the EDE. Also visible is the EDE itself, manifested as an asymmetrical bulge around the central horizontal axis of the stereograms. The appearance in the red component is mirrored in the blue component. The asymmetry of the EDE feature originates in its combination with the contours of the bipolar wind petals.

\begin{figure}[htp]
\centering
 \resizebox{4.0cm}{!}{\includegraphics{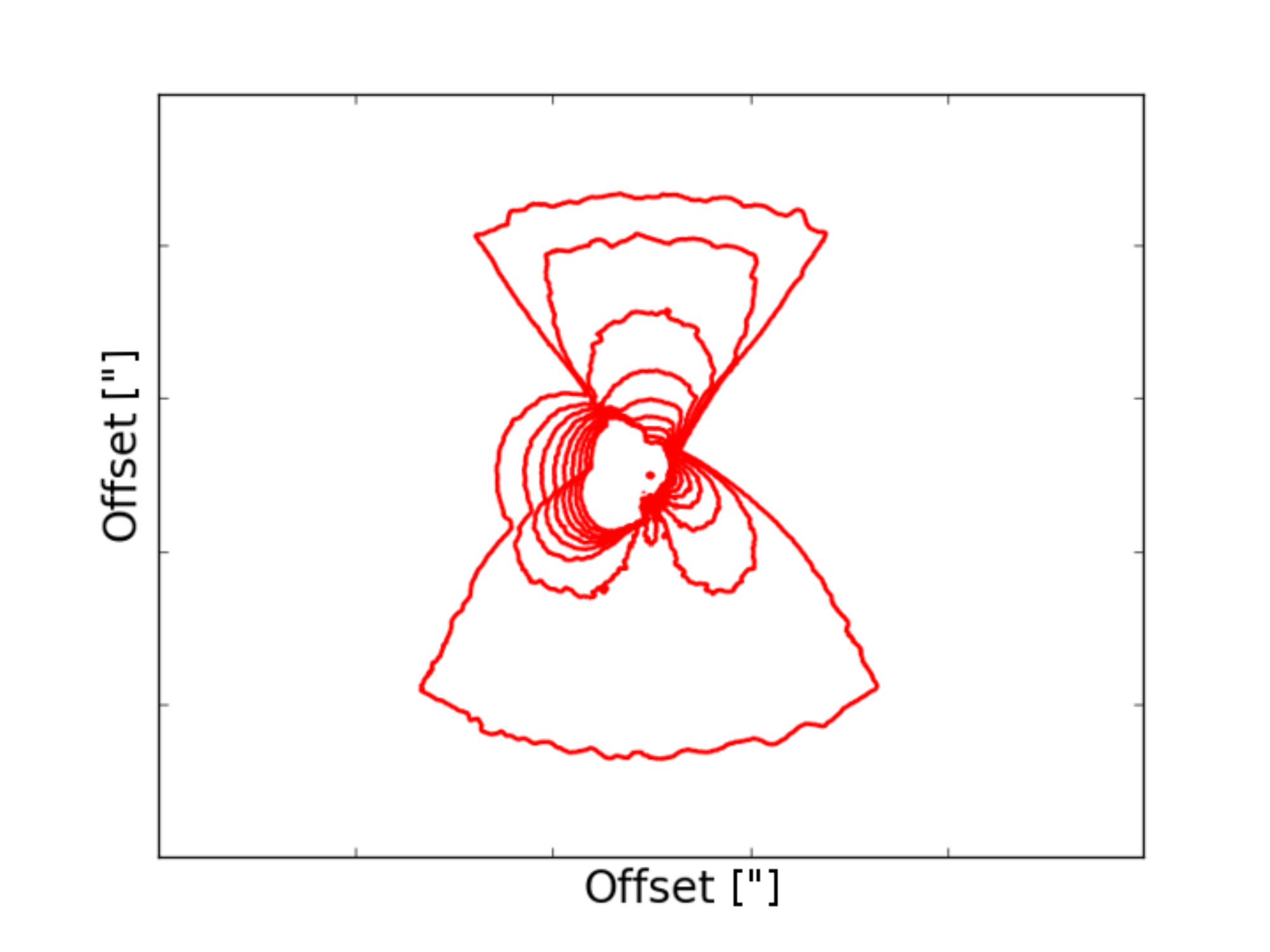}}
 \resizebox{4.0cm}{!}{\includegraphics{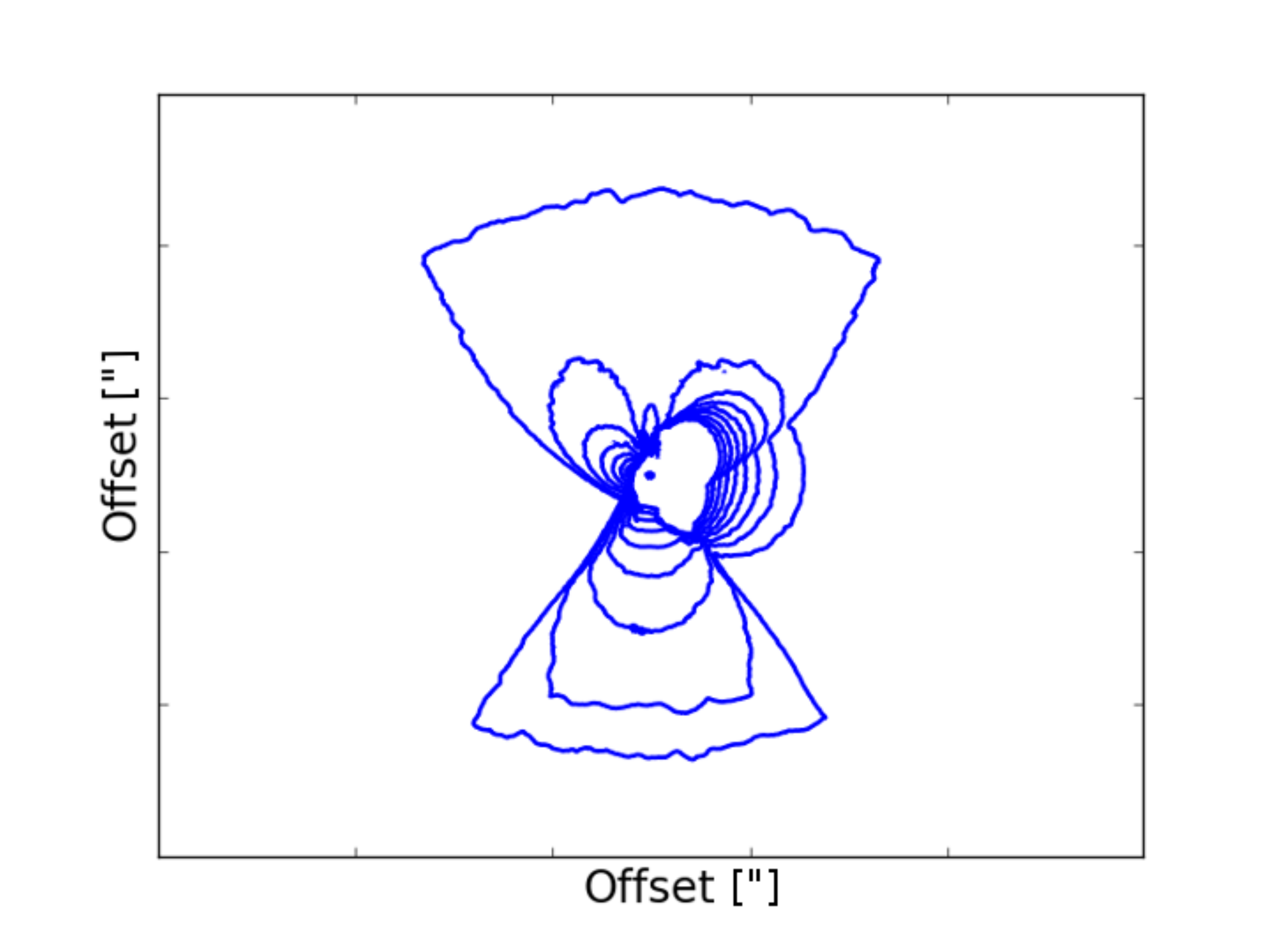}}
 \resizebox{4.0cm}{!}{\includegraphics{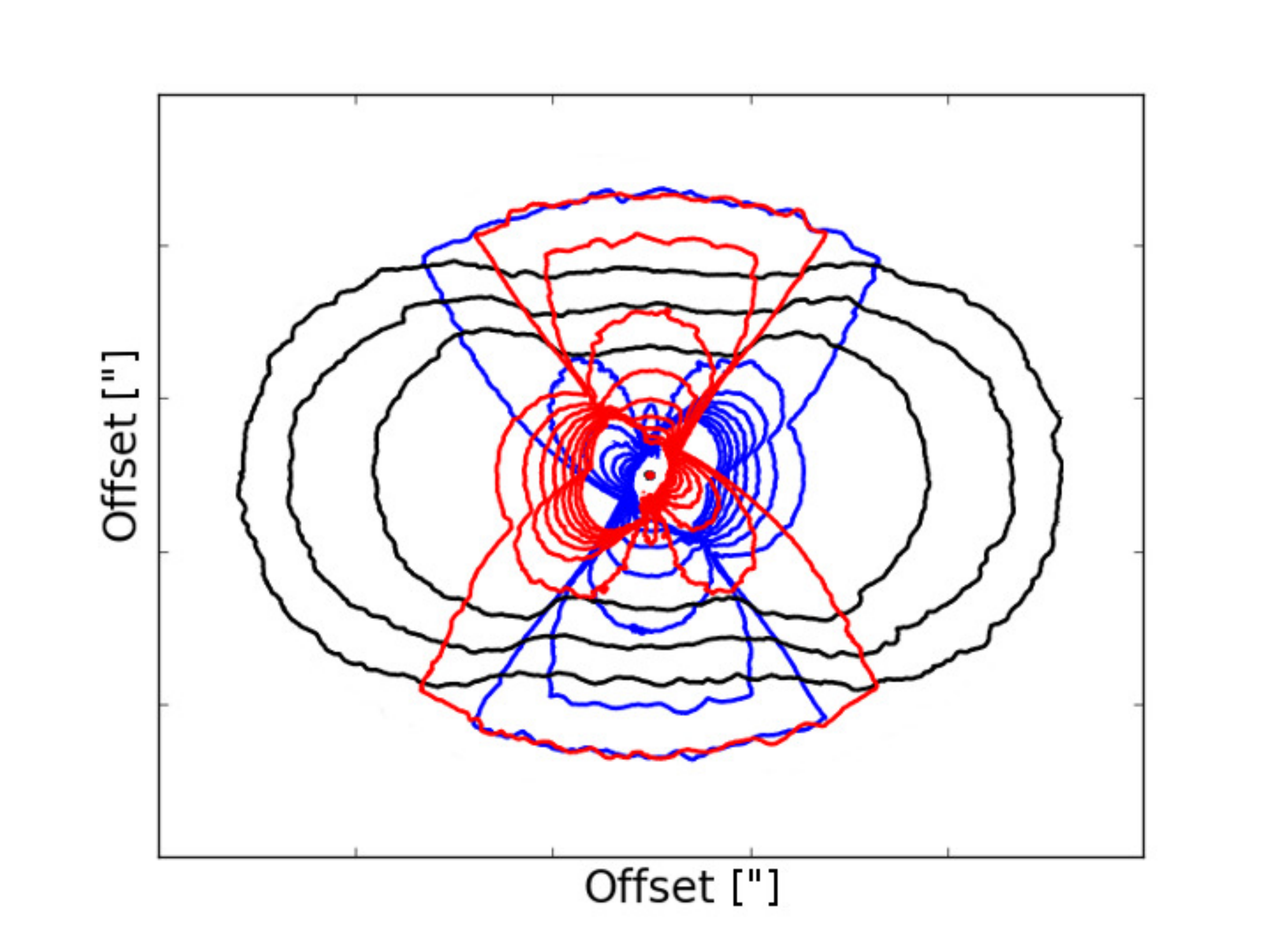}}
\caption{Stereograms of the warped EDE + bipolar outflow model. The receding (red contours) and approaching (blue contours) sides have been constructed with the outermost 49\% of the velocity channels. The remaining two percent, the material with hardly any projected velocity along the line of sight, correspond to the black contours. \label{velside2}}
\end{figure}

\section{Simulation of observations}

\subsection{Single-dish telescopes}
Simulating single-dish observations of the gas emission of a rotating EDE can be done by applying a Gaussian mask of variable width over the intrinsic 3D data. However, due to the symmetry of the modelled system, and its invariance to scale, the centering of any Gaussian filter (with an angular diameter comparable to the size of the object) over the 3D data will yield nearly identical line shapes, containing less flux.  Depending on the relative densities, one can expect the proportions of the emission contributions of both EDE and wind components to vary with the width of the Gaussian filter. \newline
Low spectral resolution data may conceal the intrinsic shape of the lines. Determining line strengths may be the only viable option to retrieve information from the molecular emission. 
\begin{figure*}[htp]
\centering
\includegraphics[width=0.9\textwidth]{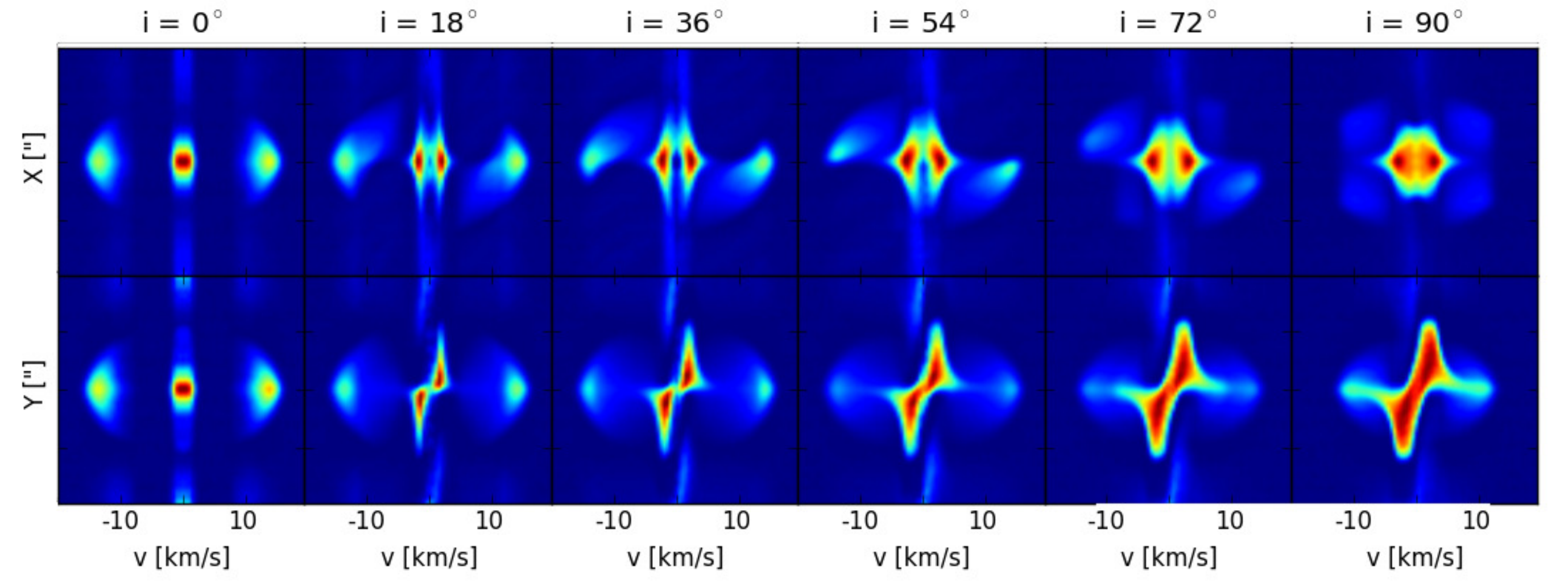}
\caption{Synthetic \emph{ALMA} observation simulation of the Keplerian disk model (with a bipolar outflow) with the C36-1 antenna configuration (res=1.5'', LAS=11''). \label{C31_ref}}
\end{figure*}

\begin{figure*}[htp]
\centering
\includegraphics[width=0.9\textwidth]{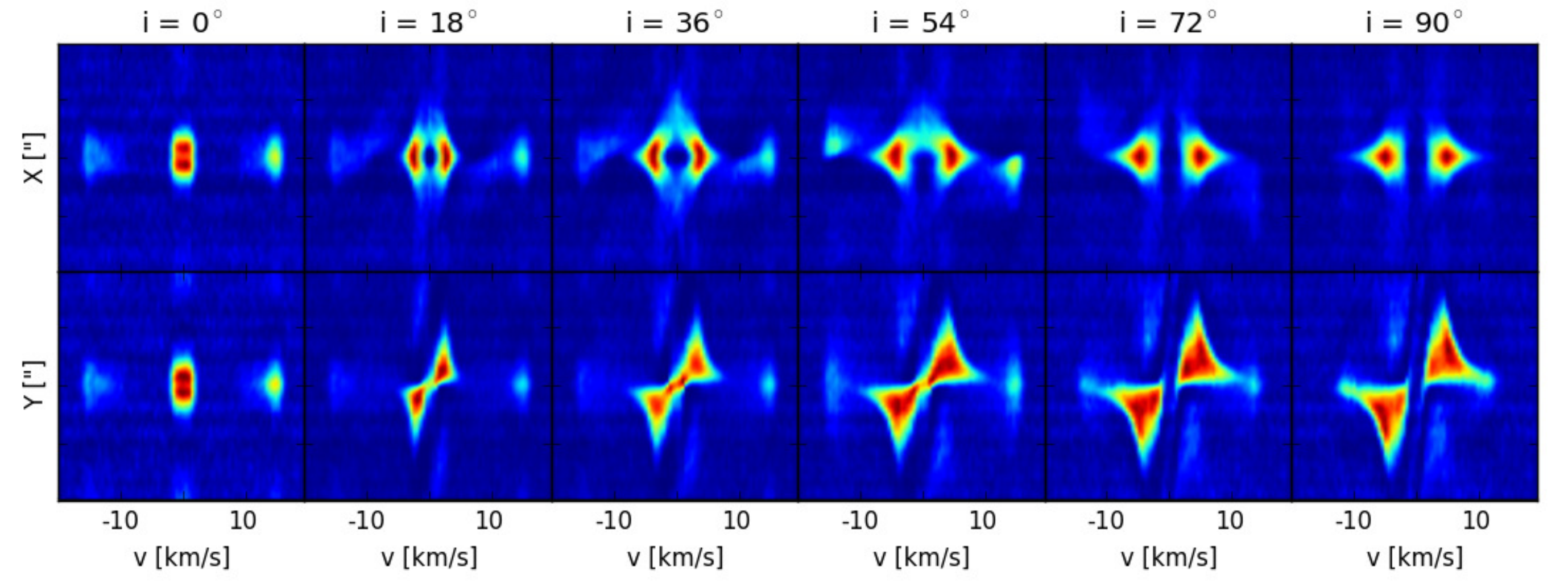}
\caption{Synthetic \emph{ALMA} observation simulation of the Keplerian disk model (with a bipolar outflow) with the C36-1 antenna configuration (res=1.5'', LAS=11''), for which the original model has been rescaled to a size 3.5 times greater than it was in Fig. \ref{C31_ref}. The effect of the largest angular scale being exceeded can be seen around zero velocity. \label{C31}}
\end{figure*}

\begin{figure*}[htp]
\centering
\includegraphics[width=0.9\textwidth]{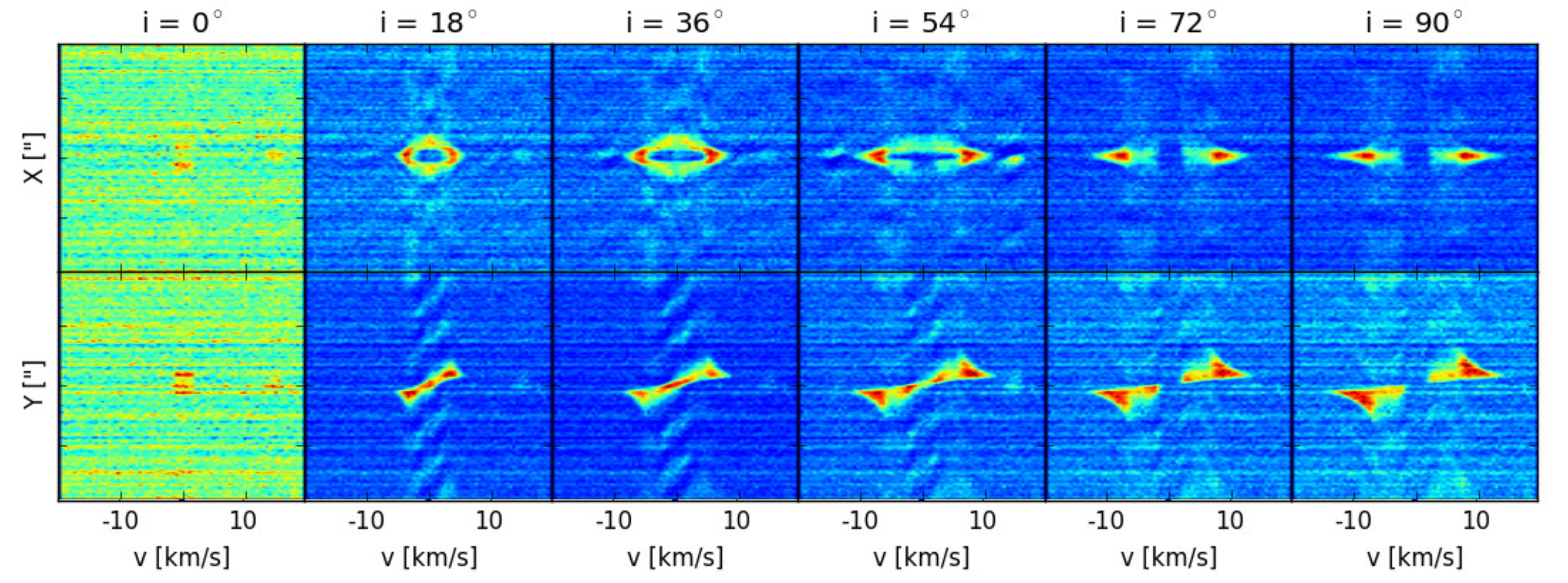}
\caption{Synthetic \emph{ALMA} observation simulation of the Keplerian disk model (with a bipolar outflow) with the C36-1 antenna configuration (res=1.5'', LAS=11''), for which the original model has been rescaled to a size 18 times greater than it was in Fig. \ref{C31_ref}. The effect of the largest angular scale being exceeded dominates the image. Most emission is resolved out. \label{C34}}
\end{figure*}

It is important to be prudent when dealing with such data, as the potential for misinterpretation is significant. For instance, erroneously assuming a spherically symmetrical outflow for a source which is intrinsically EDE-like, or has both a radial outflow and an EDE-like component, may result in serious deviations on the estimations of the physical properties of the system. This has been shown by e.g. \citet[][]{Homan2015}. The authors outline and quantify the uncertainties related to a misinterpretation of embedded spiral shapes in stellar winds. Their conclusions on the uncertainties, related to falsely interpreting the emission from a narrow spiral (which is effectively a kind of equatorial density enhancement) as originating from a simple spherically outflowing stellar wind, are also applicable here.

\subsection{\emph{ALMA}}

Using {\tt CASA} we have simulated \emph{ALMA} observations of the Keplerian disk model with a bipolar outflow. Since the considered radiative transfer model has no substructure, there is no reason to explore the effect of angular resolution. Its effect will only be of importance if the aim is to properly resolve the inner EDE, the edges of the EDE or the EDE-wind boundary. We have opted to show the effect of exceeding the chosen configuration's largest angular scale (LAS), determined by the shortest baseline of the interferometer, on the emission distribution in the PV diagrams of the Keplerian disk model with a bipolar outflow.

\begin{table}[htp]
\centering
\begin{tabular}{ l l }

\hline
\hline
\multicolumn{2}{ c }{Simulation Parameters} \\
\hline
Pixel size of original data & 0.015'' \\
Field size of original data & 7.5'' \\
Peak flux & Taken from {\tt LIME} output  \\ 
Transition & CO 3-2 (345.76599 GHz) \\
\hline
Pointing & Single \\
Channel width & 1.15MHz, centred on rest freq. \\
\hline
PWV & 0.913 mm \\
Thermal noise & standard \\
Temperature & 269 K \\
Integration time & 10 min on-source \\
\hline

\end{tabular}
\caption{The \emph{ALMA} observation simulation specifications. \label{casa}}
\end{table}

A synthetic observation is shown in Fig. \ref{C31_ref}, for which the main {\sc simobserve} parameters are presented in Table \ref{casa}. To obtain the results in Fig. \ref{C31_ref} we have simulated an observation with the most compact of the extended configurations, C36-1. This configuration has an angular resolution of 1.5'' and a largest angular scale of 11''. To properly show the effect of a decent observation, we have rescaled the pixel size to artificially place the object at the minimal distance for no EDE flux to be lost. Morphologically, the intrinsic emission and the emission in Fig. \ref{C31_ref} are very similar, as expected. The sheer number of antennae available for cycle3 ensures even for modest integration times a decent filling of the UV-plane. The high inclination models show that even though the signature of the EDE is barely unaltered, the bipolar wind emission has exceeded the LAS of the antenna configuration around zero velocity. Some of the flux of the wind is thus lost. Because our models are confined} to a numerical region, and bipolar outflows are not expected to exhibit such abrupt edges, this effect is expected to be more pronounced in real physical objects.

If the antenna configuration is chosen such that some EDE flux is lost, then this effect will target the features with the largest angular size first, which, in this case, is the emission around the rest frequency. In order to show this effect we have simulated two additional observations, shown in Fig. \ref{C31} and Fig. \ref{C34}. Both figures show synthetic observations of the same model as Fig. \ref{C31_ref}, but with a rescaled pixel size to simulate the object being even closer to the observer. This has been done to ensure the emission around zero velocity exceeds the LAS of the configuration to different degrees. As seen in Fig. \ref{C31}, some substantial differences with the intrinsic emission have arisen. The only remaining features of the bipolar wind are the high emission blobs at maximum velocity, which even disappear at high inclinations. The remaining wind emission is resolved out. The face-on model only shows the inner emission of the EDE, which remains under the LAS of the configuration. As inclination increases, the emission around the central velocity systematically exceeds the LAS, resulting in a characteristical emission deficit. In this case, however, the general morphology of the PV diagrams is not too strongly altered. The features in the PV diagrams can thus still be identified as being a rotating Keplerian EDE. Substantially reducing the LAS with respect to the pixel size augments this effect, to the point where most of the EDE emission is resolved out. Such a case can be seen in Fig. \ref{C34}, where the object has been placed even closer. The bipolar wind is undetected, and the EDE emission is observed only at the highest tangential velocities. Also, due to the weak detection, the {\tt CASA} {\sc clean} function seems to have difficulties reducing the synthetic data, resulting in a noisier datacube.

\section{Summary}

In this study we present 3D NLTE radiative transfer models of the rotational CO emission from 3D equatorial density enhancements (EDEs) with and without substructure and a variety of assumptions regarding the velocity fields in these EDEs. We have modelled five distinct velocity fields in the EDE: a Keplerian, a super-Keplerian, a sub-Keplerian EDE, a radially outflowing field, and rigid rotation. We have also modelled four distinct spatial morphologies: an axisymmetric flared EDE, one with spiral-shaped instabilities, an axisymmetric one with an annular gap, and a warped structure. We have also discussed the emission properties of a bipolar outflow, as these phenomena are strongly connected to the presence of EDEs around evolved stars.

\begin{itemize}
 
 \item For each model we have constructed channel maps, wide-slit position-velocity diagrams and spectral lines of the intrinsic emission. The velocity field has a strong effect on the PV morphology. PV diagrams enable us to constrain EDE height and inclination. Some degeneracy exists in the shapes of individual PV diagrams for different global velocity fields. Making use of the orthogonal PV diagrams may completely eliminate this ambiguity, and facilitates the identification of the nature of the global velocity field.
 
 \item Having introduced substructure (warps, spiral instabilities and annular gaps) into the models we found that the orthogonal wide-slit PV diagrams prove quasi-unable to recover and identify these peculiarities. Gaps and spirals are readily detected in the channel maps. The warped EDE is undetectable in either channel maps, PV diagrams or spectral lines, unless the warp consists of an even number of undulations and the EDE is oriented such that the height is constant along the line-of-sight. However, a substantial morphological difference is found between the stereogram of the simple EDE and the warped EDE, providing a method to identify warping.
 
 \item Generally, the spectral lines are unaffected by the introduction of density substructure in the EDE. When considering different velocity fields, we conclude that for most inclinations the EDEs produce spectral lines which are relatively broad, making it difficult to distinguish from a possible superposed bipolar outflow component. When the velocity fields contains a strong radial velocity component, these could even be misinterpreted as spherical outflows. Only when observed under low inclination angles can the EDE be distinguished from the bipolar outflow. In this case the dual nature of the observed object may be recognised in the spectral lines as a narrow central spike on top of the broader emission feature of the outflow.
 
 
 \item We found that the optically thick rotational transitions of the ground vibrational state of CO form $J$=3$-$2 up to $J$=7$-$6 have relative line strengths which are independent of inclination, resulting in a robust way of probing the radial density profile of the EDE.
 
 \item Simulations of synthetic ALMA observations show how the visual aspect of the emission signatures can change when the largest angular scale (imposed by the smallest baseline of the interferometer) is exceeded. For a rotating EDE the emission around zero velocity will be first to be discarded by the destructive interference.
 
\end{itemize}


\begin{acknowledgements} 
 W.H. acknowledges support from the Fonds voor Wetenschappelijk Onderzoek Vlaanderen (FWO). We would also like to express sincere thanks to Michiel Hogerheijde, Christian Brinch and Wouter Vlemmings for their support with the {\tt LIME} code, as well as to Markus Schmalzl of the Allegro node \emph{ALMA} support community, and to Adam Avison of the UK ARC node for their assistance with {\tt CASA}. Sincere thanks to Allard Jan van Marle for his countless helpful contributions to this research. We would also like to present thanks to Shazrene Mohamed for her insightful input, and to Hans Van Winckel for proofreading.
\end{acknowledgements}

\bibliographystyle{aa}
\bibliography{wardhoman_biblio}

\begin{thebibliography}{50}
\expandafter\ifx\csname natexlab\endcsname\relax\def\natexlab#1{#1}\fi

\bibitem[{{Andrews} \& {Williams}(2007)}]{Andrews2007}
{Andrews}, S.~M. \& {Williams}, J.~P. 2007, \apj, 659, 705

\bibitem[{{Andrews} {et~al.}(2009){Andrews}, {Wilner}, {Hughes}, {Qi}, \&
  {Dullemond}}]{Andrews2009}
{Andrews}, S.~M., {Wilner}, D.~J., {Hughes}, A.~M., {Qi}, C., \& {Dullemond},
  C.~P. 2009, \apj, 700, 1502

\bibitem[{{Andrews} {et~al.}(2010){Andrews}, {Wilner}, {Hughes}, {Qi}, \&
  {Dullemond}}]{Andrews2010}
{Andrews}, S.~M., {Wilner}, D.~J., {Hughes}, A.~M., {Qi}, C., \& {Dullemond},
  C.~P. 2010, \apj, 723, 1241

\bibitem[{{Bekiaris} {et~al.}(2016){Bekiaris}, {Glazebrook}, {Fluke}, \&
  {Abraham}}]{Bekiaris2016}
{Bekiaris}, G., {Glazebrook}, K., {Fluke}, C.~J., \& {Abraham}, R. 2016,
  \mnras, 455, 754

\bibitem[{{Boccaletti} {et~al.}(2015){Boccaletti}, {Thalmann}, {Lagrange},
  {Janson}, {Augereau}, {Schneider}, {Milli}, {Grady}, {Debes}, {Langlois},
  {Mouillet}, {Henning}, {Dominik}, {Maire}, {Beuzit}, {Carson}, {Dohlen},
  {Engler}, {Feldt}, {Fusco}, {Ginski}, {Girard}, {Hines}, {Kasper}, {Mawet},
  {M{\'e}nard}, {Meyer}, {Moutou}, {Olofsson}, {Rodigas}, {Sauvage},
  {Schlieder}, {Schmid}, {Turatto}, {Udry}, {Vakili}, {Vigan}, {Wahhaj}, \&
  {Wisniewski}}]{Boccaletti2015}
{Boccaletti}, A., {Thalmann}, C., {Lagrange}, A.-M., {et~al.} 2015, \nat, 526,
  230

\bibitem[{{Brinch} \& {Hogerheijde}(2010)}]{Brinch2010}
{Brinch}, C. \& {Hogerheijde}, M.~R. 2010, \aap, 523, A25

\bibitem[{{Bujarrabal} \& {Alcolea}(2013)}]{Bujarrabal2013b}
{Bujarrabal}, V. \& {Alcolea}, J. 2013, \aap, 552, A116

\bibitem[{{Bujarrabal} {et~al.}(2016){Bujarrabal}, {Castro-Carrizo}, {Alcolea},
  {Santander-Garcia}, {Van Winckel}, \& {Sanchez Contreras}}]{Bujarrabal2016}
{Bujarrabal}, V., {Castro-Carrizo}, A., {Alcolea}, J., {et~al.} 2016, ArXiv
  e-prints

\bibitem[{{Bujarrabal} {et~al.}(2015){Bujarrabal}, {Castro-Carrizo}, {Alcolea},
  \& {Van Winckel}}]{Bujarrabal2015}
{Bujarrabal}, V., {Castro-Carrizo}, A., {Alcolea}, J., \& {Van Winckel}, H.
  2015, \aap, 575, L7

\bibitem[{{Casassus} {et~al.}(2015){Casassus}, {Marino}, {P{\'e}rez}, {Roman},
  {Dunhill}, {Armitage}, {Cuadra}, {Wootten}, {van der Plas}, {Cieza}, {Moral},
  {Christiaens}, \& {Montesinos}}]{Casassus2015}
{Casassus}, S., {Marino}, S., {P{\'e}rez}, S., {et~al.} 2015, \apj, 811, 92

\bibitem[{{Cassen} \& {Moosman}(1981)}]{Cassen1981}
{Cassen}, P. \& {Moosman}, A. 1981, \icarus, 48, 353

\bibitem[{{Chiang} \& {Goldreich}(1997)}]{Chiang1997}
{Chiang}, E.~I. \& {Goldreich}, P. 1997, \apj, 490, 368

\bibitem[{{Chiu} {et~al.}(2006){Chiu}, {Hoang}, {Dinh-V-Trung}, {Lim}, {Kwok},
  {Hirano}, \& {Muthu}}]{Chiu2006}
{Chiu}, P.-J., {Hoang}, C.-T., {Dinh-V-Trung}, {et~al.} 2006, \apj, 645, 605

\bibitem[{{Christiaens} {et~al.}(2014){Christiaens}, {Casassus}, {Perez}, {van
  der Plas}, \& {M{\'e}nard}}]{Christiaens2014}
{Christiaens}, V., {Casassus}, S., {Perez}, S., {van der Plas}, G., \&
  {M{\'e}nard}, F. 2014, \apjl, 785, L12

\bibitem[{{Decin} {et~al.}(2015){Decin}, {Richards}, {Neufeld}, {Steffen},
  {Melnick}, \& {Lombaert}}]{Decin2015}
{Decin}, L., {Richards}, A.~M.~S., {Neufeld}, D., {et~al.} 2015, \aap, 574, A5

\bibitem[{{Dipierro} {et~al.}(2014){Dipierro}, {Lodato}, {Testi}, \& {de
  Gregorio Monsalvo}}]{Dipierro2014}
{Dipierro}, G., {Lodato}, G., {Testi}, L., \& {de Gregorio Monsalvo}, I. 2014,
  \mnras, 444, 1919

\bibitem[{{Dipierro} {et~al.}(2015){Dipierro}, {Pinilla}, {Lodato}, \&
  {Testi}}]{Dipierro2015}
{Dipierro}, G., {Pinilla}, P., {Lodato}, G., \& {Testi}, L. 2015, \mnras, 451,
  974

\bibitem[{{Dong} {et~al.}(2015){Dong}, {Zhu}, \& {Whitney}}]{Dong2015}
{Dong}, R., {Zhu}, Z., \& {Whitney}, B. 2015, \apj, 809, 93

\bibitem[{{Duch{\^e}ne} {et~al.}(2013){Duch{\^e}ne}, {Bouvier}, {Moraux},
  {Bouy}, {Konopacky}, \& {Ghez}}]{Duchene2013}
{Duch{\^e}ne}, G., {Bouvier}, J., {Moraux}, E., {et~al.} 2013, \aap, 555, A137

\bibitem[{{Flock} {et~al.}(2015){Flock}, {Ruge}, {Dzyurkevich}, {Henning},
  {Klahr}, \& {Wolf}}]{Flock2015}
{Flock}, M., {Ruge}, J.~P., {Dzyurkevich}, N., {et~al.} 2015, \aap, 574, A68

\bibitem[{{Hartmann} {et~al.}(1998){Hartmann}, {Calvet}, {Gullbring}, \&
  {D'Alessio}}]{Hartmann1998}
{Hartmann}, L., {Calvet}, N., {Gullbring}, E., \& {D'Alessio}, P. 1998, \apj,
  495, 385

\bibitem[{{Hirano} {et~al.}(2004){Hirano}, {Shinnaga}, {Dinh-V-Trung}, {Fong},
  {Keto}, {Patel}, {Qi}, {Young}, {Zhang}, \& {Zhao}}]{Hirano2004}
{Hirano}, N., {Shinnaga}, H., {Dinh-V-Trung}, {et~al.} 2004, \apjl, 616, L43

\bibitem[{{Homan} {et~al.}(2015){Homan}, {Decin}, {de Koter}, {van Marle},
  {Lombaert}, \& {Vlemmings}}]{Homan2015}
{Homan}, W., {Decin}, L., {de Koter}, A., {et~al.} 2015, \aap, 579, A118

\bibitem[{{Juh{\'a}sz} {et~al.}(2015){Juh{\'a}sz}, {Benisty}, {Pohl},
  {Dullemond}, {Dominik}, \& {Paardekooper}}]{Juhasz2015}
{Juh{\'a}sz}, A., {Benisty}, M., {Pohl}, A., {et~al.} 2015, \mnras, 451, 1147

\bibitem[{{Kato} {et~al.}(2009){Kato}, {Nakamura}, {Tandokoro}, {Fujimoto}, \&
  {Ida}}]{Kato2009}
{Kato}, M.~T., {Nakamura}, K., {Tandokoro}, R., {Fujimoto}, M., \& {Ida}, S.
  2009, \apj, 691, 1697

\bibitem[{{Kenyon} \& {Hartmann}(1987)}]{Kenyon1987}
{Kenyon}, S.~J. \& {Hartmann}, L. 1987, \apj, 323, 714

\bibitem[{{Kervella} {et~al.}(2015){Kervella}, {Montarg{\`e}s}, {Lagadec},
  {Ridgway}, {Haubois}, {Girard}, {Ohnaka}, {Perrin}, \&
  {Gallenne}}]{Kervella2015}
{Kervella}, P., {Montarg{\`e}s}, M., {Lagadec}, E., {et~al.} 2015, \aap, 578,
  A77

\bibitem[{{Kitamura} {et~al.}(2002){Kitamura}, {Momose}, {Yokogawa}, {Kawabe},
  {Tamura}, \& {Ida}}]{Kitamura2002}
{Kitamura}, Y., {Momose}, M., {Yokogawa}, S., {et~al.} 2002, \apj, 581, 357

\bibitem[{{Lee}(2013)}]{Lee2013}
{Lee}, U. 2013, \pasj, 65, 122

\bibitem[{{Les} \& {Lin}(2015)}]{Les2015}
{Les}, R. \& {Lin}, M.-K. 2015, \mnras, 450, 1503

\bibitem[{{Lin}(2015)}]{Lin2015}
{Lin}, M.-K. 2015, \mnras, 448, 3806

\bibitem[{{Montgomery}(2012)}]{Montgomery2012}
{Montgomery}, M.~M. 2012, \apjl, 753, L27

\bibitem[{{Nixon}(2015)}]{Nixon2015}
{Nixon}, C. 2015, \mnras, 450, 2459

\bibitem[{{Nixon} \& {King}(2015)}]{Nixon2015b}
{Nixon}, C. \& {King}, A. 2015, ArXiv e-prints

\bibitem[{{Ogilvie} \& {Latter}(2013)}]{Ogilvie2013}
{Ogilvie}, G.~I. \& {Latter}, H.~N. 2013, \mnras, 433, 2403

\bibitem[{{Pavlyuchenkov} {et~al.}(2007){Pavlyuchenkov}, {Semenov}, {Henning},
  {Guilloteau}, {Pi{\'e}tu}, {Launhardt}, \& {Dutrey}}]{Pavlyuchenkov2007}
{Pavlyuchenkov}, Y., {Semenov}, D., {Henning}, T., {et~al.} 2007, \apj, 669,
  1262

\bibitem[{{Raghavan} {et~al.}(2010){Raghavan}, {McAlister}, {Henry}, {Latham},
  {Marcy}, {Mason}, {Gies}, {White}, \& {ten Brummelaar}}]{Raghavan2010}
{Raghavan}, D., {McAlister}, H.~A., {Henry}, T.~J., {et~al.} 2010, \apjs, 190,
  1

\bibitem[{{Rebollido} {et~al.}(2015){Rebollido}, {Mer{\'{\i}}n}, {Ribas},
  {Bustamante}, {Bouy}, {Riviere-Marichalar}, {Prusti}, {Pilbratt},
  {Andr{\'e}}, \& {{\'A}brah{\'a}m}}]{Rebollido2015}
{Rebollido}, I., {Mer{\'{\i}}n}, B., {Ribas}, {\'A}., {et~al.} 2015, \aap, 581,
  A30

\bibitem[{{Sch{\"o}del} {et~al.}(2002){Sch{\"o}del}, {Ott}, {Genzel},
  {Hofmann}, {Lehnert}, {Eckart}, {Mouawad}, {Alexander}, {Reid}, {Lenzen},
  {Hartung}, {Lacombe}, {Rouan}, {Gendron}, {Rousset}, {Lagrange}, {Brandner},
  {Ageorges}, {Lidman}, {Moorwood}, {Spyromilio}, {Hubin}, \&
  {Menten}}]{Schodel2002}
{Sch{\"o}del}, R., {Ott}, T., {Genzel}, R., {et~al.} 2002, \nat, 419, 694

\bibitem[{{Sch{\"o}ier} {et~al.}(2005){Sch{\"o}ier}, {van der Tak}, {van
  Dishoeck}, \& {Black}}]{Schoier2005}
{Sch{\"o}ier}, F.~L., {van der Tak}, F.~F.~S., {van Dishoeck}, E.~F., \&
  {Black}, J.~H. 2005, \aap, 432, 369

\bibitem[{{Terquem} {et~al.}(2015){Terquem}, {S{\o}rensen-Clark}, \&
  {Bouvier}}]{Terquem2015}
{Terquem}, C., {S{\o}rensen-Clark}, P.~M., \& {Bouvier}, J. 2015, \mnras, 454,
  3472

\bibitem[{{Toomre}(1964)}]{Toomre1964}
{Toomre}, A. 1964, \apj, 139, 1217

\bibitem[{{Ud-Doula} {et~al.}(2008){Ud-Doula}, {Owocki}, \&
  {Townsend}}]{UdDoula2008}
{Ud-Doula}, A., {Owocki}, S.~P., \& {Townsend}, R.~H.~D. 2008, \mnras, 385, 97

\bibitem[{{van Winckel}(2003)}]{VanWinckel2003}
{van Winckel}, H. 2003, \araa, 41, 391

\bibitem[{{Van Winckel} {et~al.}(2006){Van Winckel}, {Lloyd Evans}, {Reyniers},
  {Deroo}, \& {Gielen}}]{VanWinckel2006}
{Van Winckel}, H., {Lloyd Evans}, T., {Reyniers}, M., {Deroo}, P., \& {Gielen},
  C. 2006, \memsai, 77, 943

\bibitem[{{Visser} \& {Dullemond}(2010)}]{Visser2010}
{Visser}, R. \& {Dullemond}, C.~P. 2010, \aap, 519, A28

\bibitem[{{Vorobyov} \& {Basu}(2015)}]{Vorobyov2015}
{Vorobyov}, E.~I. \& {Basu}, S. 2015, \apj, 805, 115

\bibitem[{{Wilner} \& {Lay}(2000)}]{Wilner2000}
{Wilner}, D.~J. \& {Lay}, O.~P. 2000, Protostars and Planets IV, 509

\bibitem[{{Yang} {et~al.}(2010){Yang}, {Stancil}, {Balakrishnan}, \&
  {Forrey}}]{Yang2010}
{Yang}, B., {Stancil}, P.~C., {Balakrishnan}, N., \& {Forrey}, R.~C. 2010,
  \apj, 718, 1062

\bibitem[{{Zhu} {et~al.}(2015){Zhu}, {Dong}, {Stone}, \& {Rafikov}}]{Zhu2015}
{Zhu}, Z., {Dong}, R., {Stone}, J.~M., \& {Rafikov}, R.~R. 2015, ArXiv e-prints

\end{thebibliography}

\clearpage

\begin{appendix}

\section{Channel maps of the calculated models}

In this section we show the channel maps of the radiative transfer models which have been discussed in the paper above. To this end, we have respresented the synthetic 3D data by means of a 4 by 4 tile plot, showing slices throughout velocity space. The used colour coding is dictated by the linear colourbar shown in Fig. \ref{cmap}.

\clearpage

\begin{figure*}[htp]
 \centering
 \resizebox{7.5cm}{!}{\includegraphics{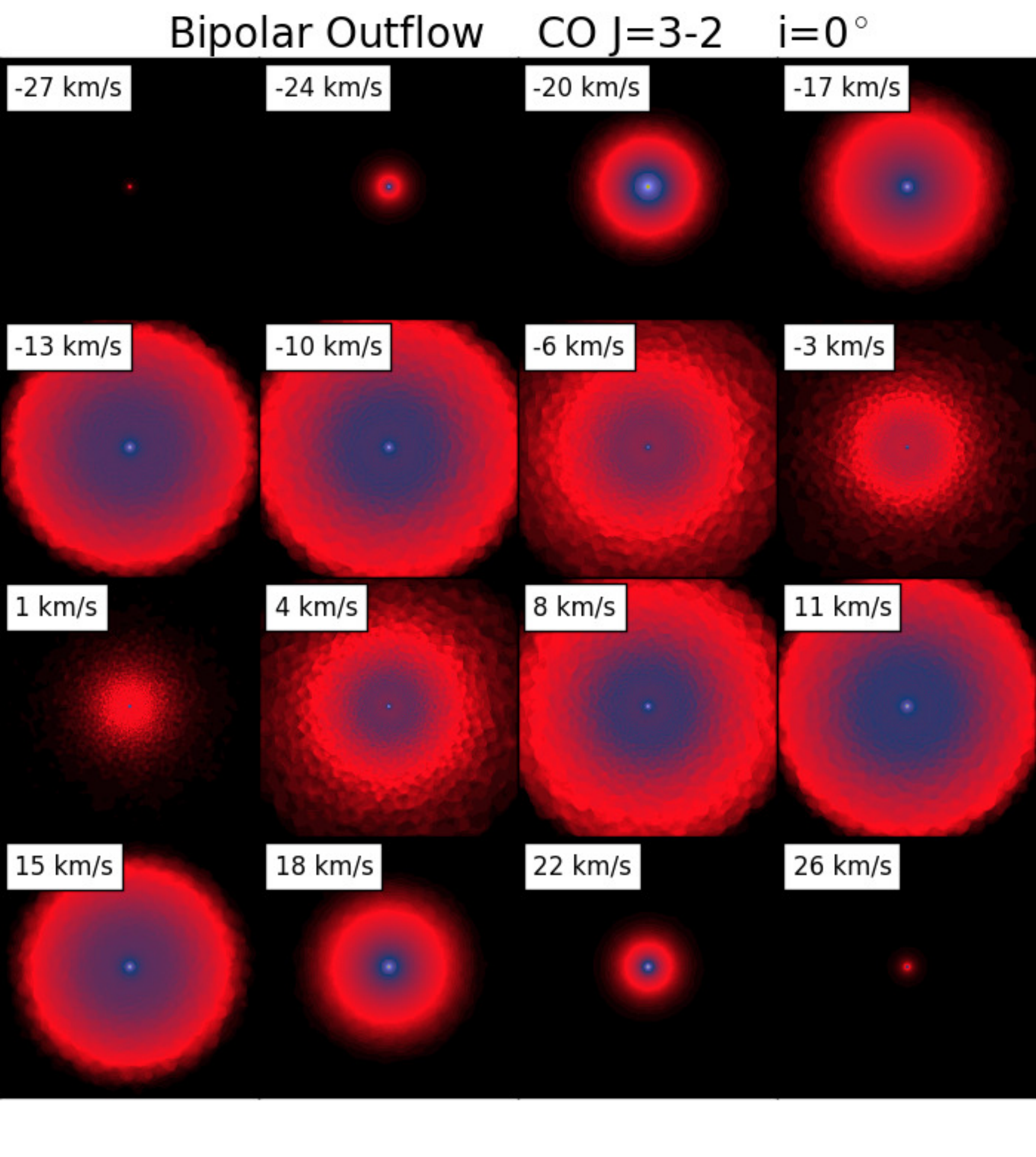}}
 \resizebox{7.5cm}{!}{\includegraphics{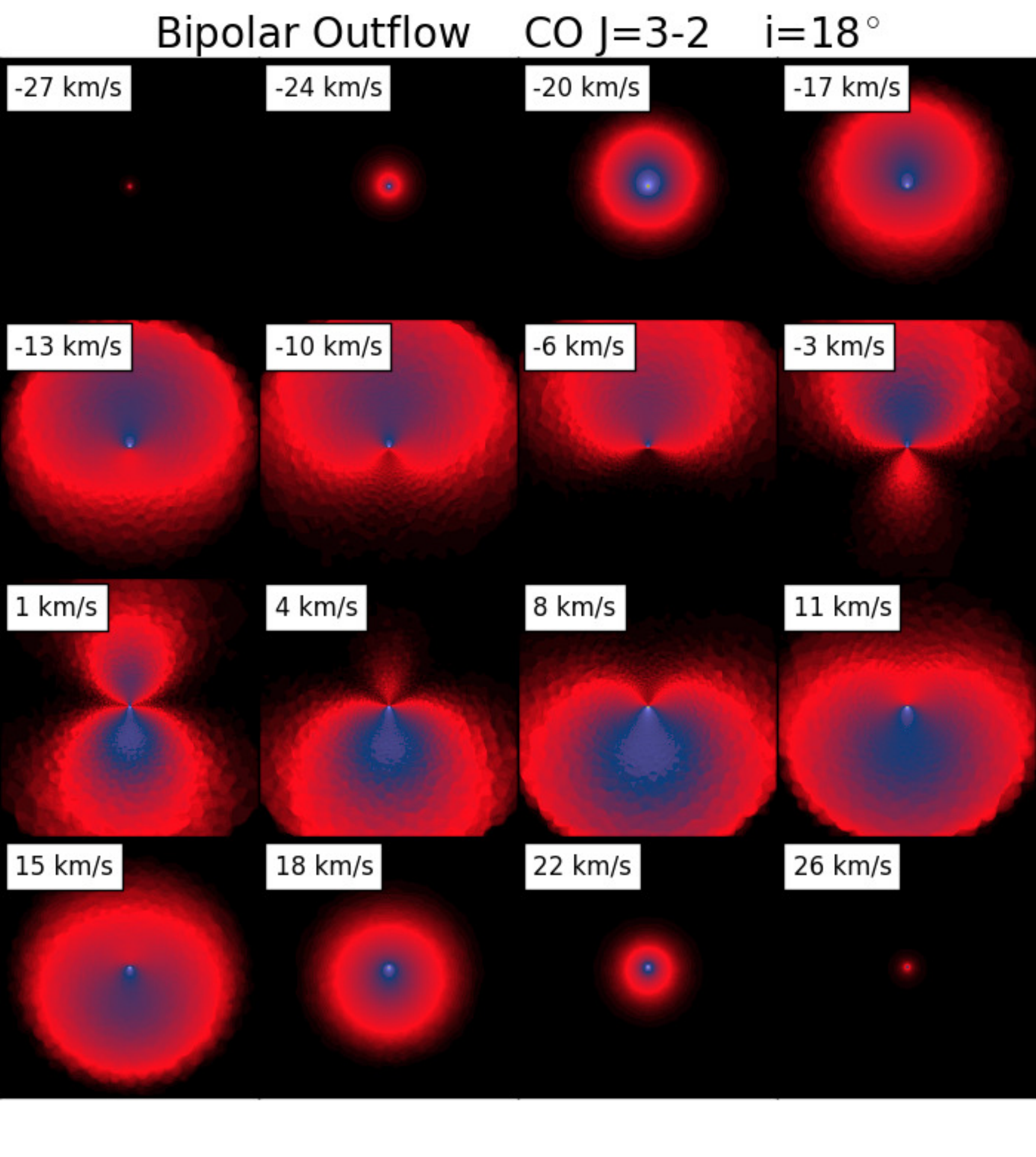}}
 \resizebox{7.5cm}{!}{\includegraphics{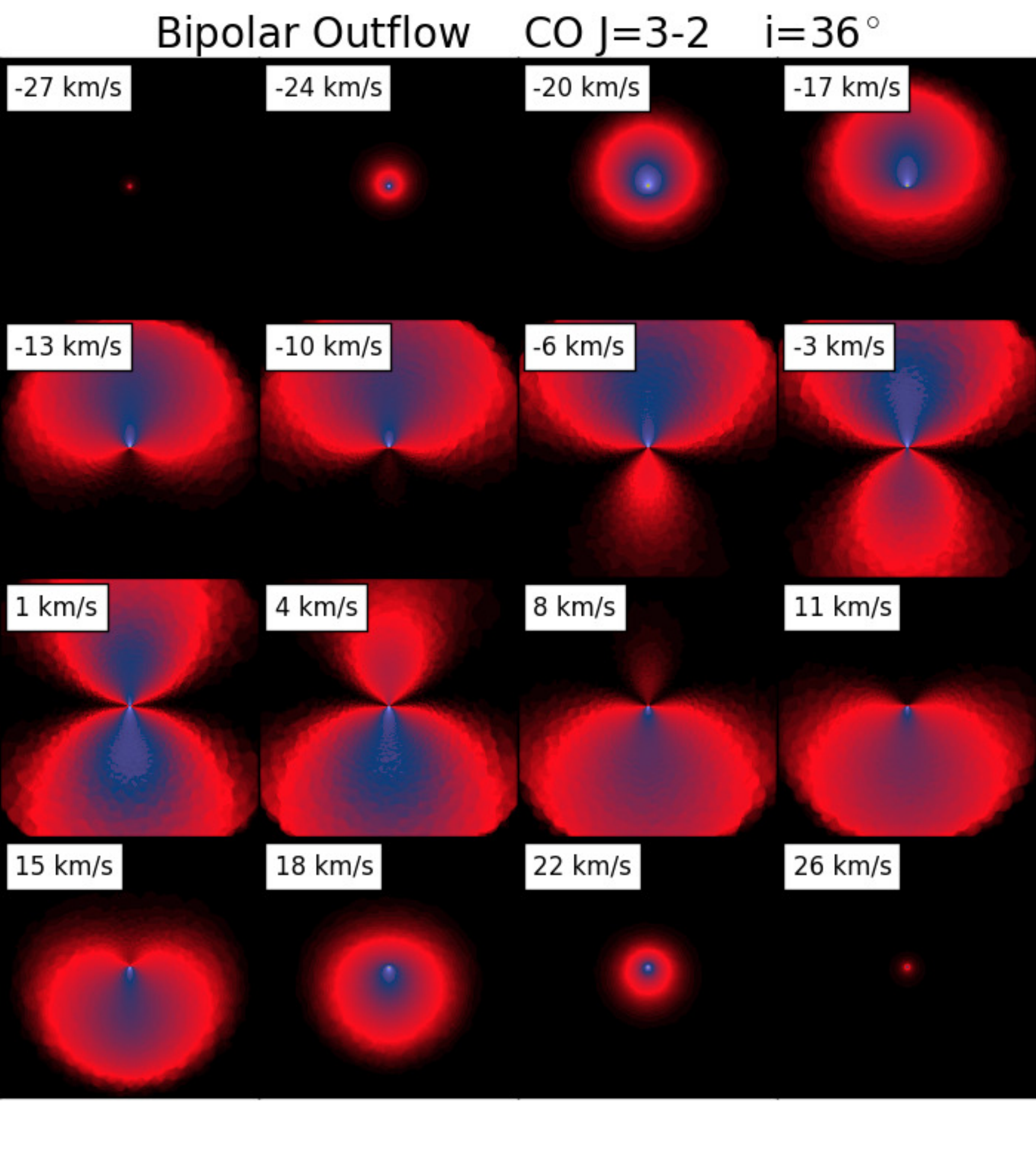}}
 \resizebox{7.5cm}{!}{\includegraphics{Bipolar_i54-eps-converted-to.pdf}}
 \resizebox{7.5cm}{!}{\includegraphics{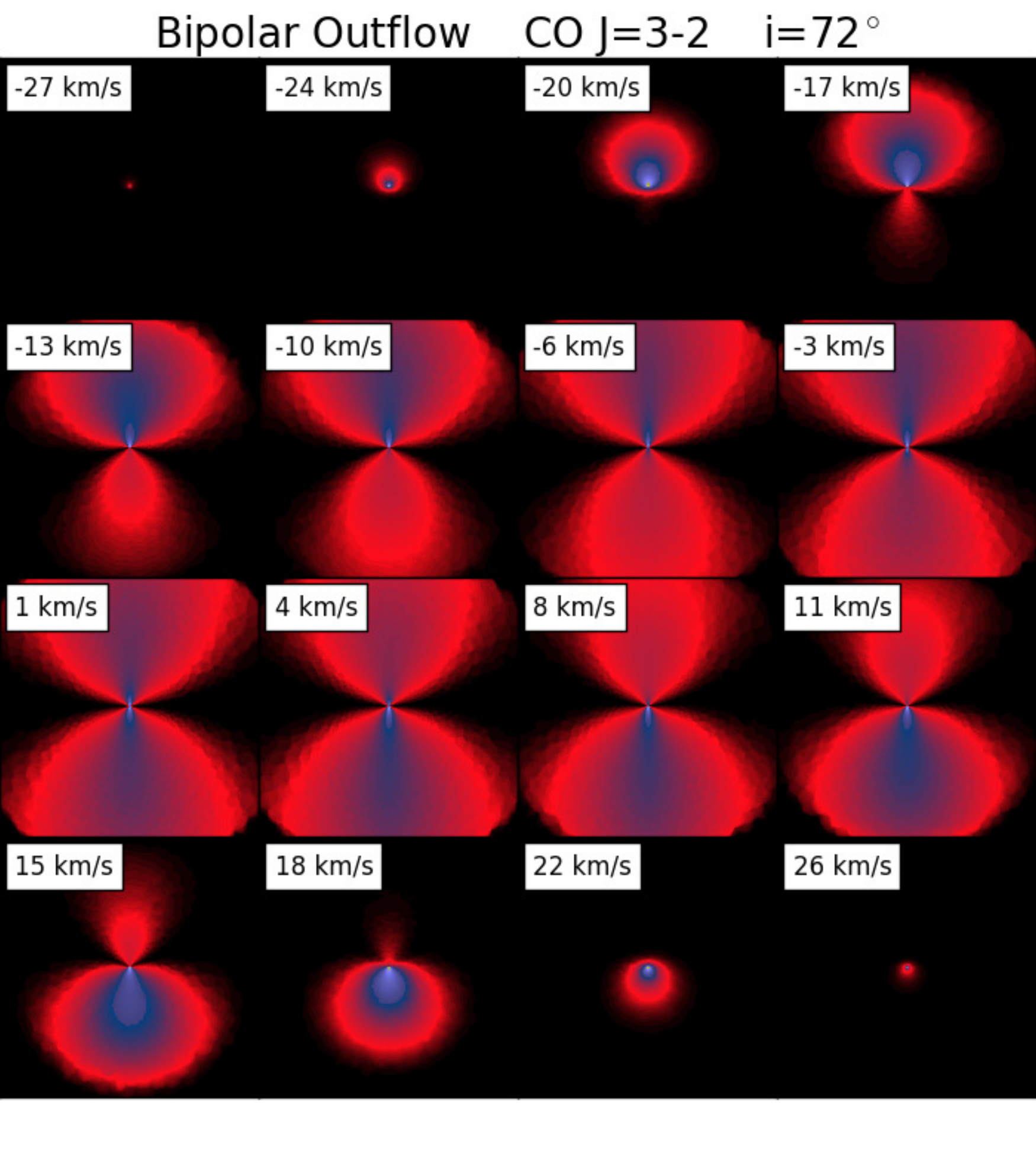}}
 \resizebox{7.5cm}{!}{\includegraphics{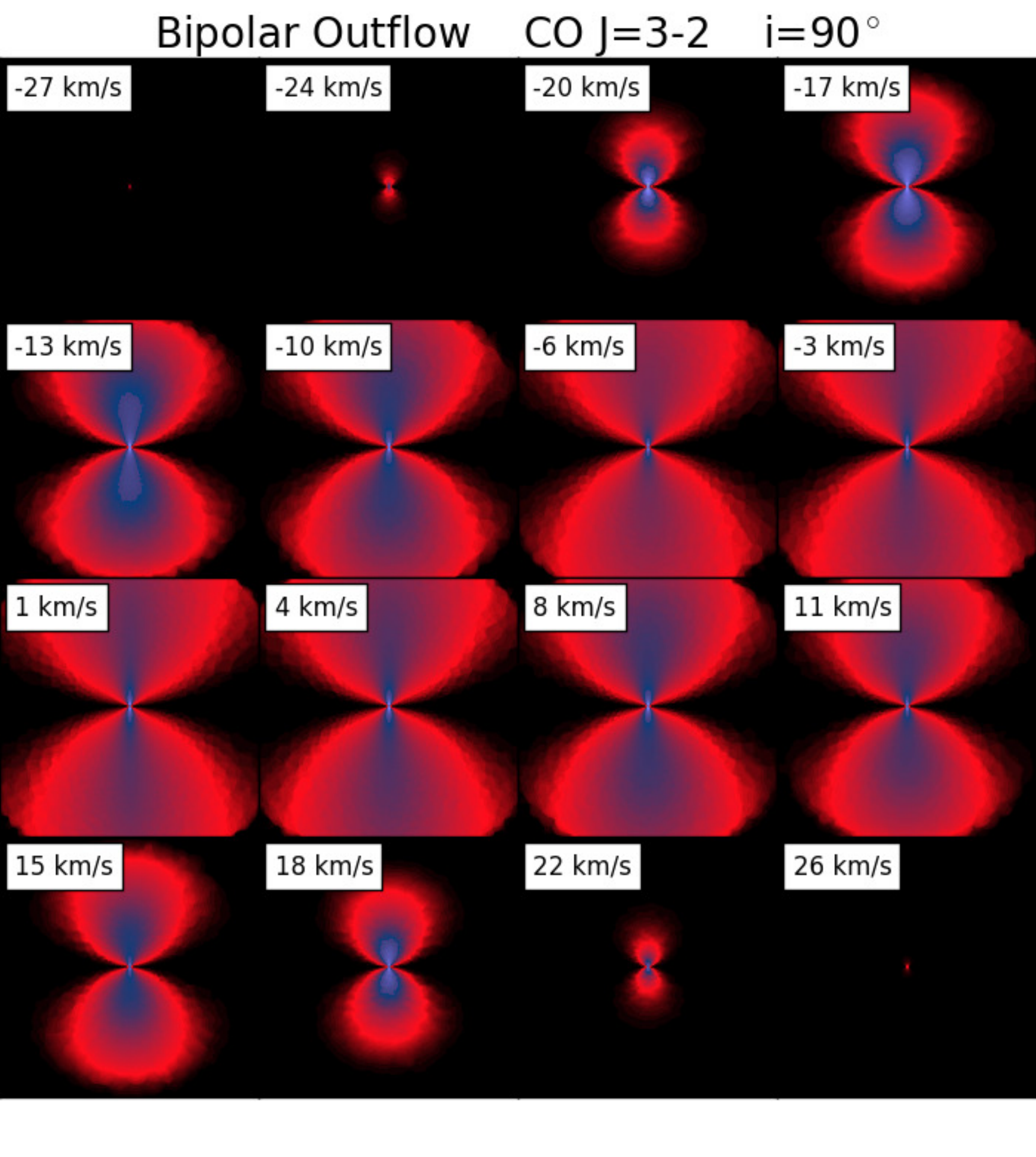}}
 \caption{Channel maps of the model described in the title of each panel plot.}
\end{figure*}

\begin{figure*}[htp]
 \centering
 \resizebox{7.5cm}{!}{\includegraphics{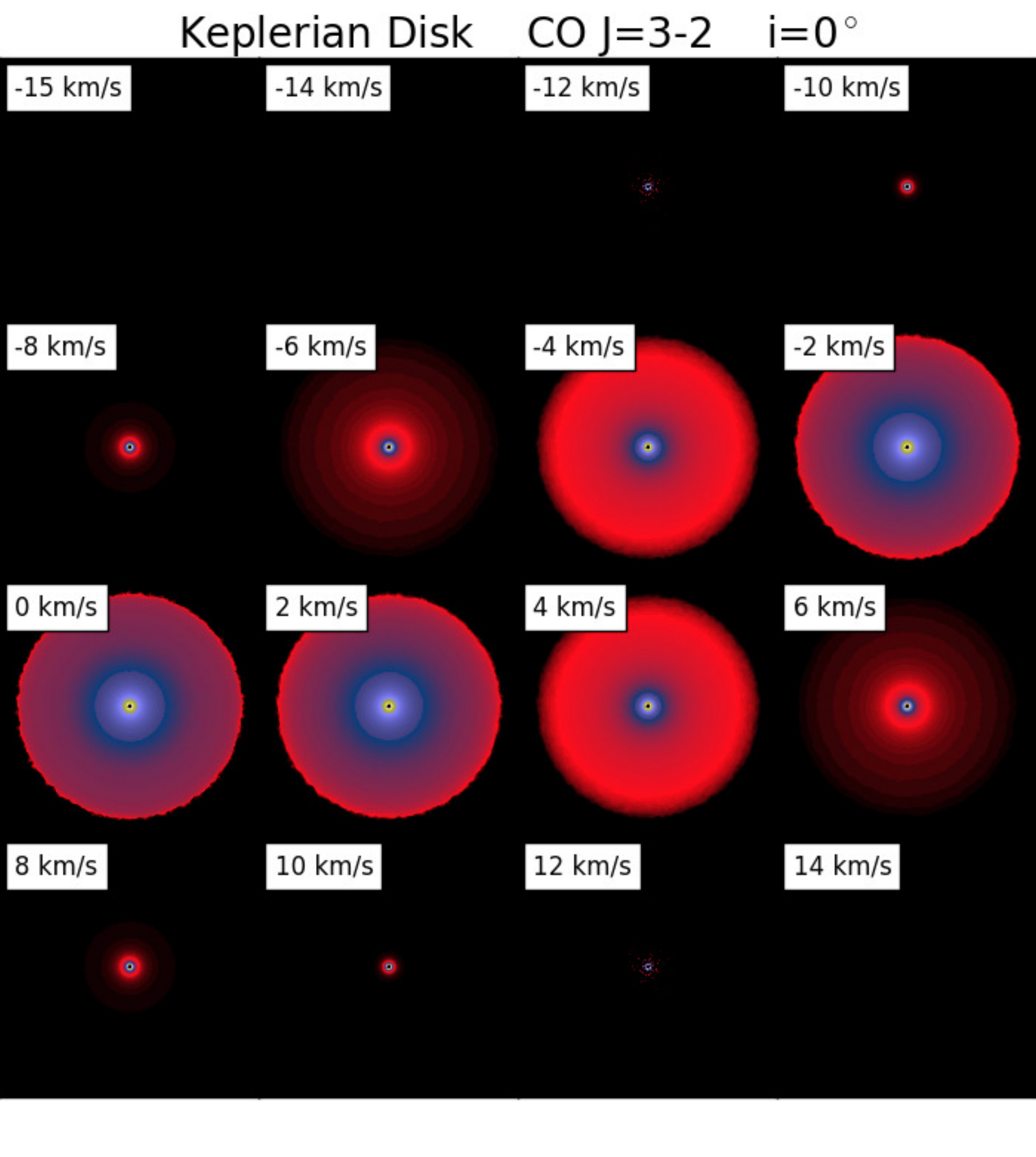}}
 \resizebox{7.5cm}{!}{\includegraphics{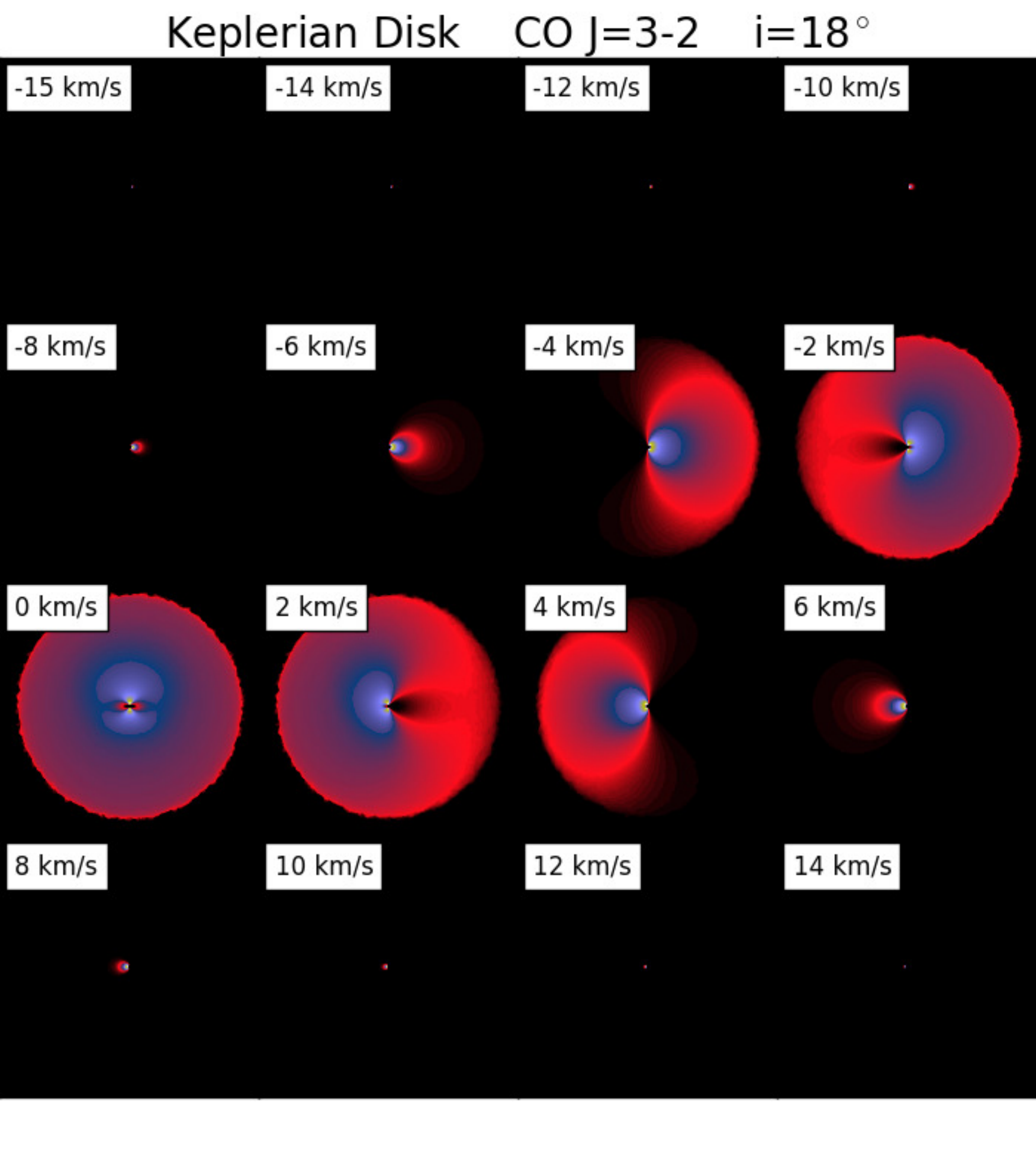}}
 \resizebox{7.5cm}{!}{\includegraphics{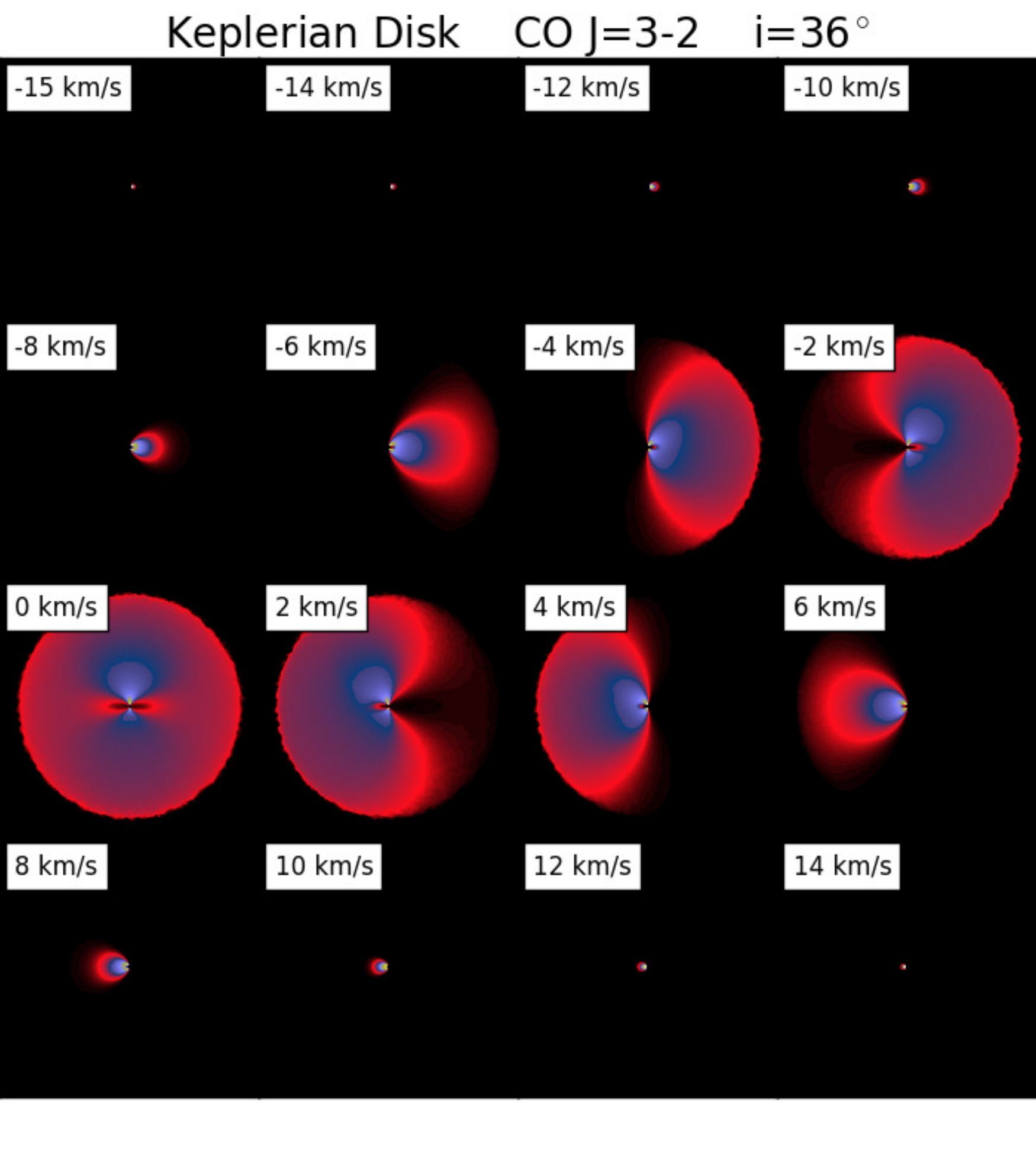}}
 \resizebox{7.5cm}{!}{\includegraphics{Keplerian_i54-eps-converted-to.pdf}}
 \resizebox{7.5cm}{!}{\includegraphics{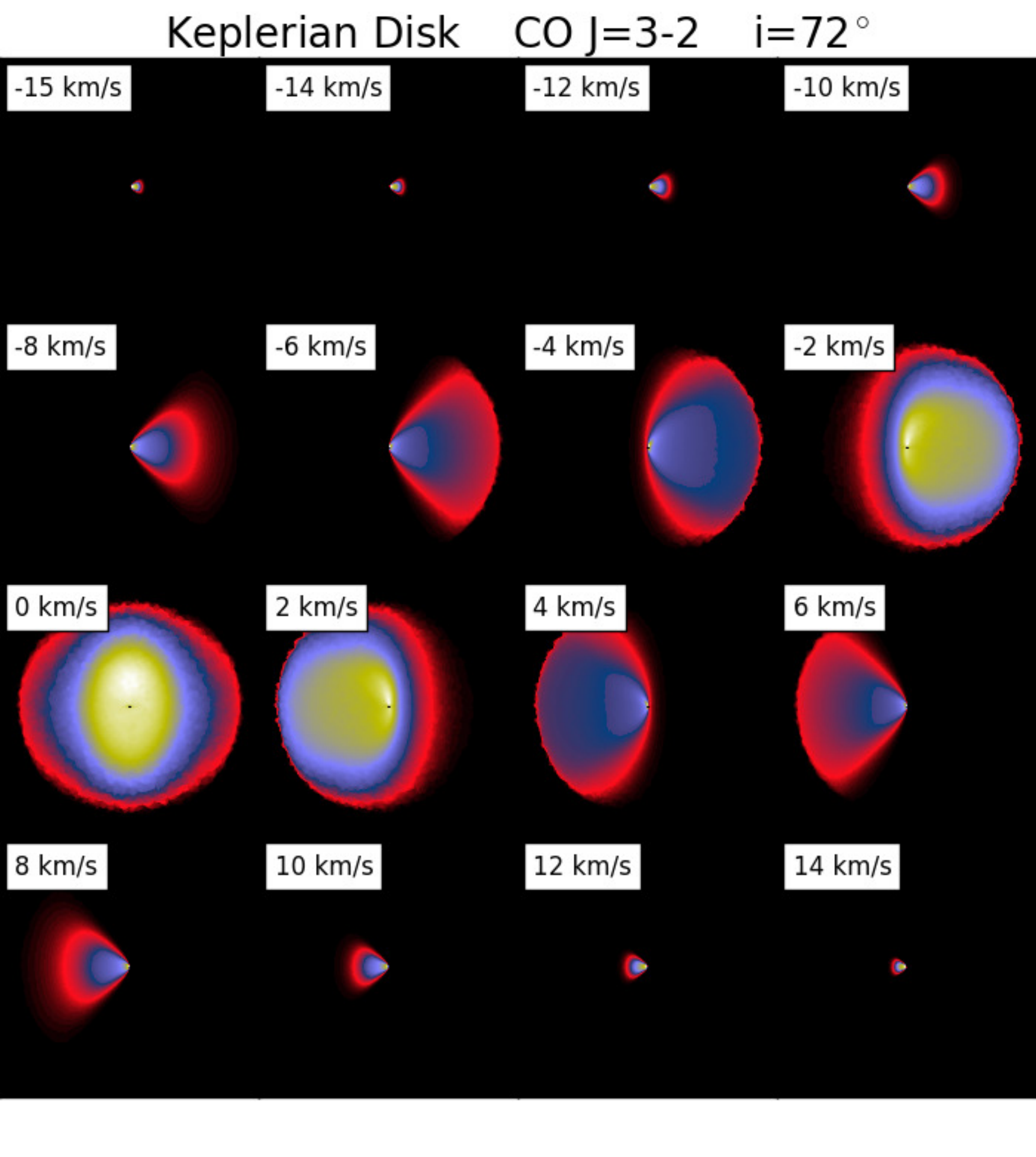}}
 \resizebox{7.5cm}{!}{\includegraphics{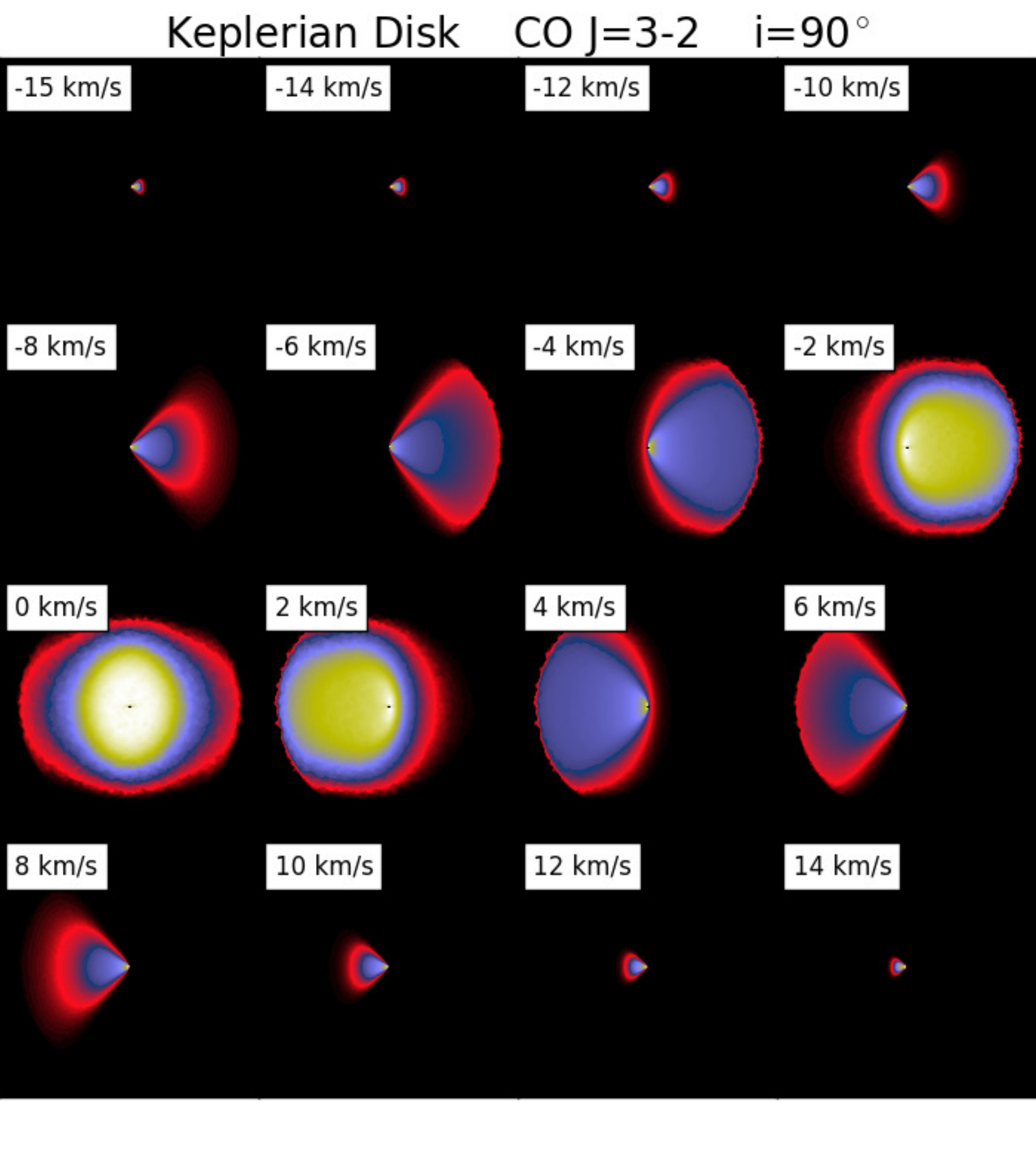}}
 \caption{Channel maps of the model described in the title of each panel plot.}
\end{figure*}

\begin{figure*}[htp]
 \centering
 \resizebox{7.5cm}{!}{\includegraphics{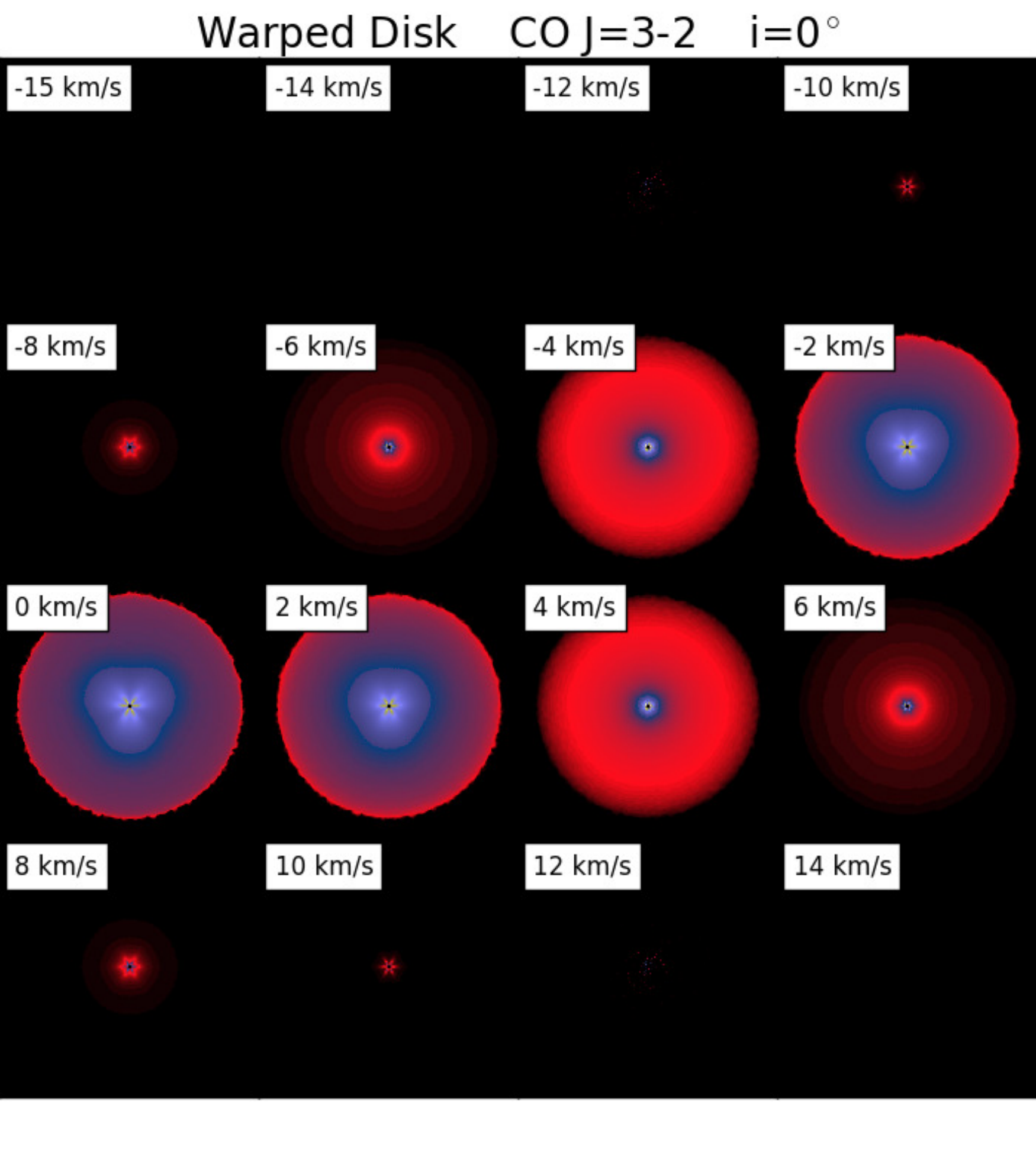}}
 \resizebox{7.5cm}{!}{\includegraphics{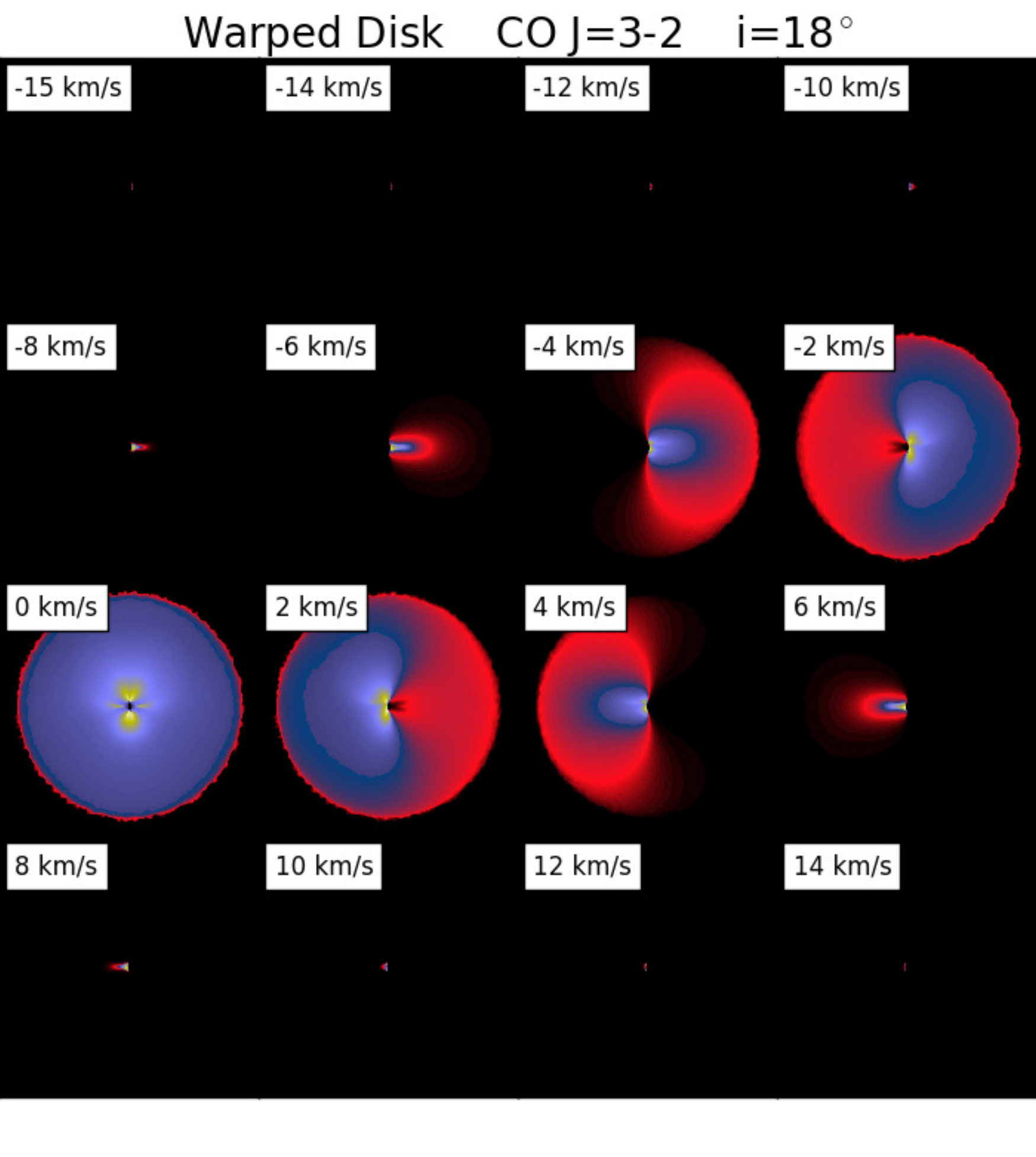}}
 \resizebox{7.5cm}{!}{\includegraphics{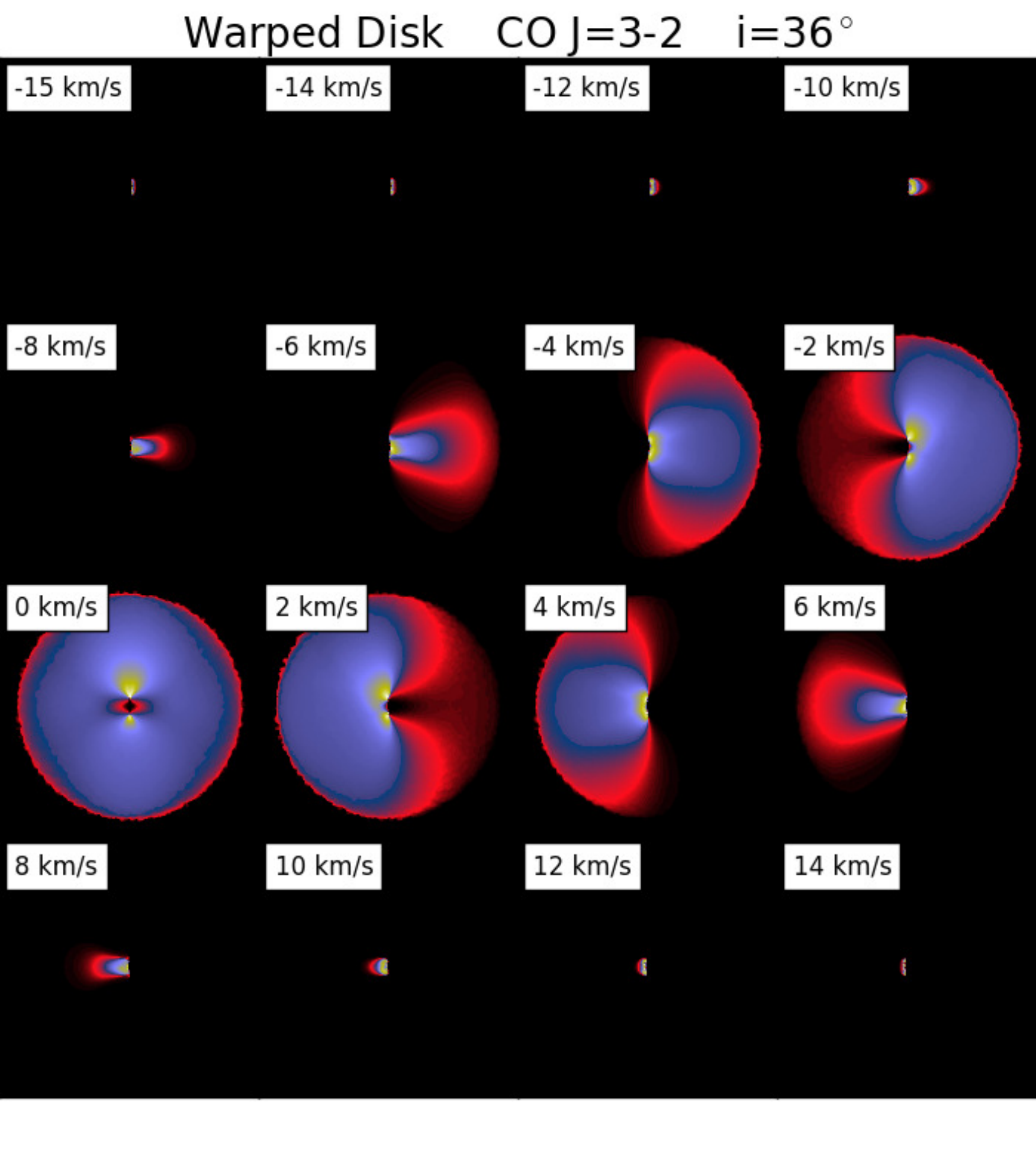}}
 \resizebox{7.5cm}{!}{\includegraphics{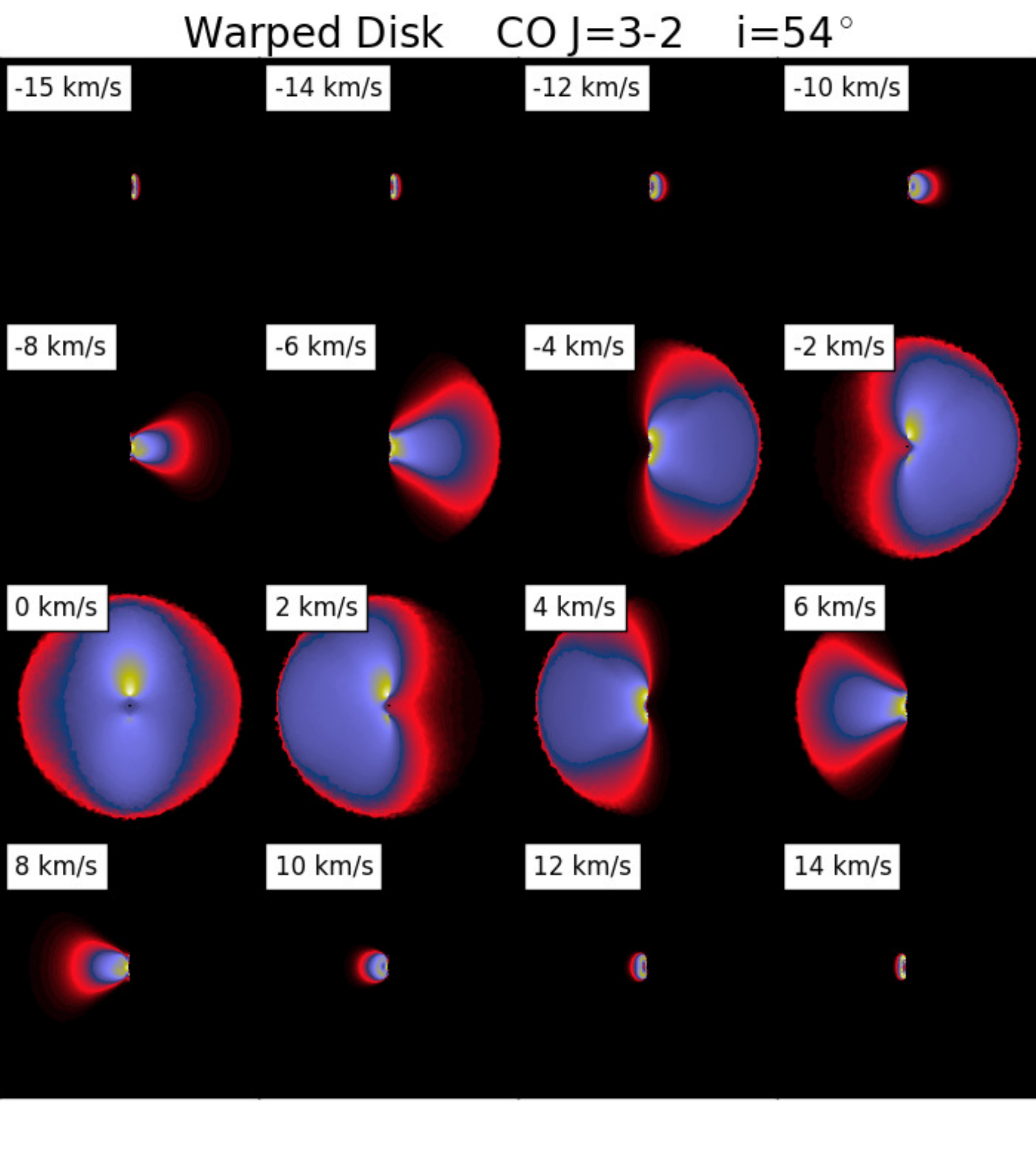}}
 \resizebox{7.5cm}{!}{\includegraphics{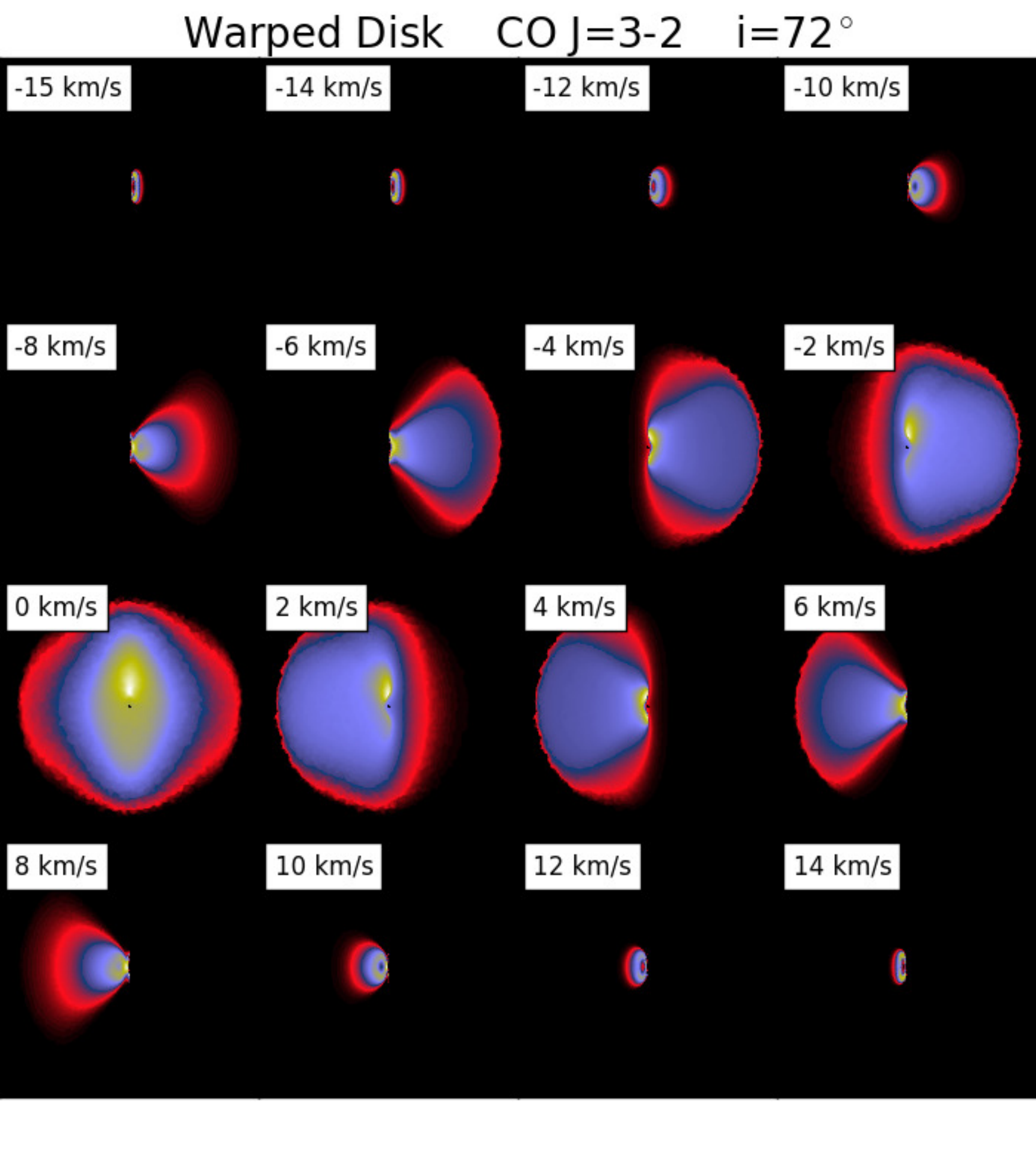}}
 \resizebox{7.5cm}{!}{\includegraphics{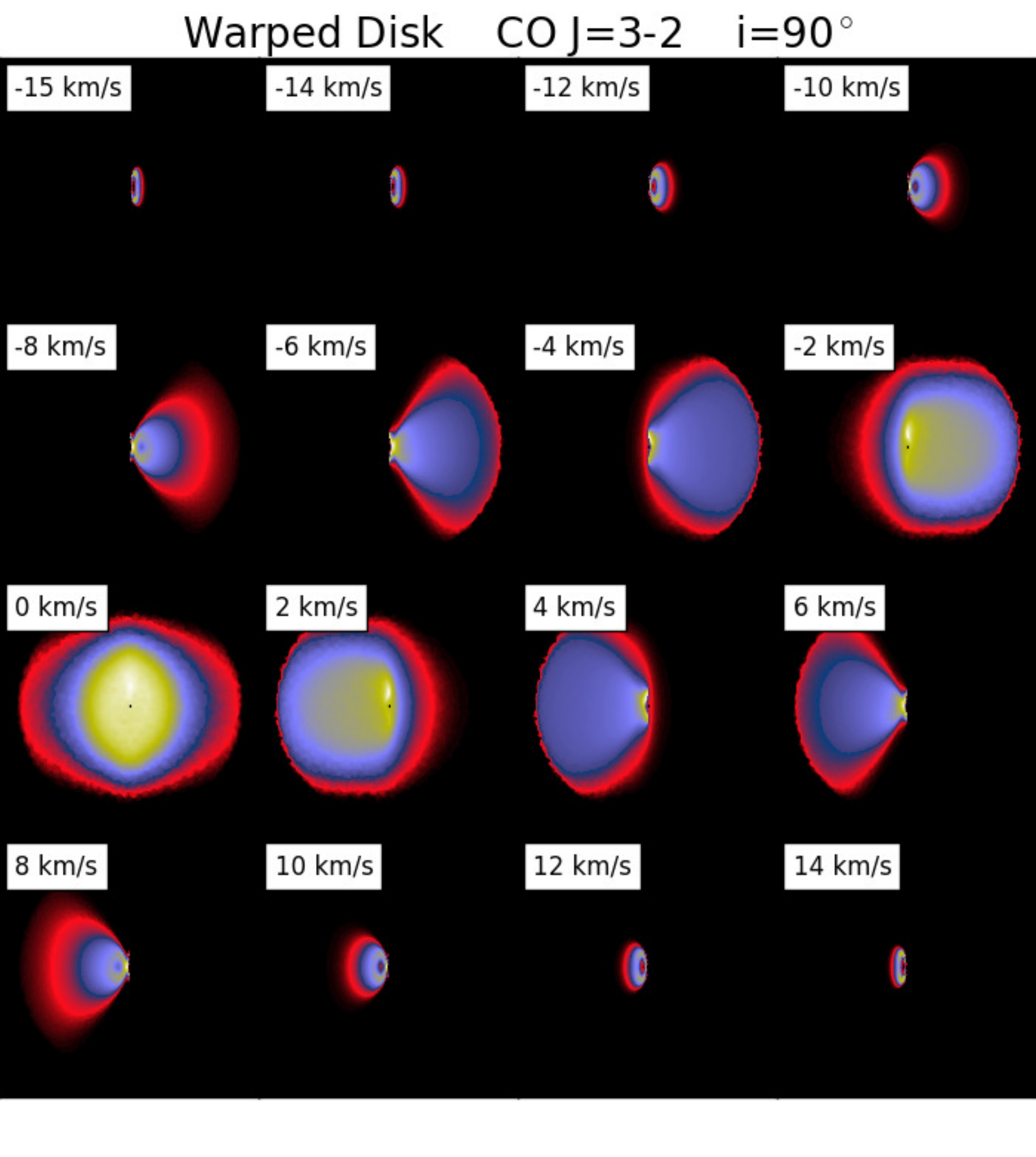}}
 \caption{Channel maps of the model described in the title of each panel plot. \label{warp_chan}}
\end{figure*}

\begin{figure*}[htp]
 \centering
 \resizebox{7.5cm}{!}{\includegraphics{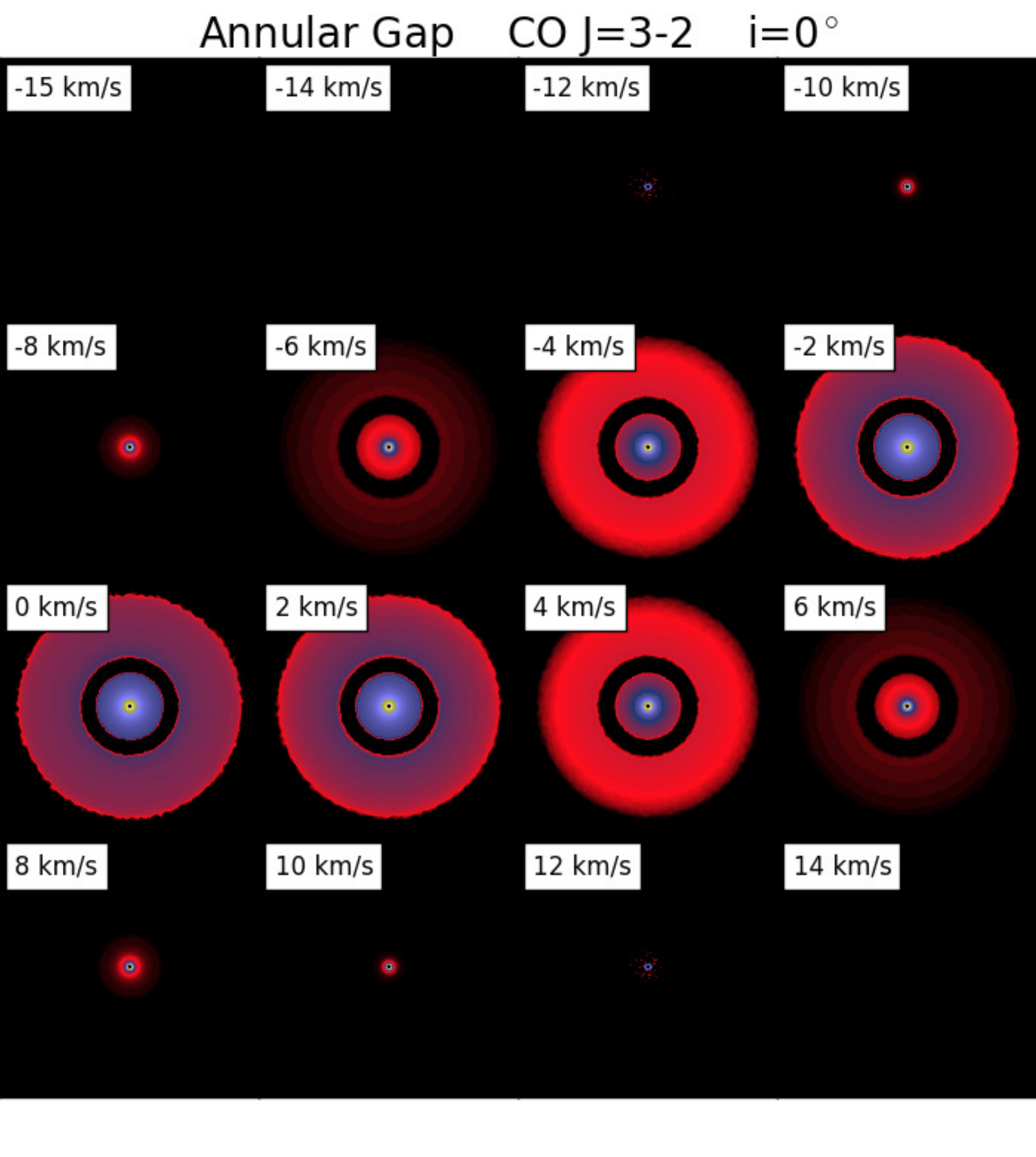}}
 \resizebox{7.5cm}{!}{\includegraphics{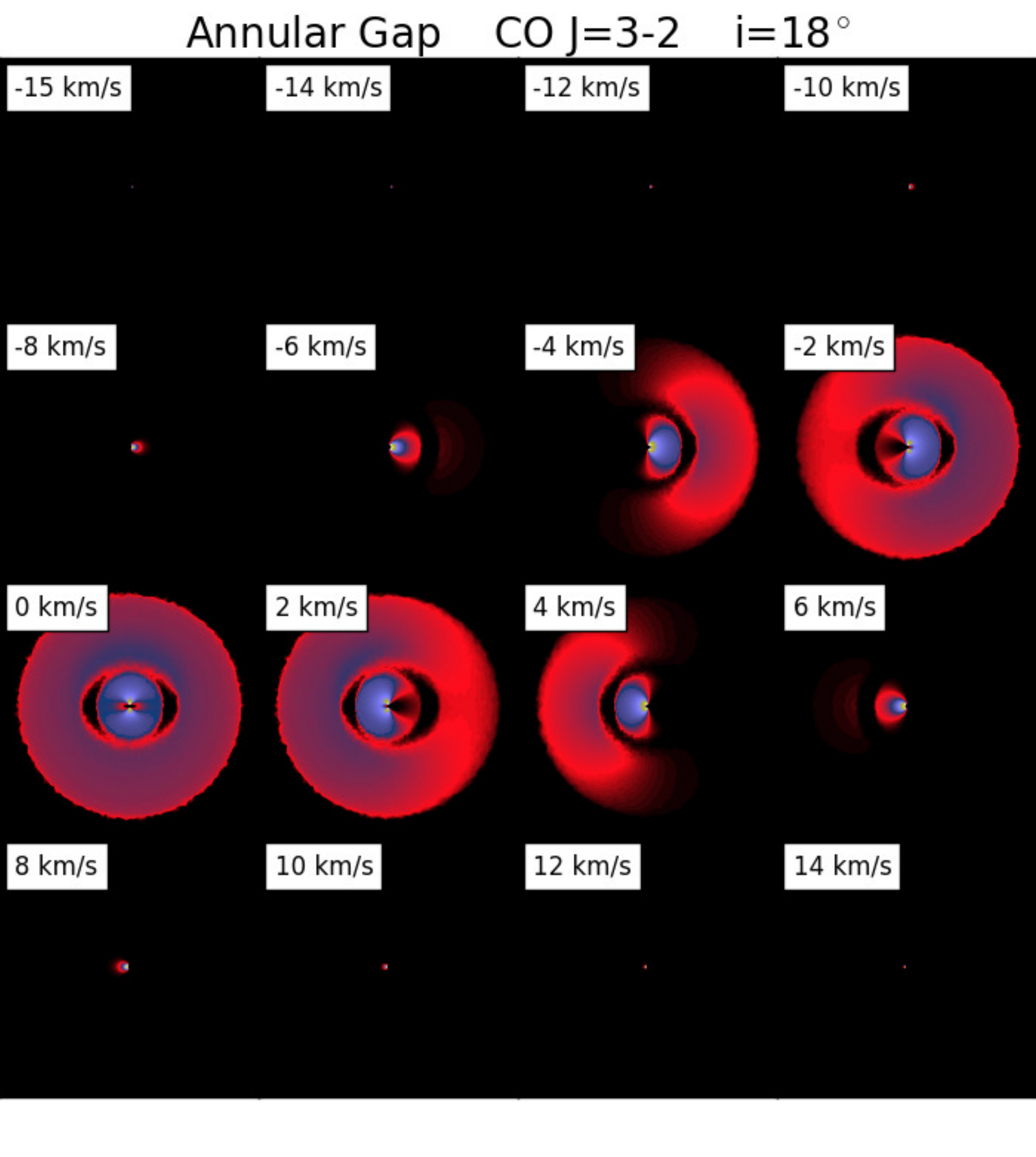}}
 \resizebox{7.5cm}{!}{\includegraphics{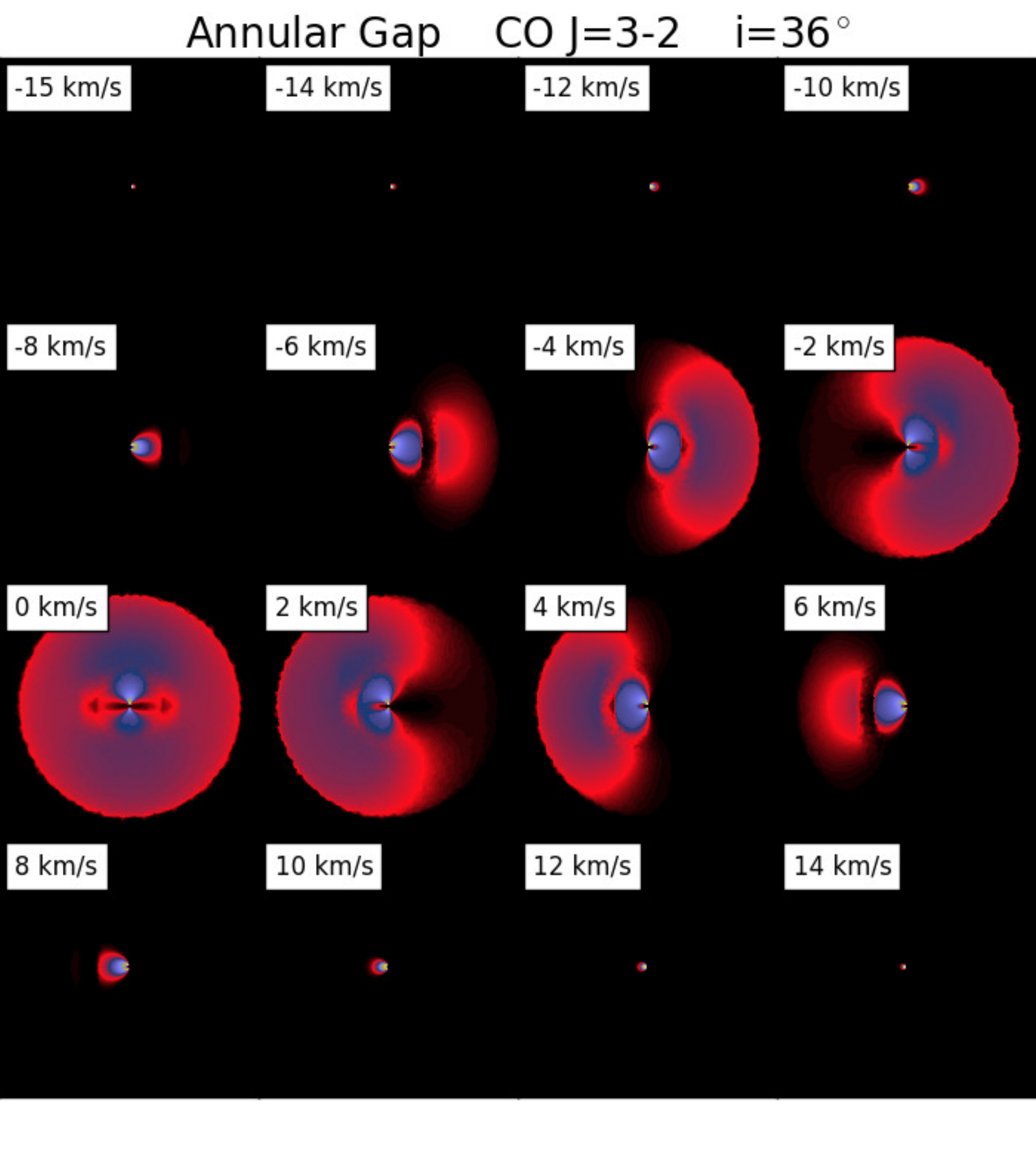}}
 \resizebox{7.5cm}{!}{\includegraphics{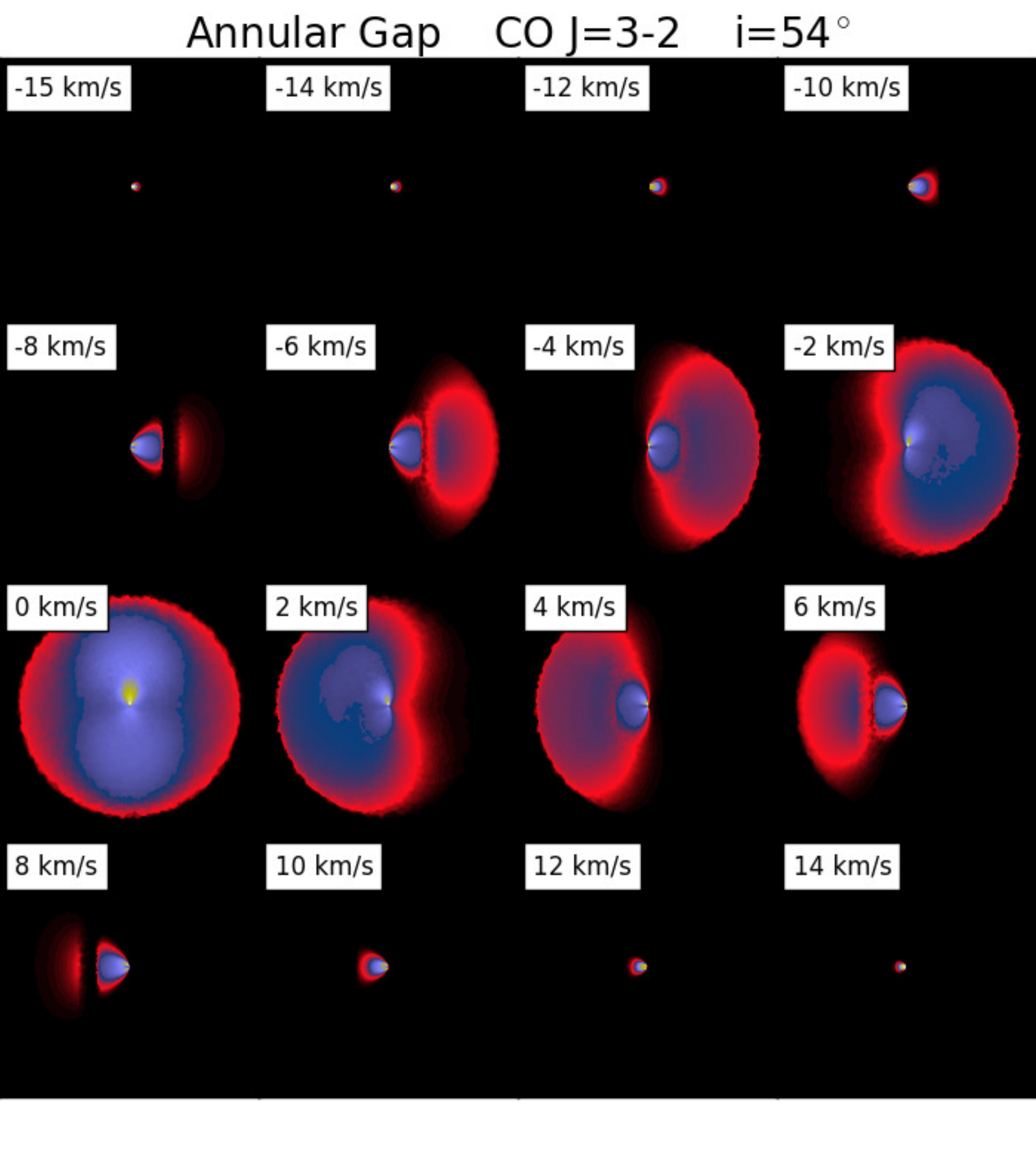}}
 \resizebox{7.5cm}{!}{\includegraphics{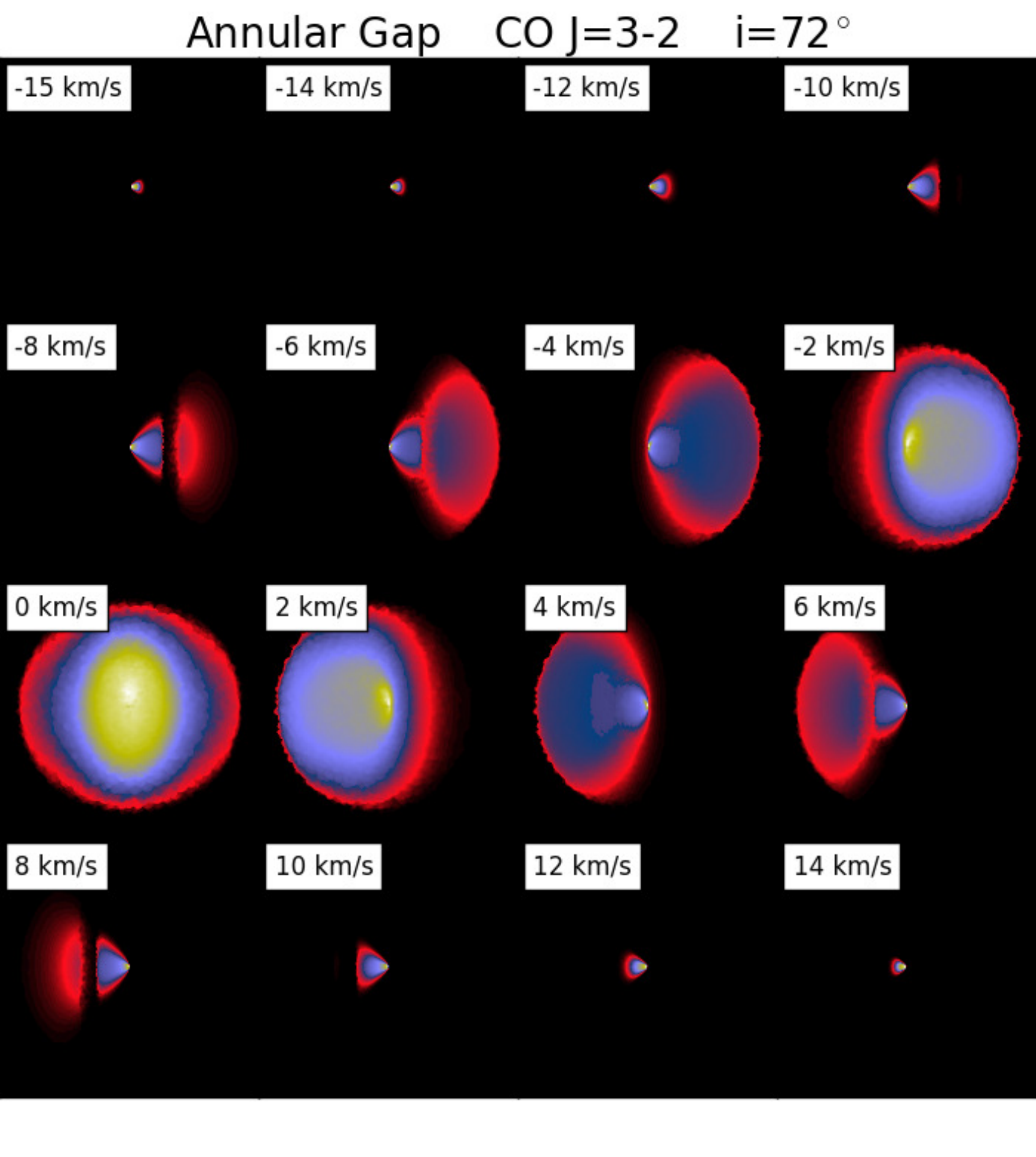}}
 \resizebox{7.5cm}{!}{\includegraphics{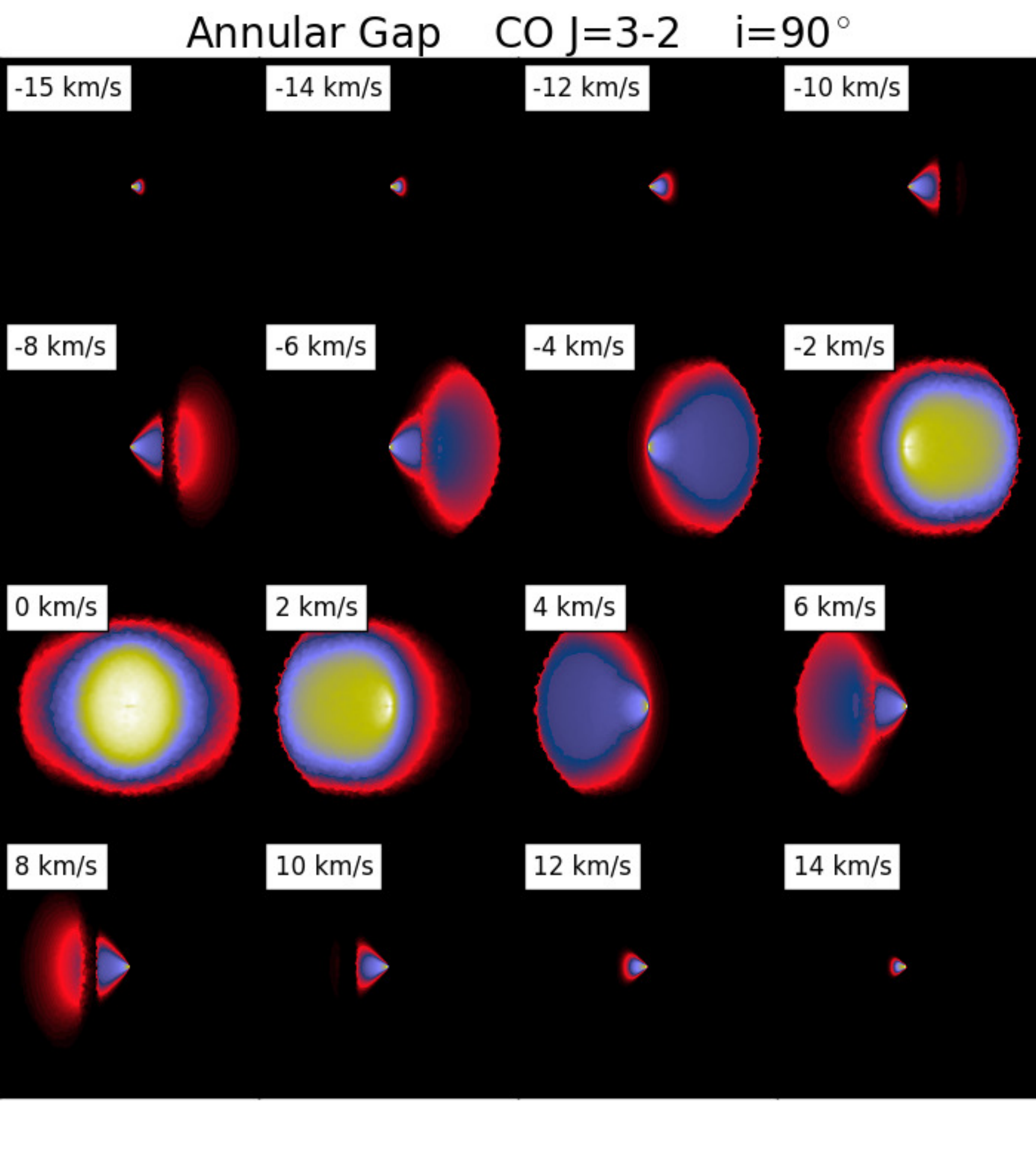}}
 \caption{Channel maps of the model described in the title of each panel plot. \label{gap_chan}}
\end{figure*}

\begin{figure*}[htp]
 \centering
 \resizebox{7.5cm}{!}{\includegraphics{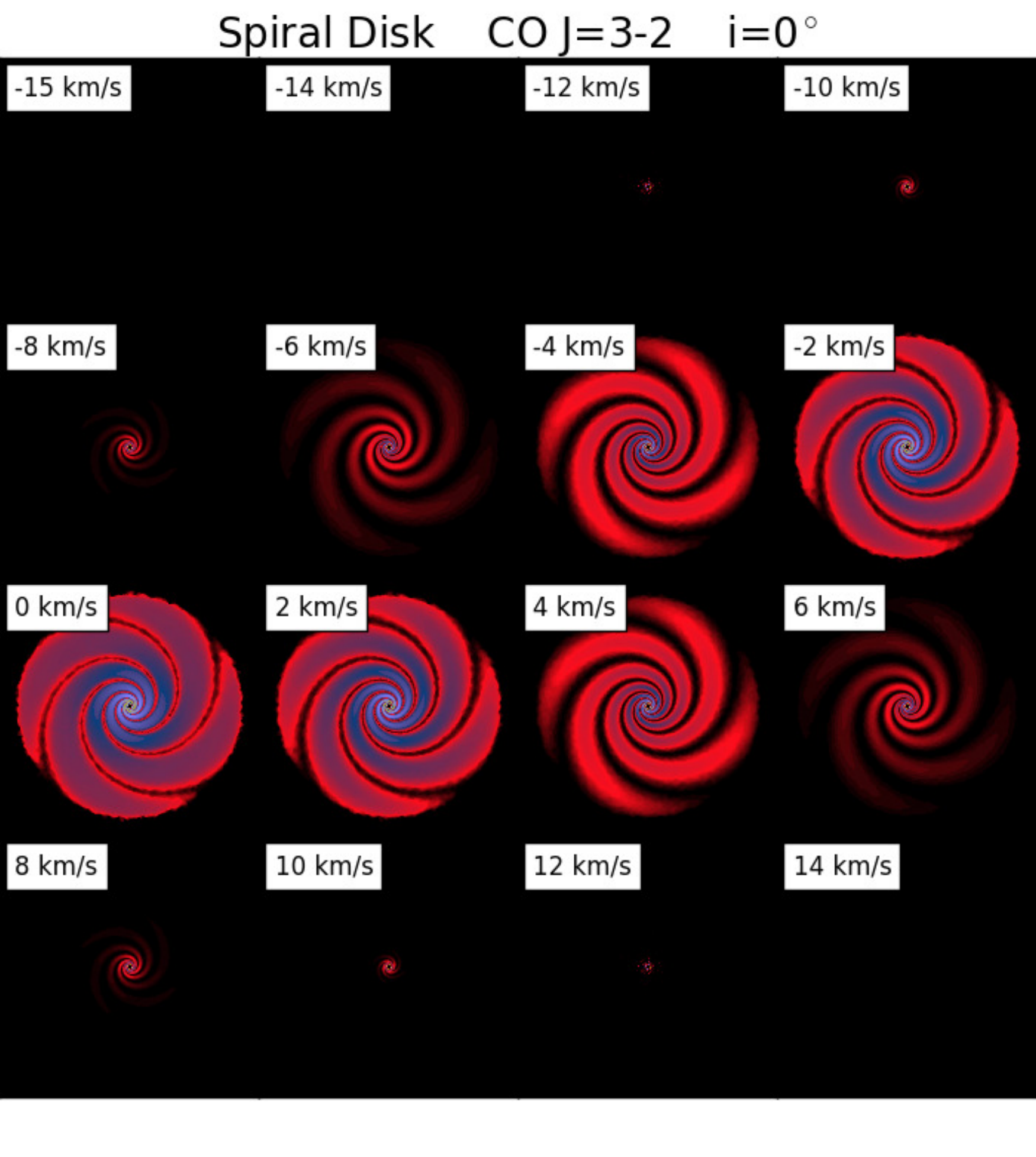}}
 \resizebox{7.5cm}{!}{\includegraphics{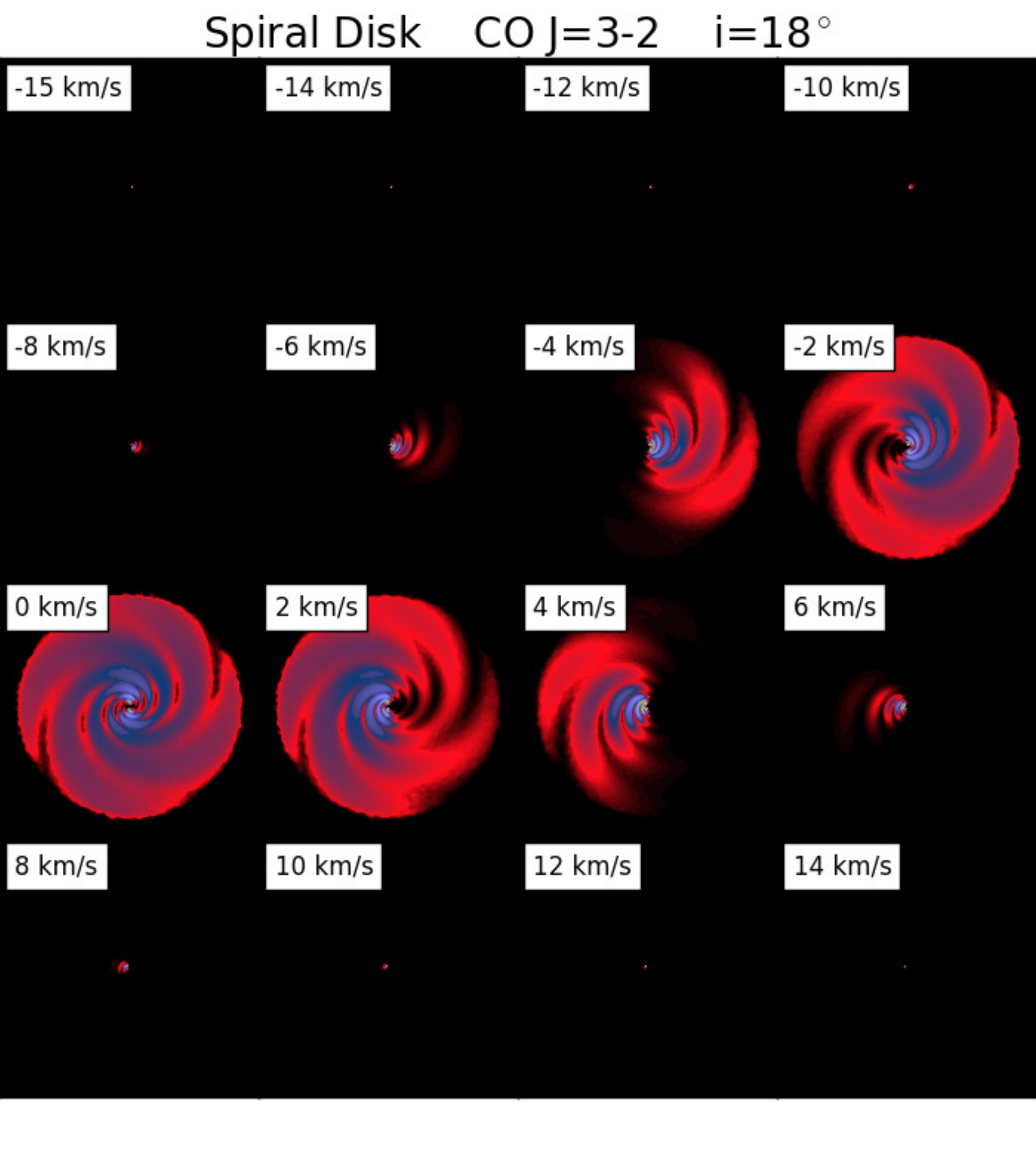}}
 \resizebox{7.5cm}{!}{\includegraphics{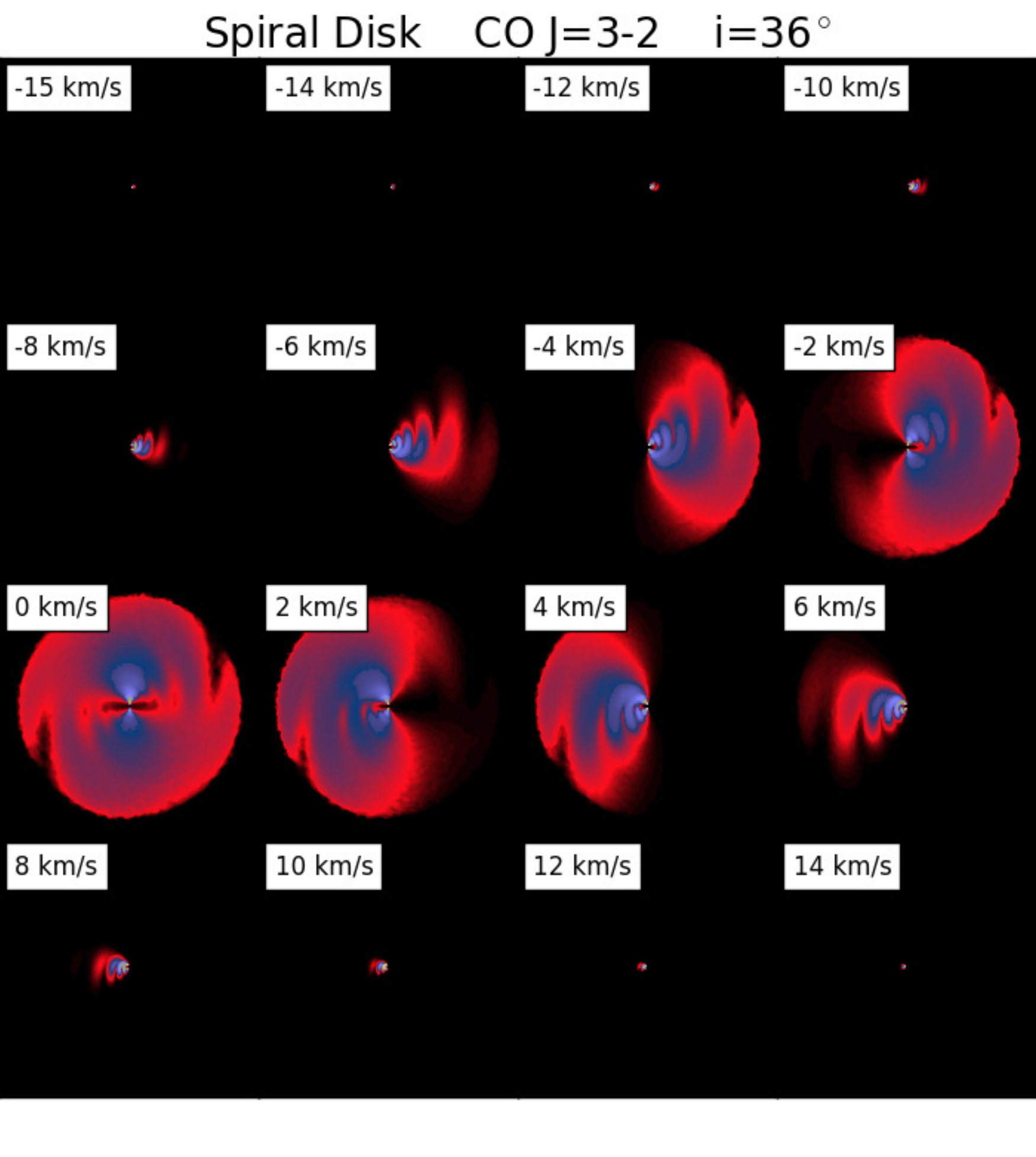}}
 \resizebox{7.5cm}{!}{\includegraphics{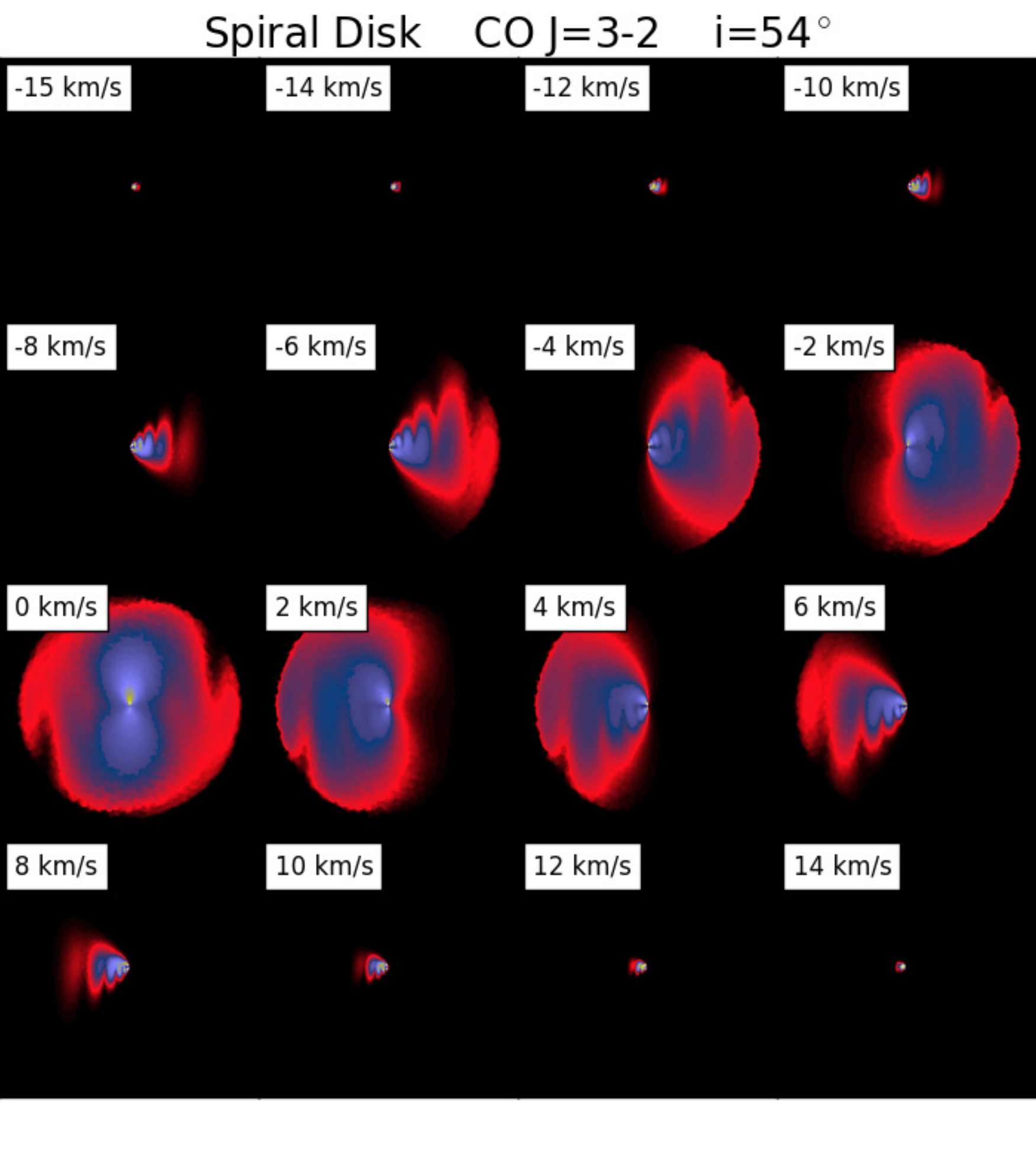}}
 \resizebox{7.5cm}{!}{\includegraphics{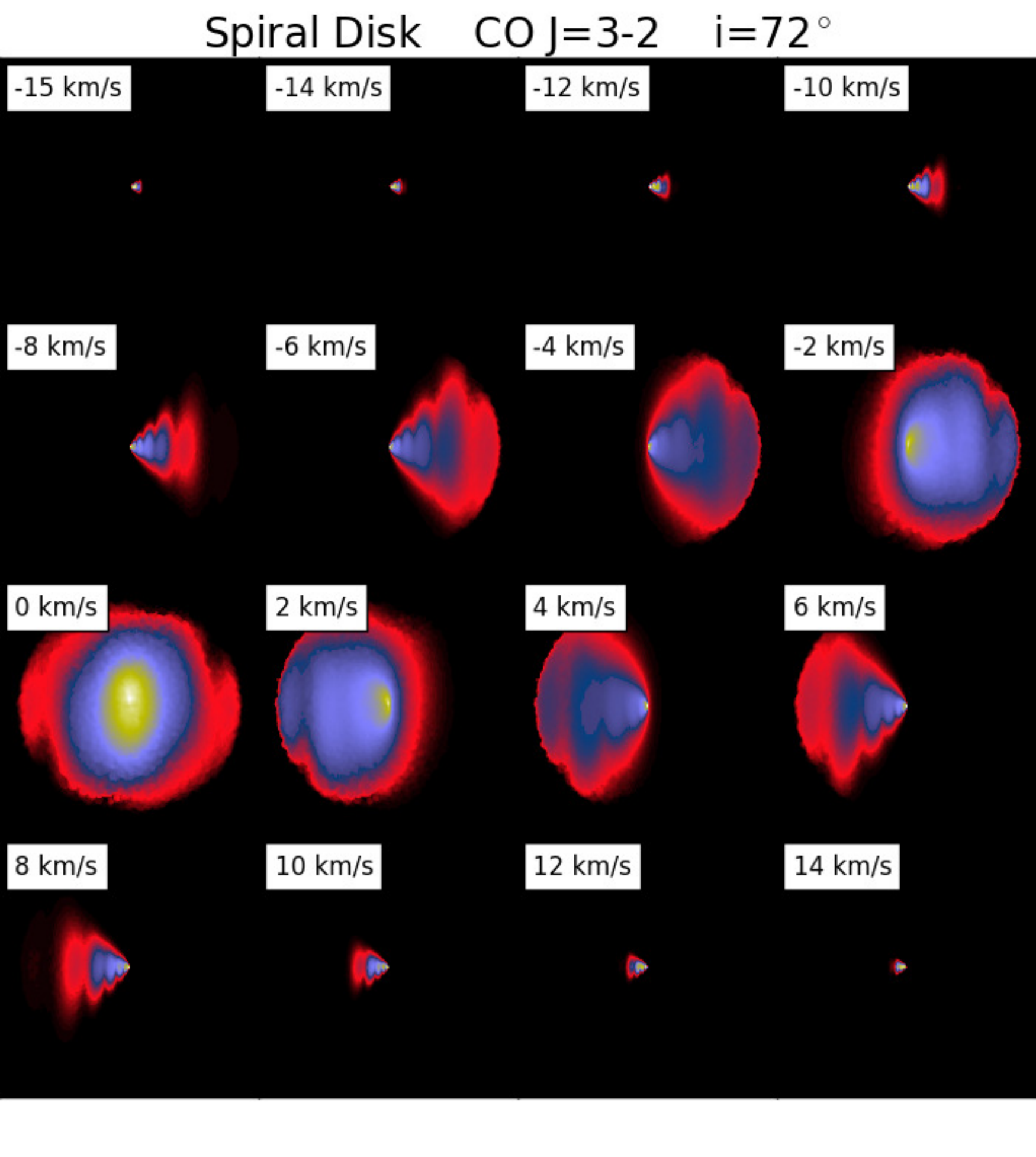}}
 \resizebox{7.5cm}{!}{\includegraphics{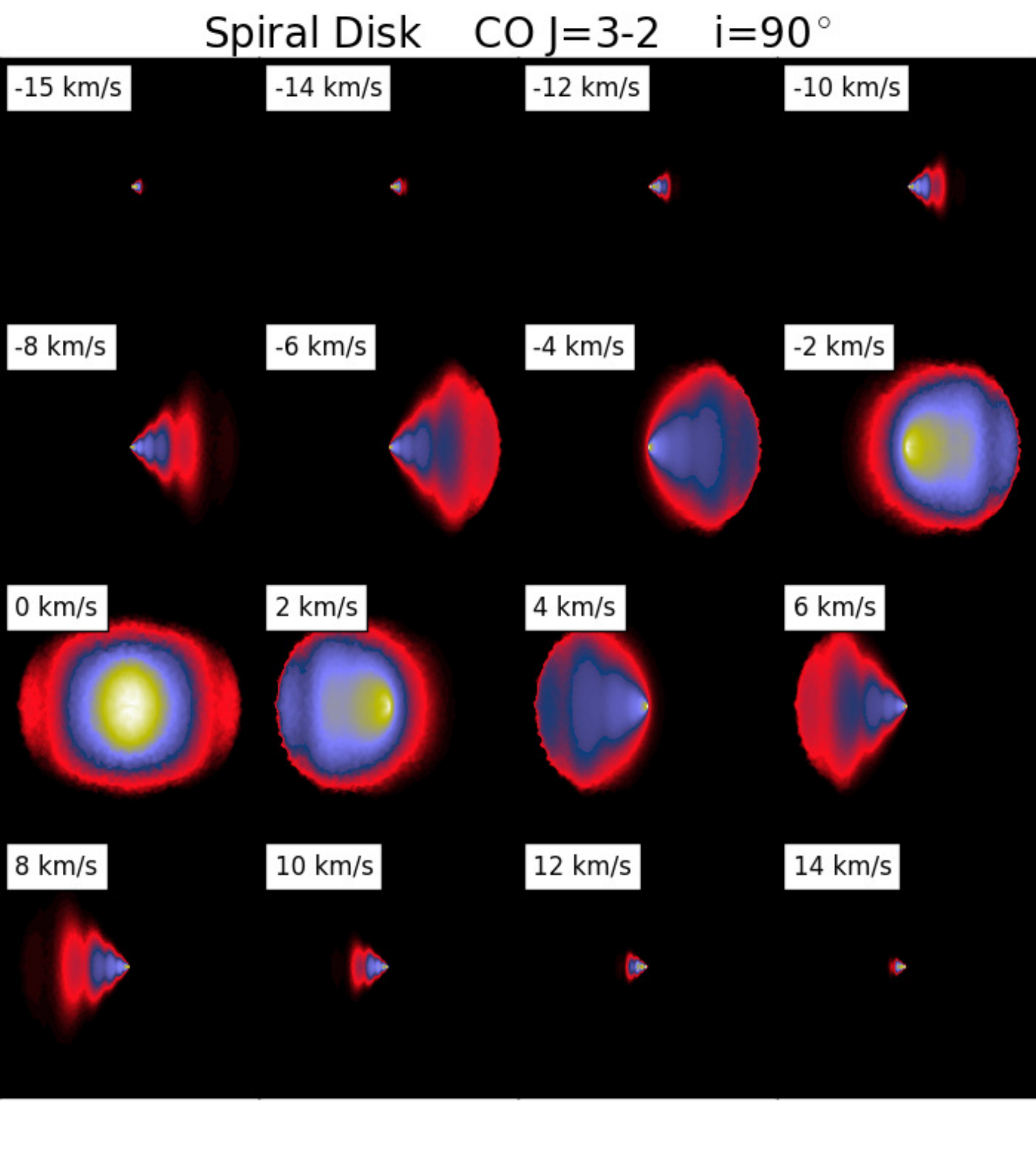}}
 \caption{Channel maps of the model described in the title of each panel plot. \label{spir_chan}}
\end{figure*}

\begin{figure*}[htp]
 \centering
 \resizebox{7.5cm}{!}{\includegraphics{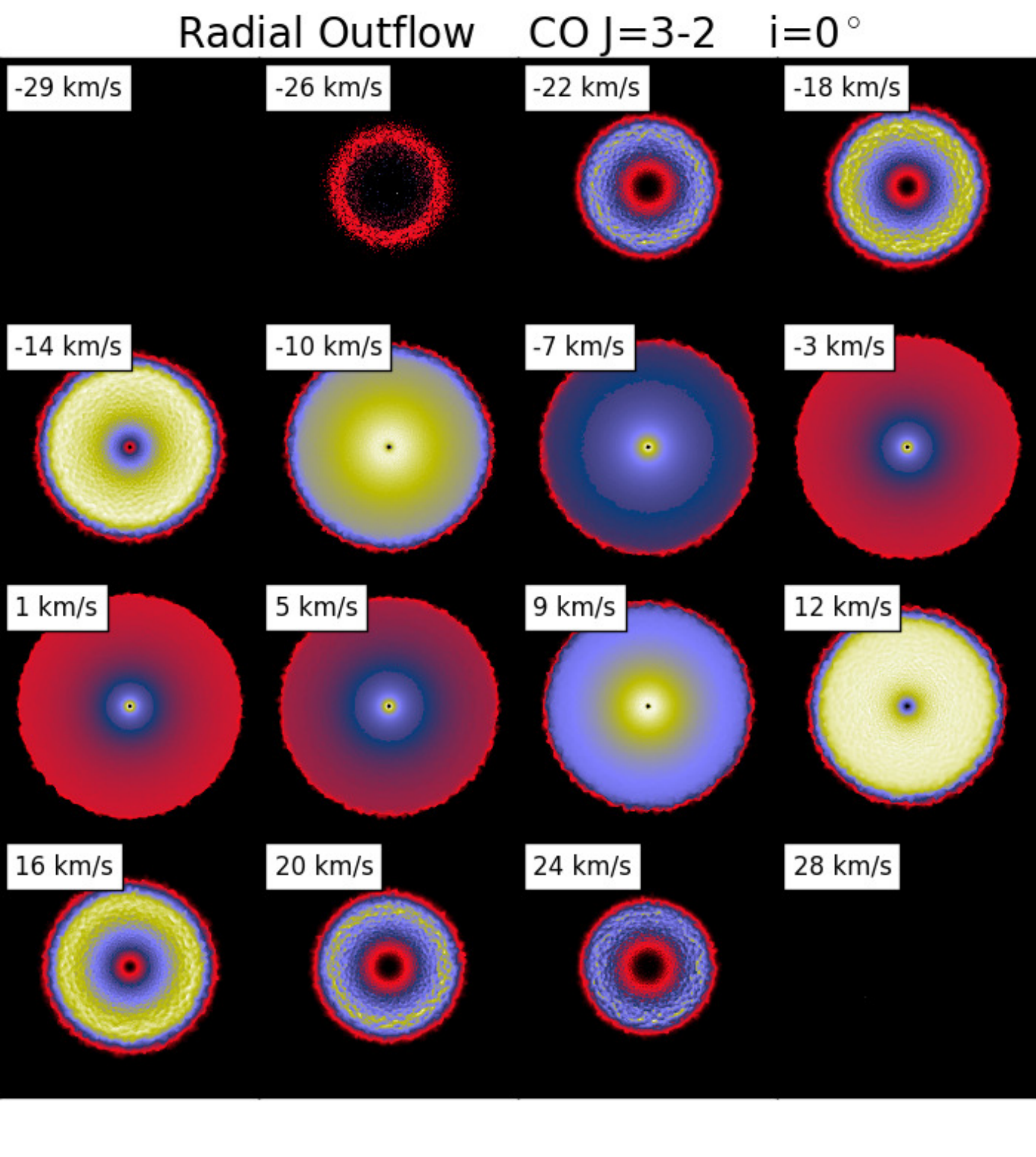}}
 \resizebox{7.5cm}{!}{\includegraphics{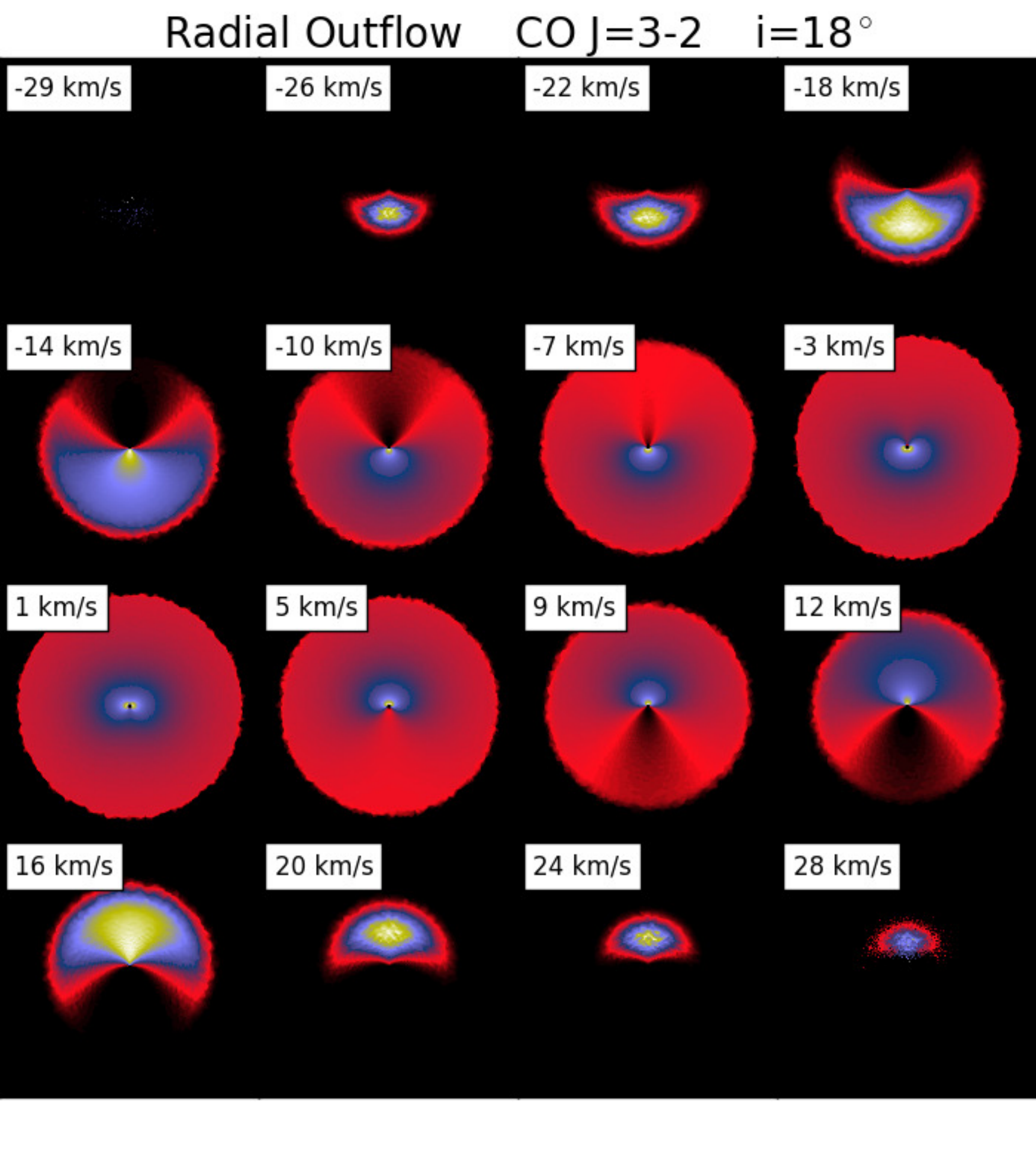}}
 \resizebox{7.5cm}{!}{\includegraphics{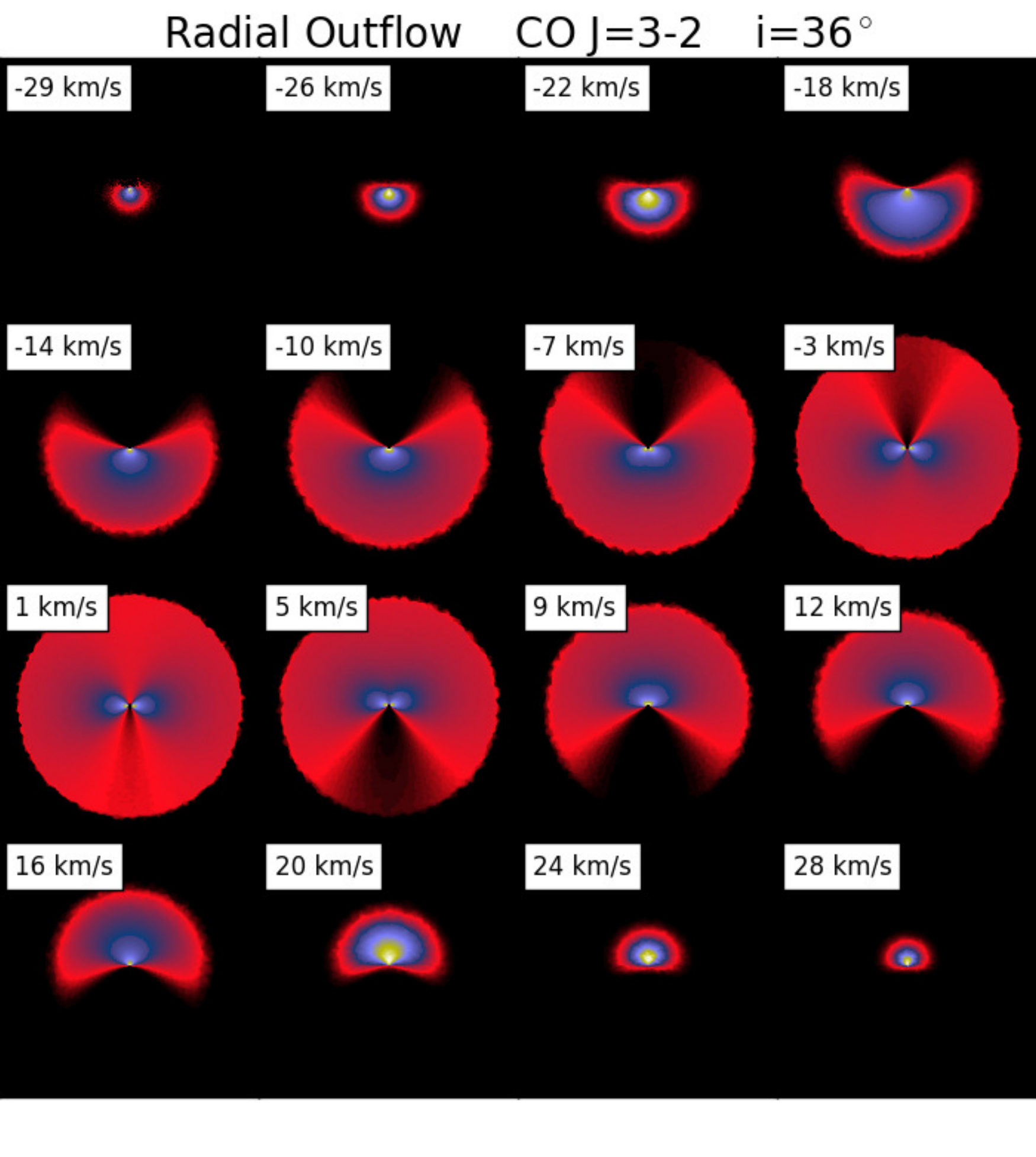}}
 \resizebox{7.5cm}{!}{\includegraphics{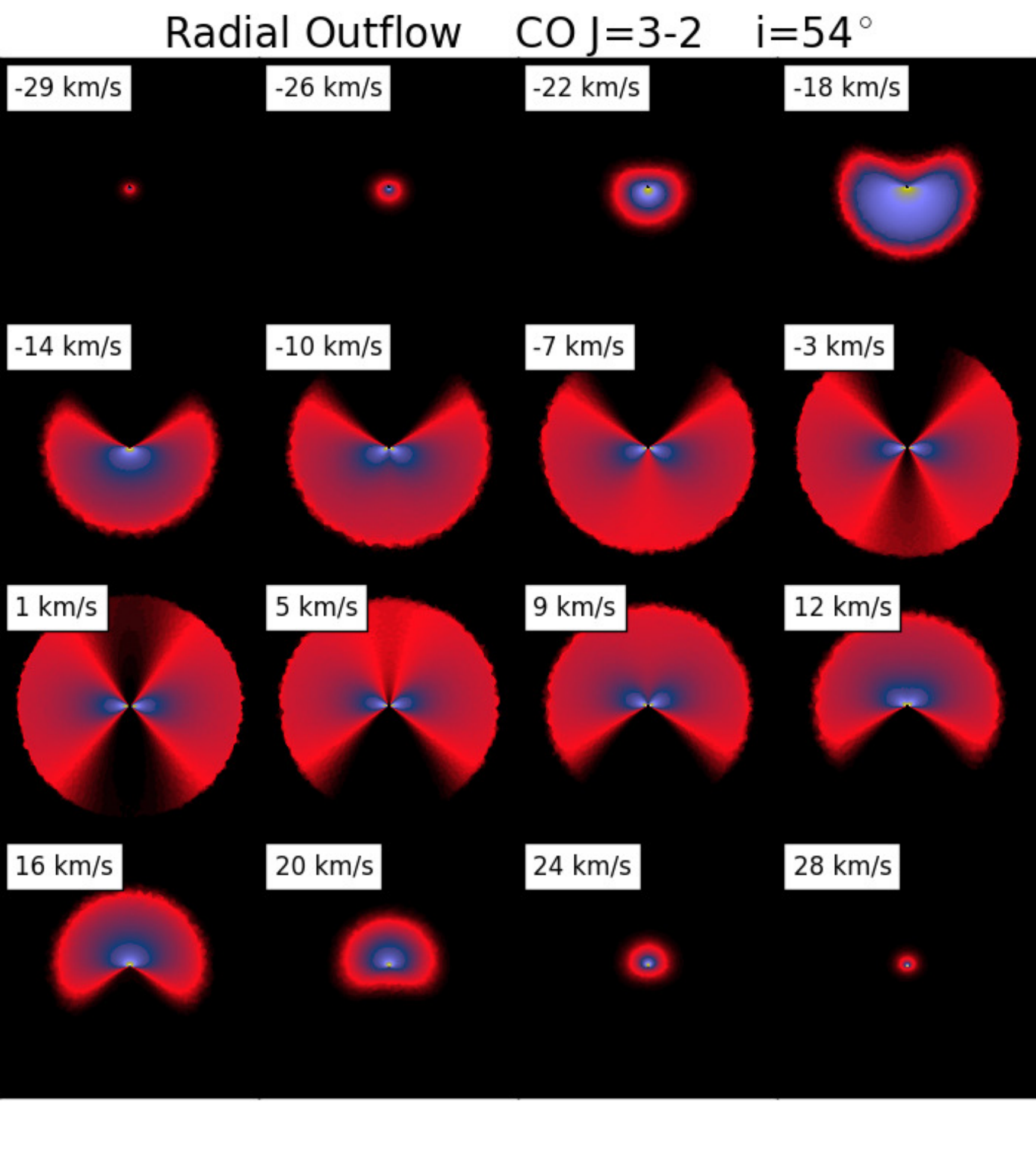}}
 \resizebox{7.5cm}{!}{\includegraphics{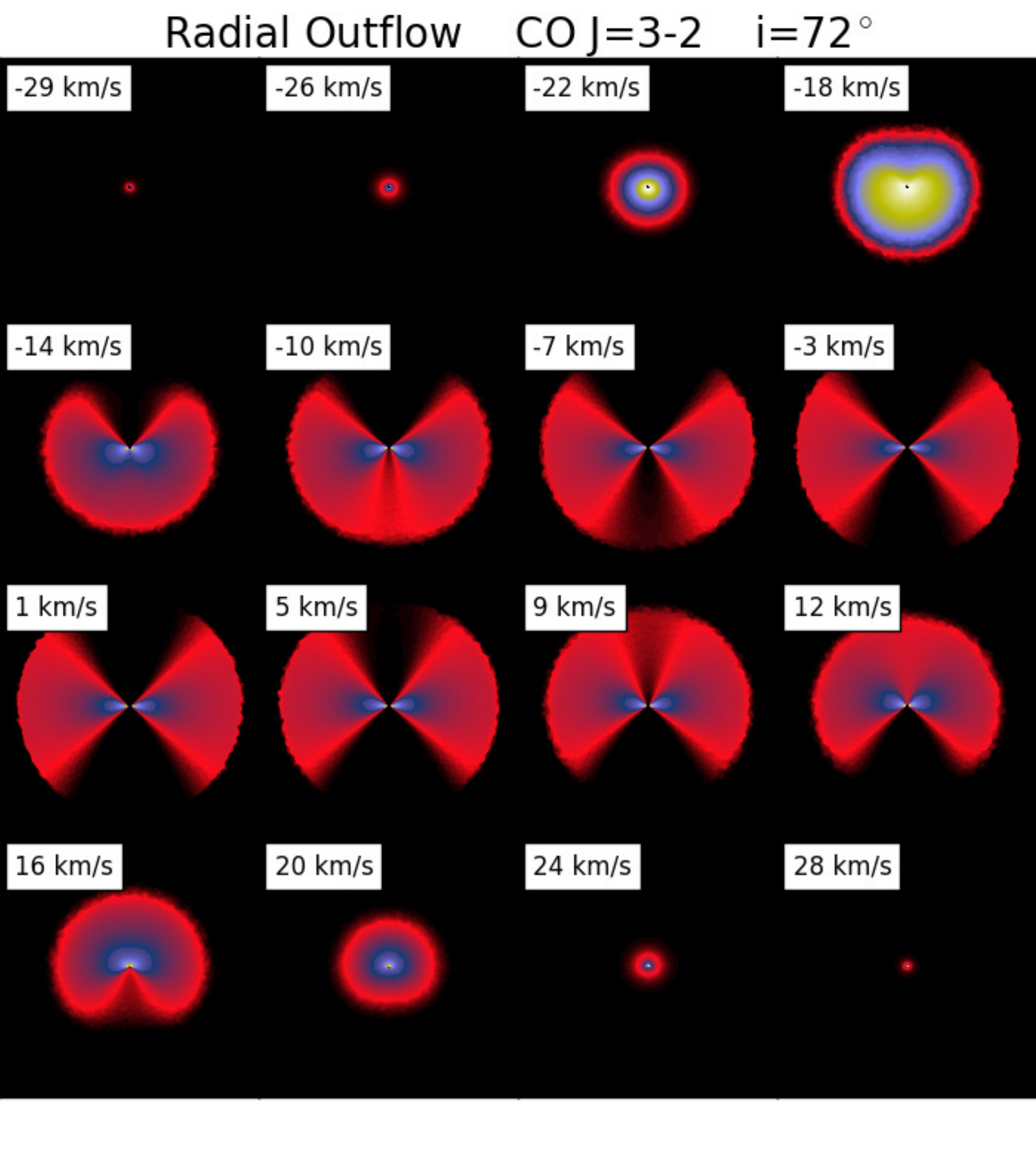}}
 \resizebox{7.5cm}{!}{\includegraphics{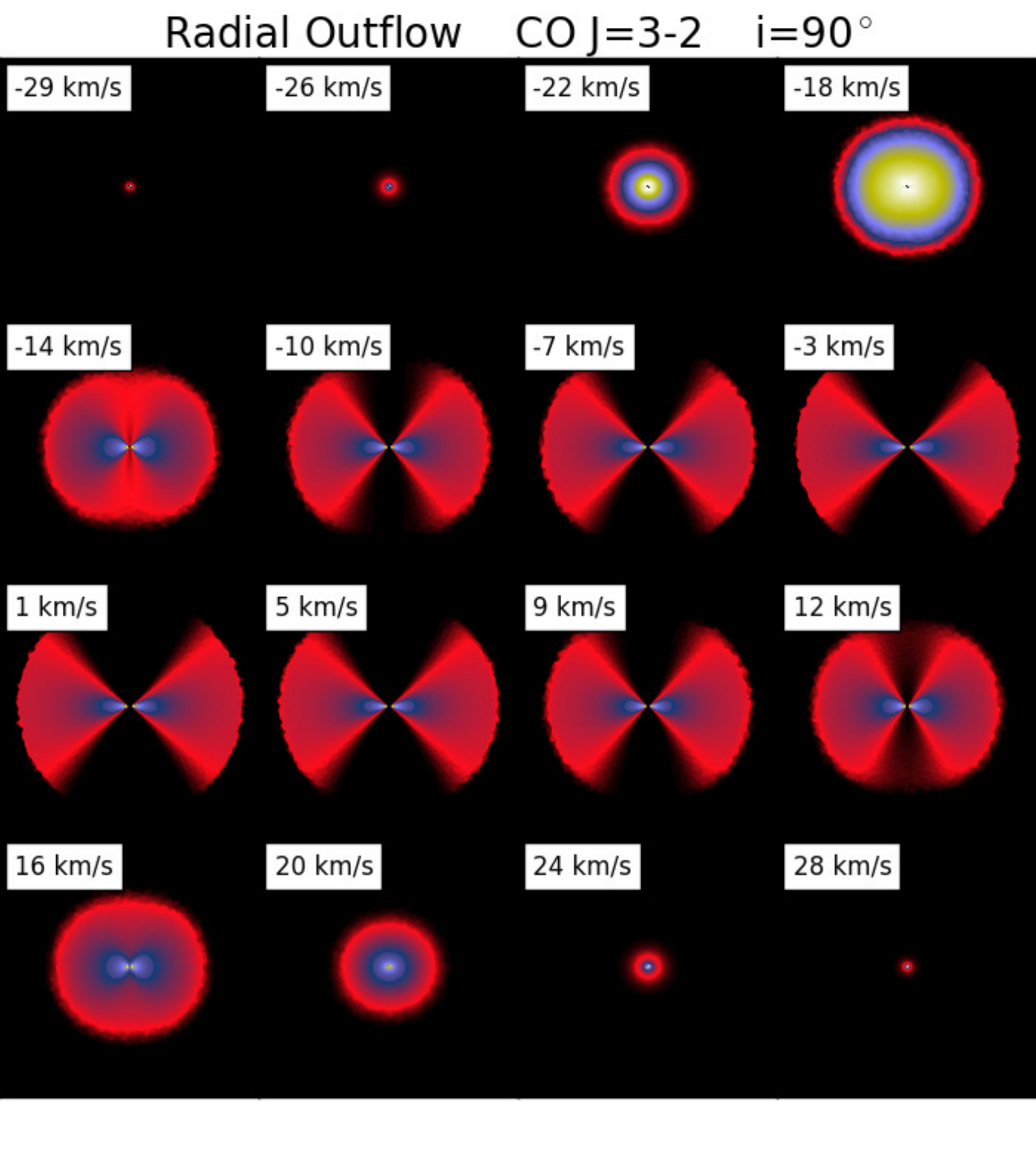}}
 \caption{Channel maps of the model described in the title of each panel plot.}
\end{figure*}

\begin{figure*}[htp]
 \centering
 \resizebox{7.5cm}{!}{\includegraphics{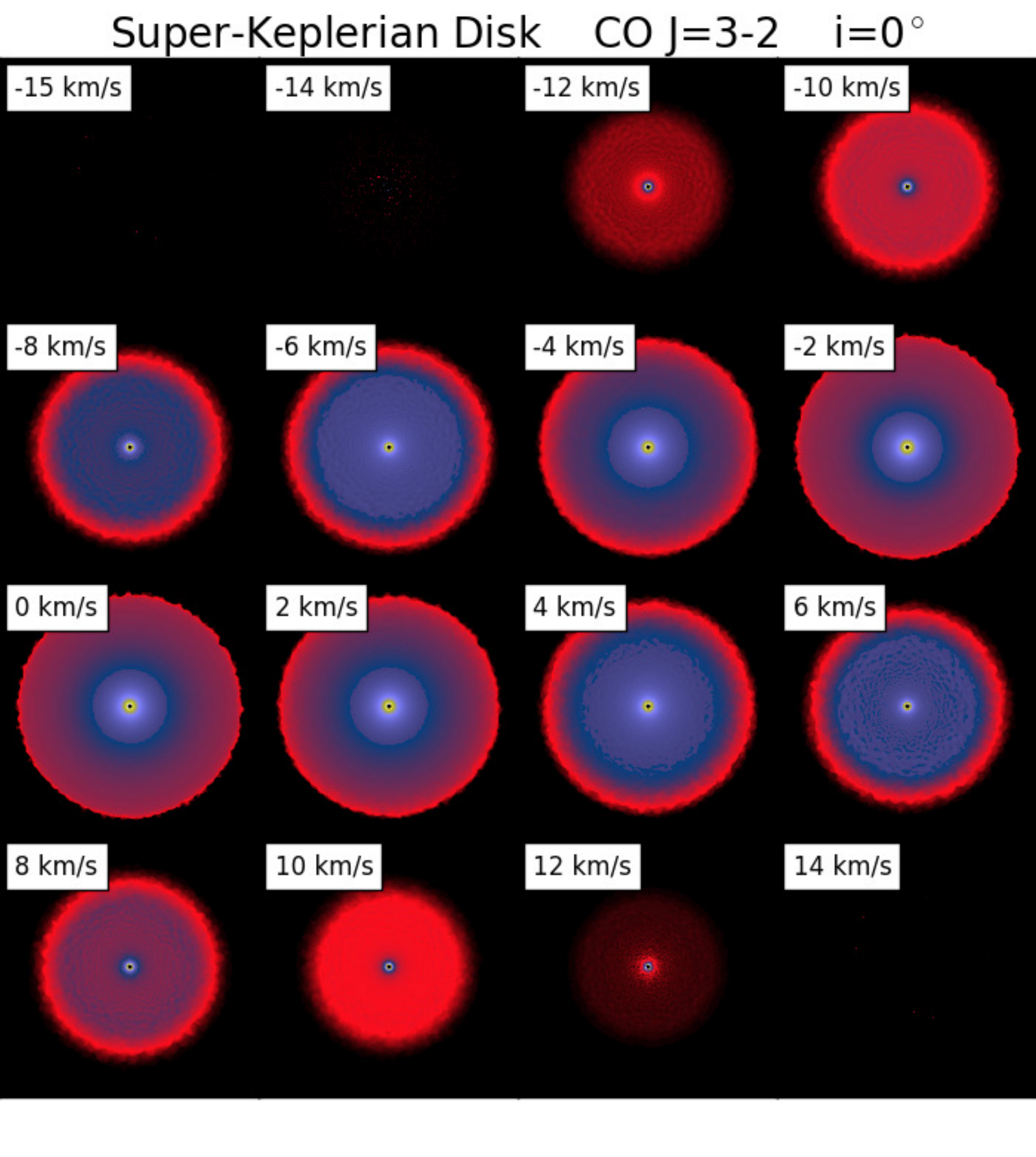}}
 \resizebox{7.5cm}{!}{\includegraphics{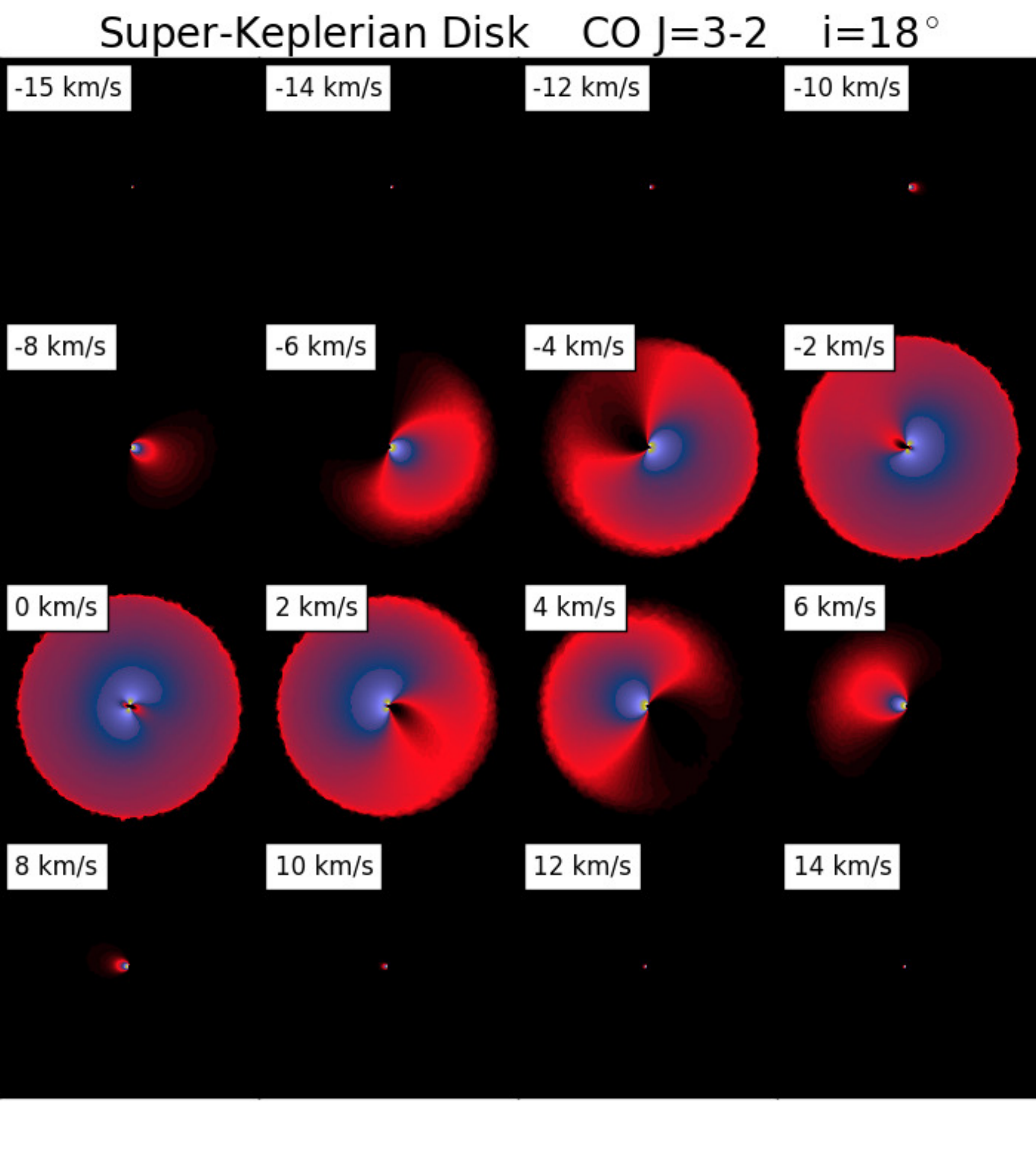}}
 \resizebox{7.5cm}{!}{\includegraphics{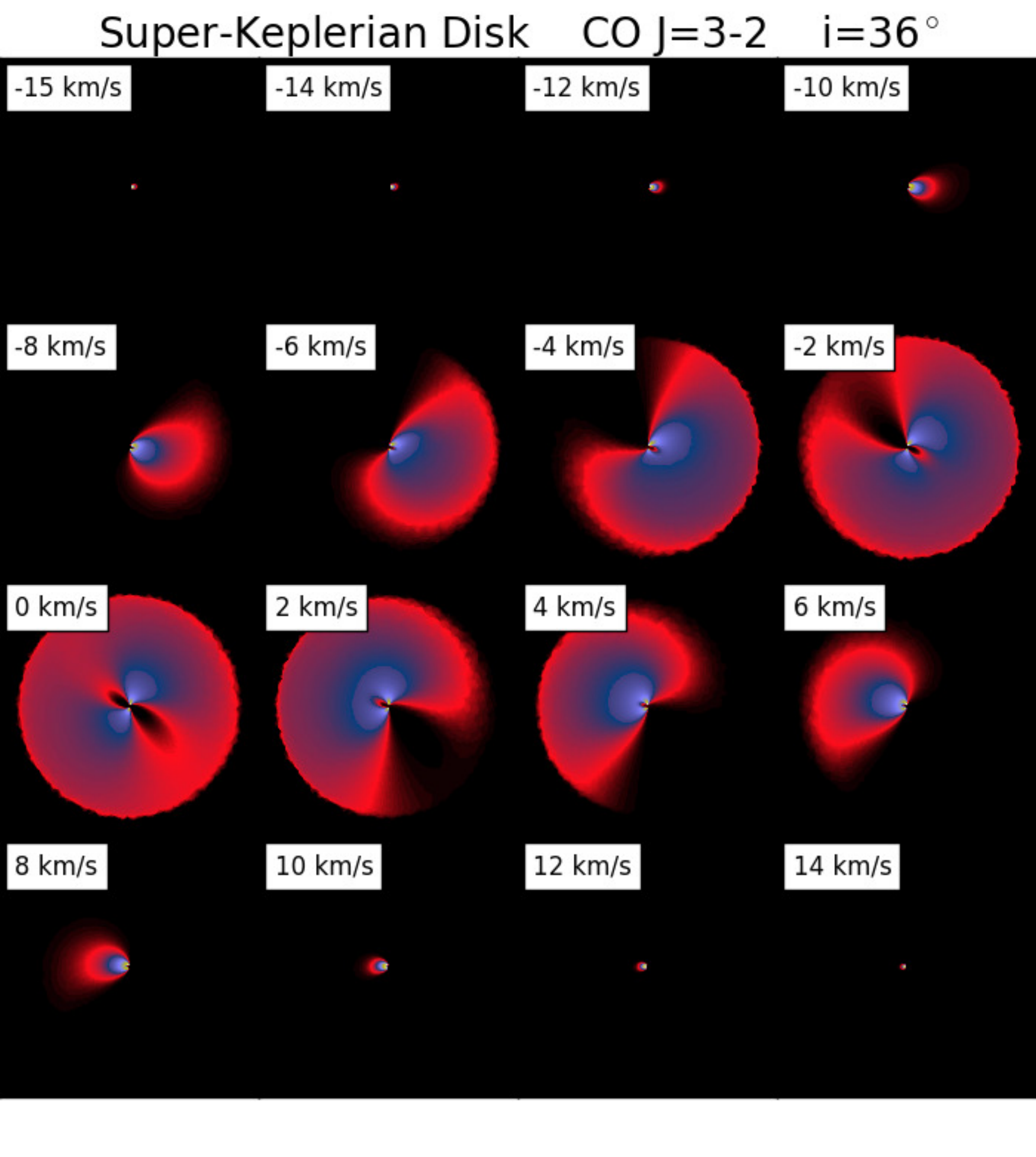}}
 \resizebox{7.5cm}{!}{\includegraphics{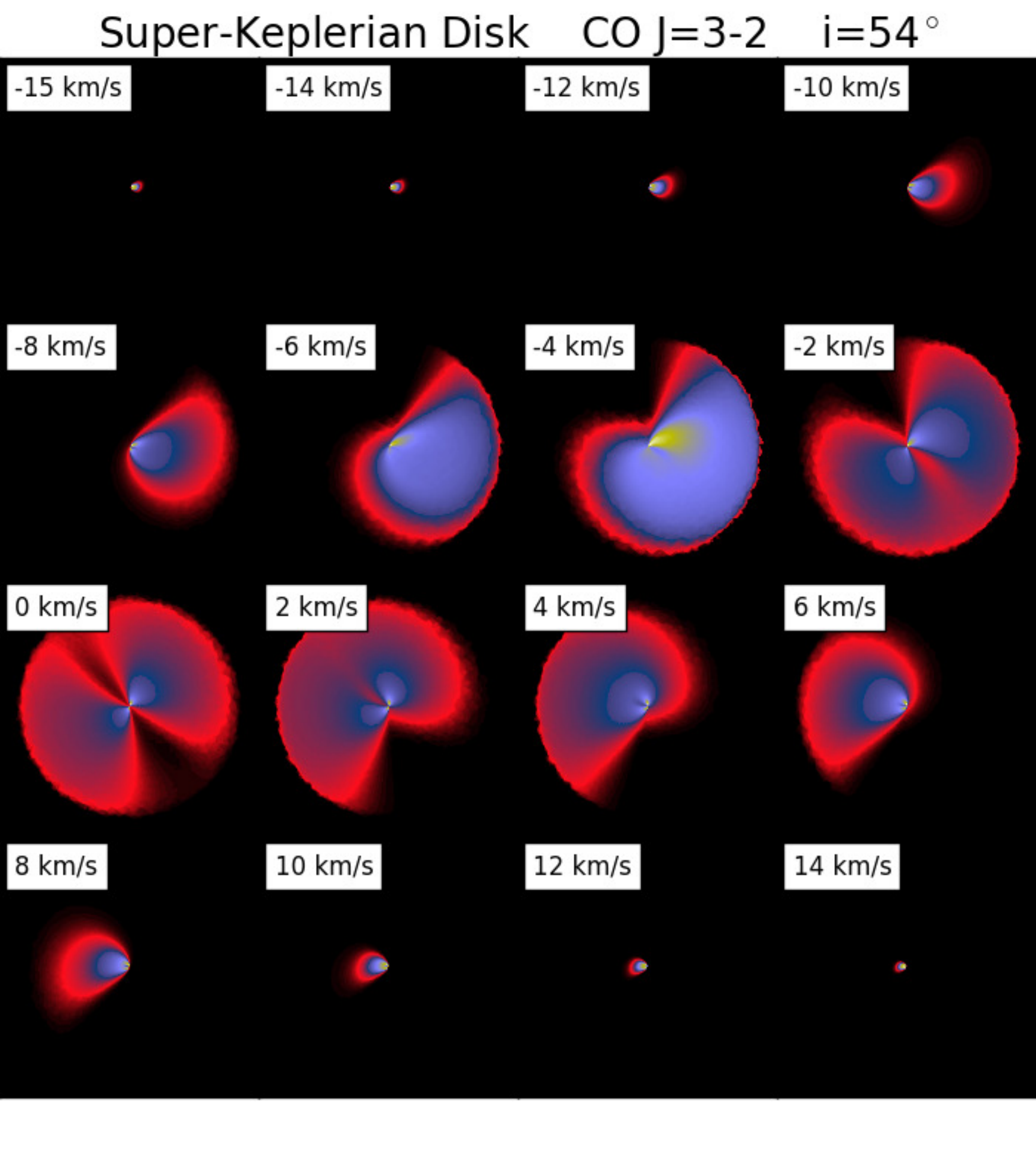}}
 \resizebox{7.5cm}{!}{\includegraphics{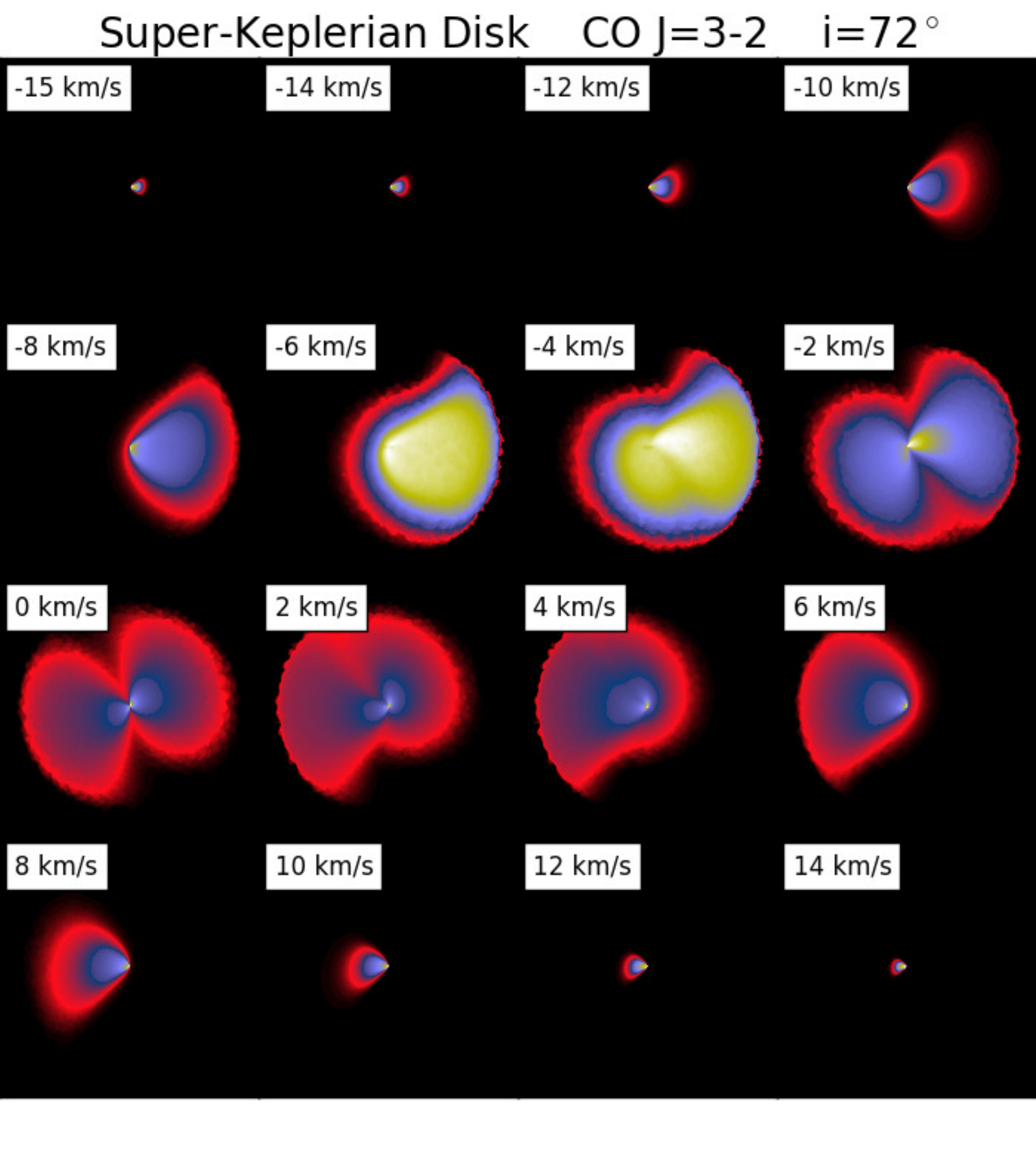}}
 \resizebox{7.5cm}{!}{\includegraphics{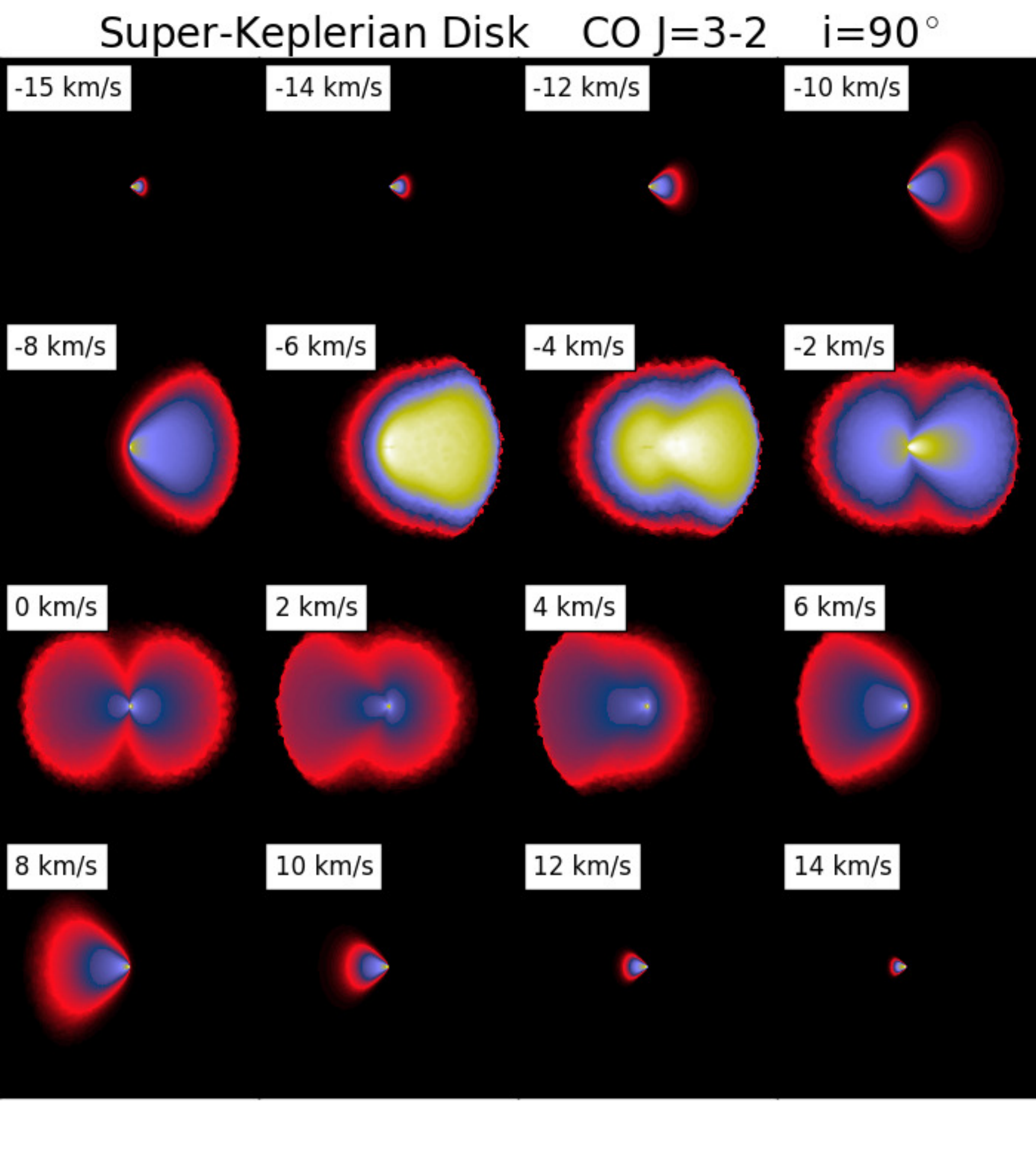}}
 \caption{Channel maps of the model described in the title of each panel plot.}
\end{figure*}

\begin{figure*}[htp]
 \centering
 \resizebox{7.5cm}{!}{\includegraphics{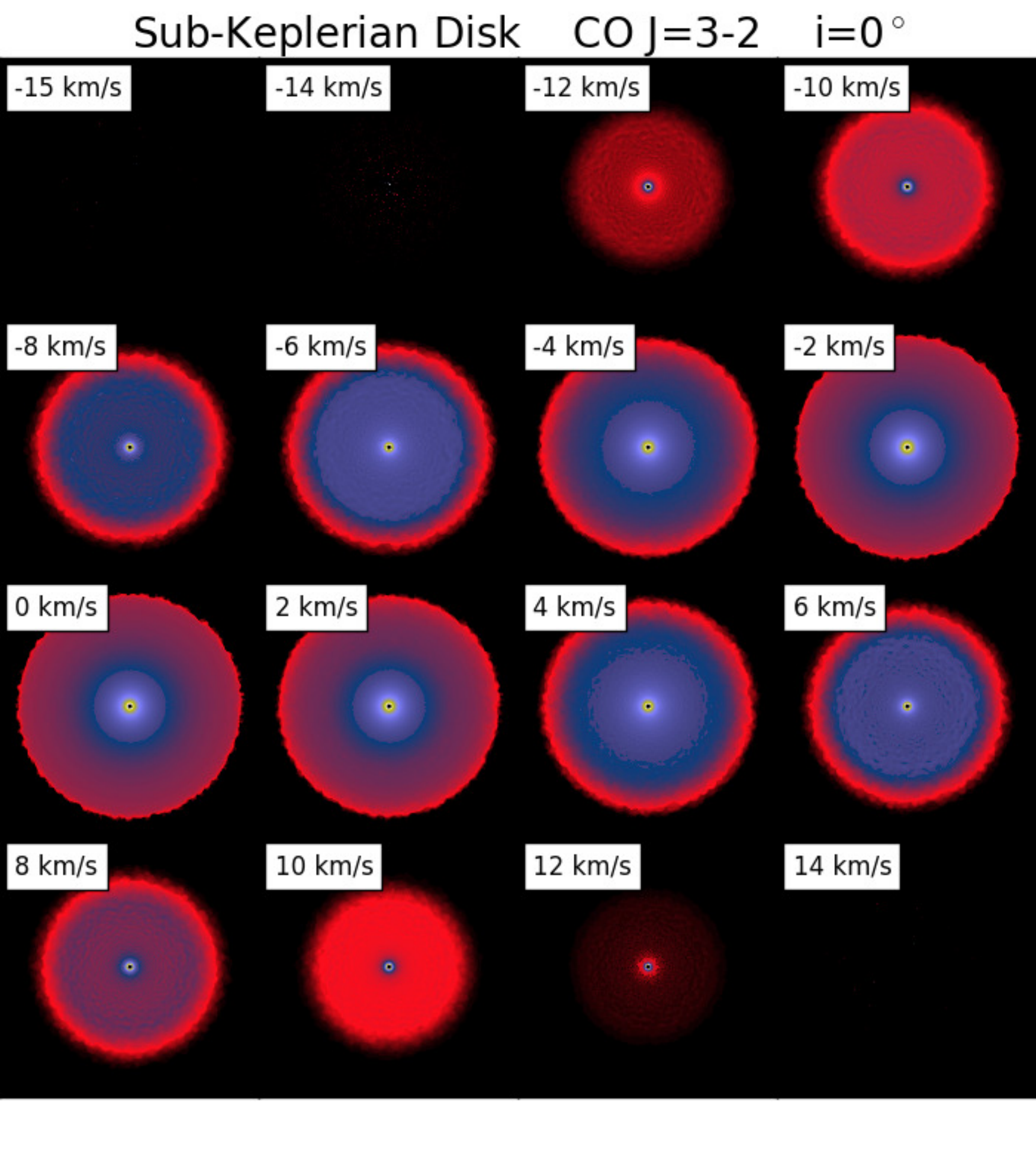}}
 \resizebox{7.5cm}{!}{\includegraphics{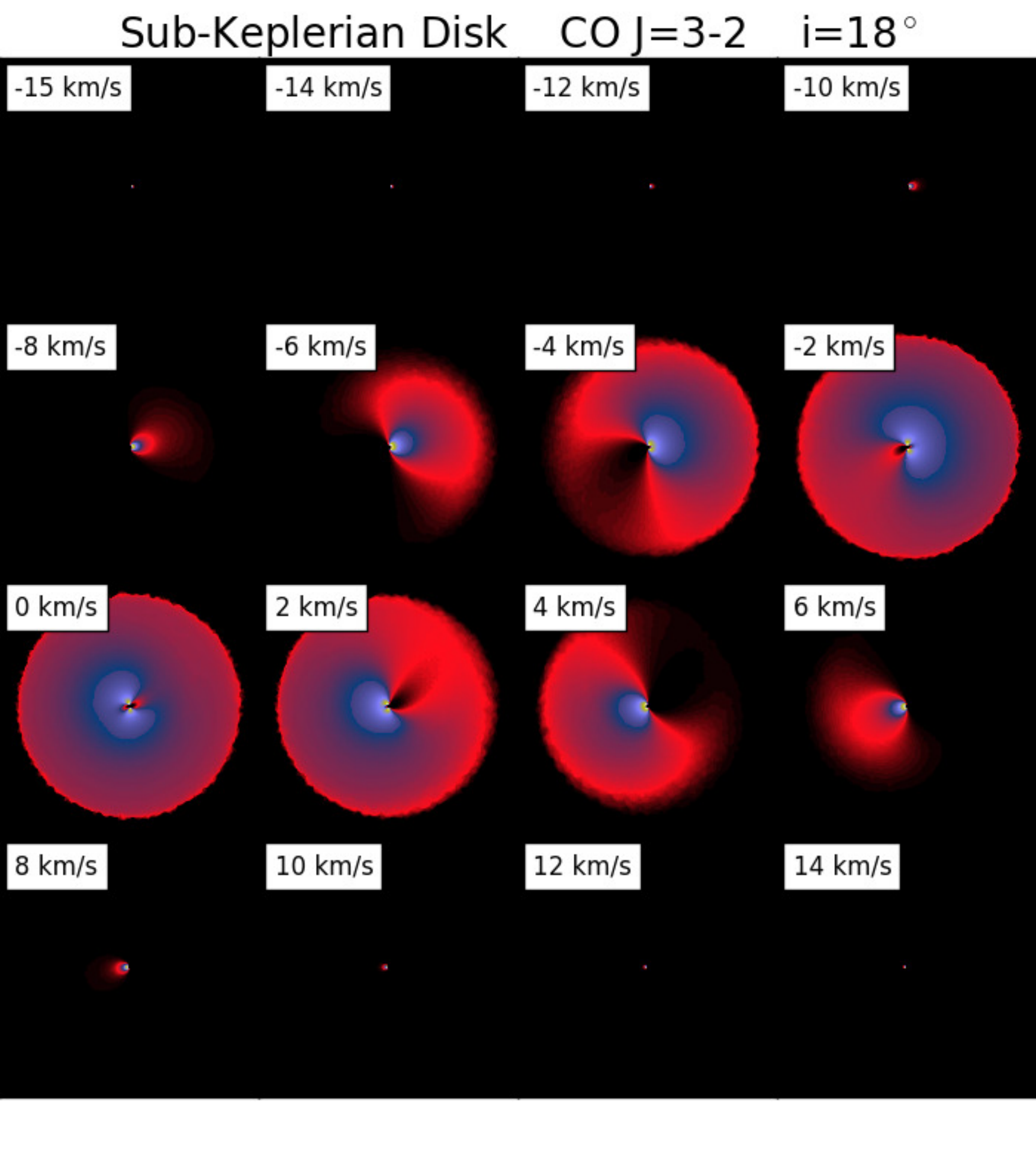}}
 \resizebox{7.5cm}{!}{\includegraphics{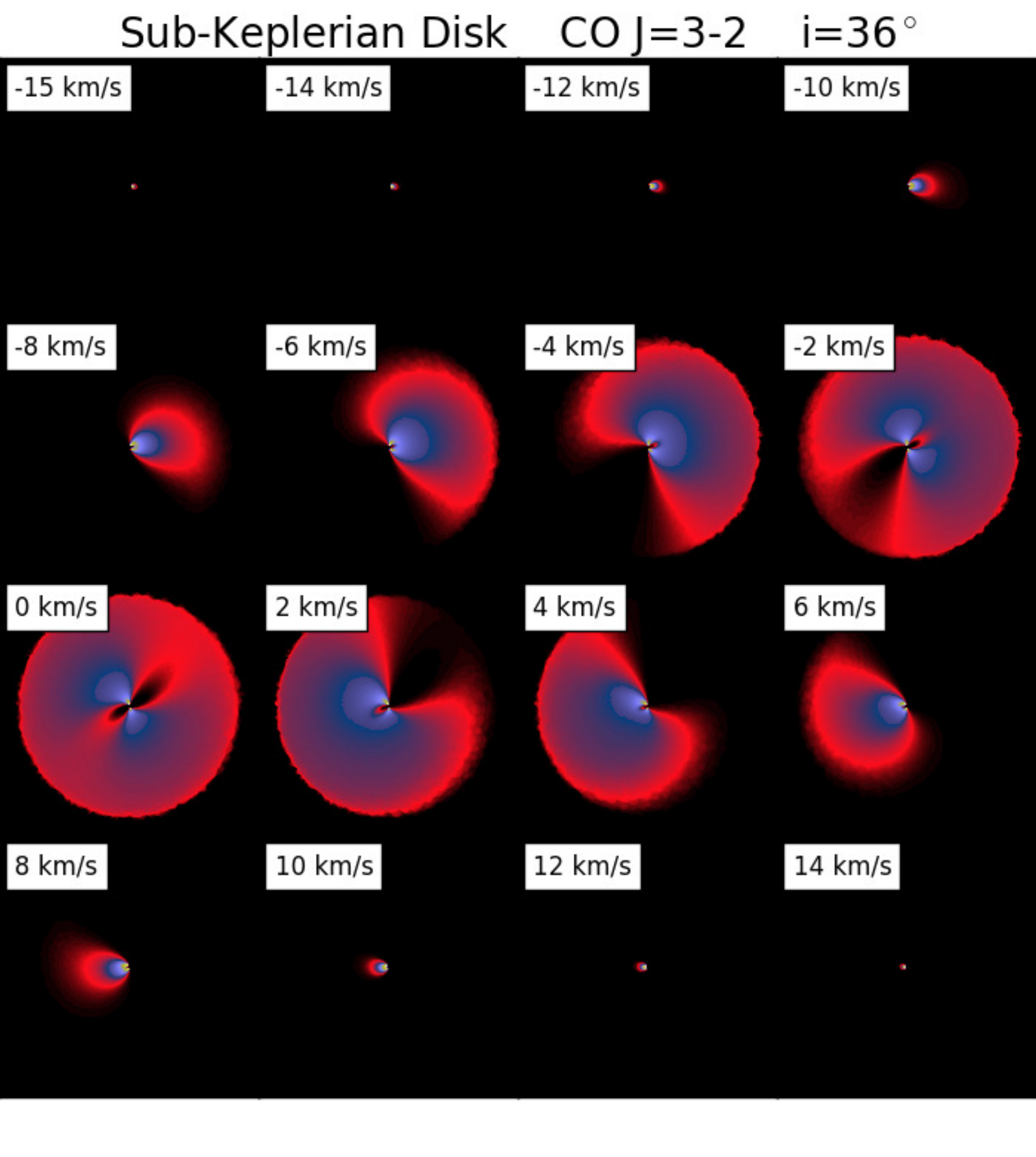}}
 \resizebox{7.5cm}{!}{\includegraphics{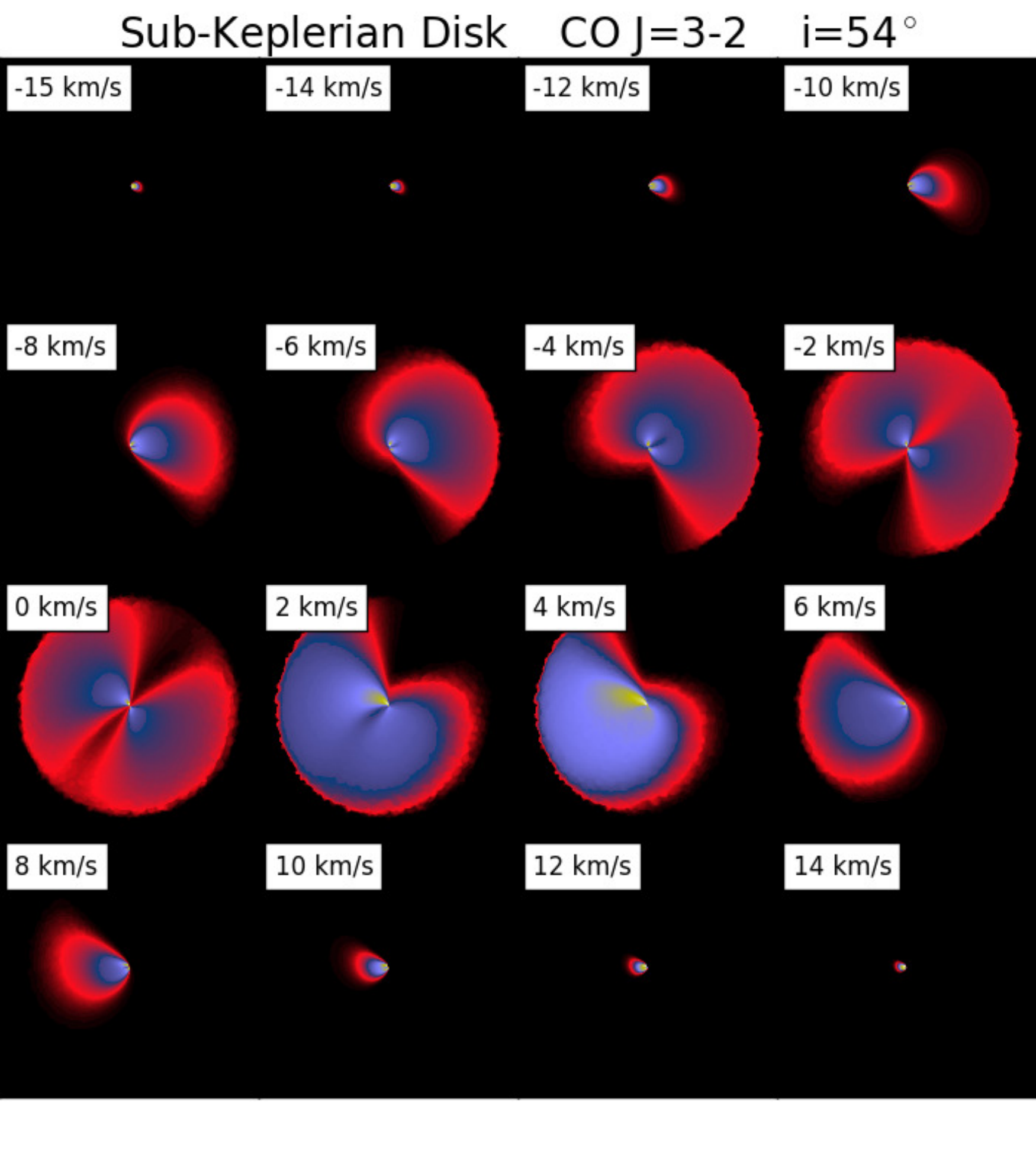}}
 \resizebox{7.5cm}{!}{\includegraphics{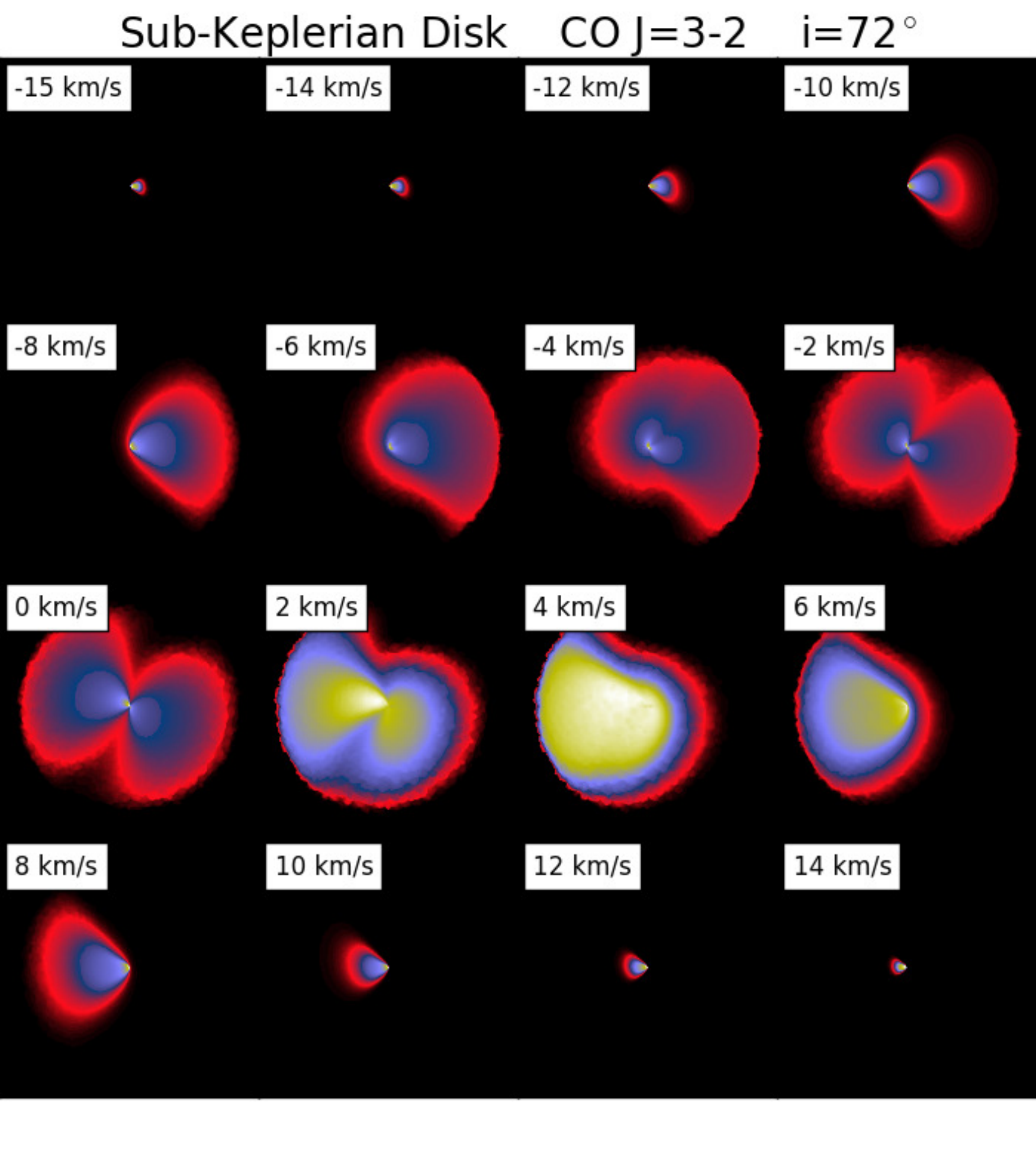}}
 \resizebox{7.5cm}{!}{\includegraphics{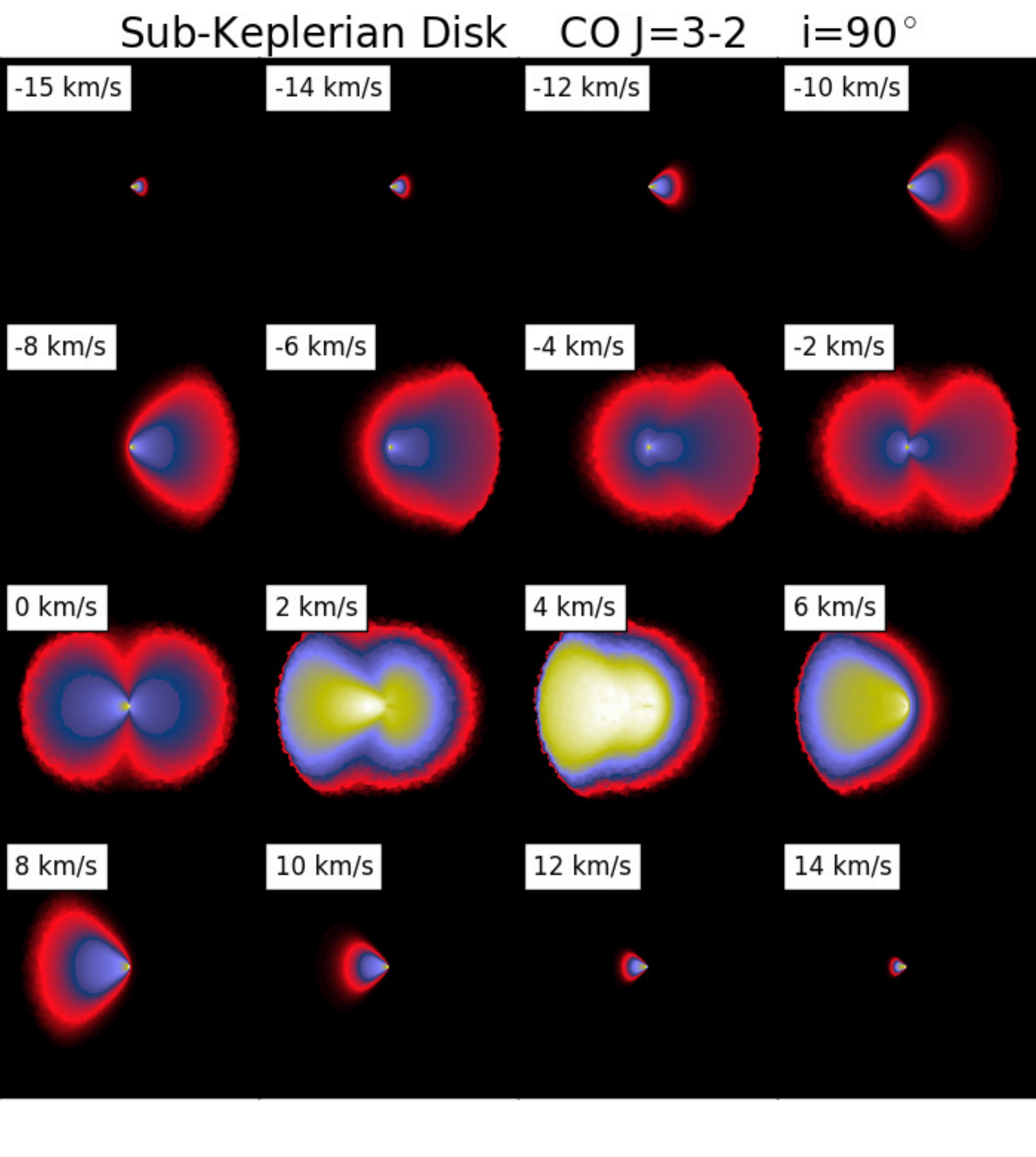}}
 \caption{Channel maps of the model described in the title of each panel plot.}
\end{figure*}

\begin{figure*}[htp]
 \centering
 \resizebox{7.5cm}{!}{\includegraphics{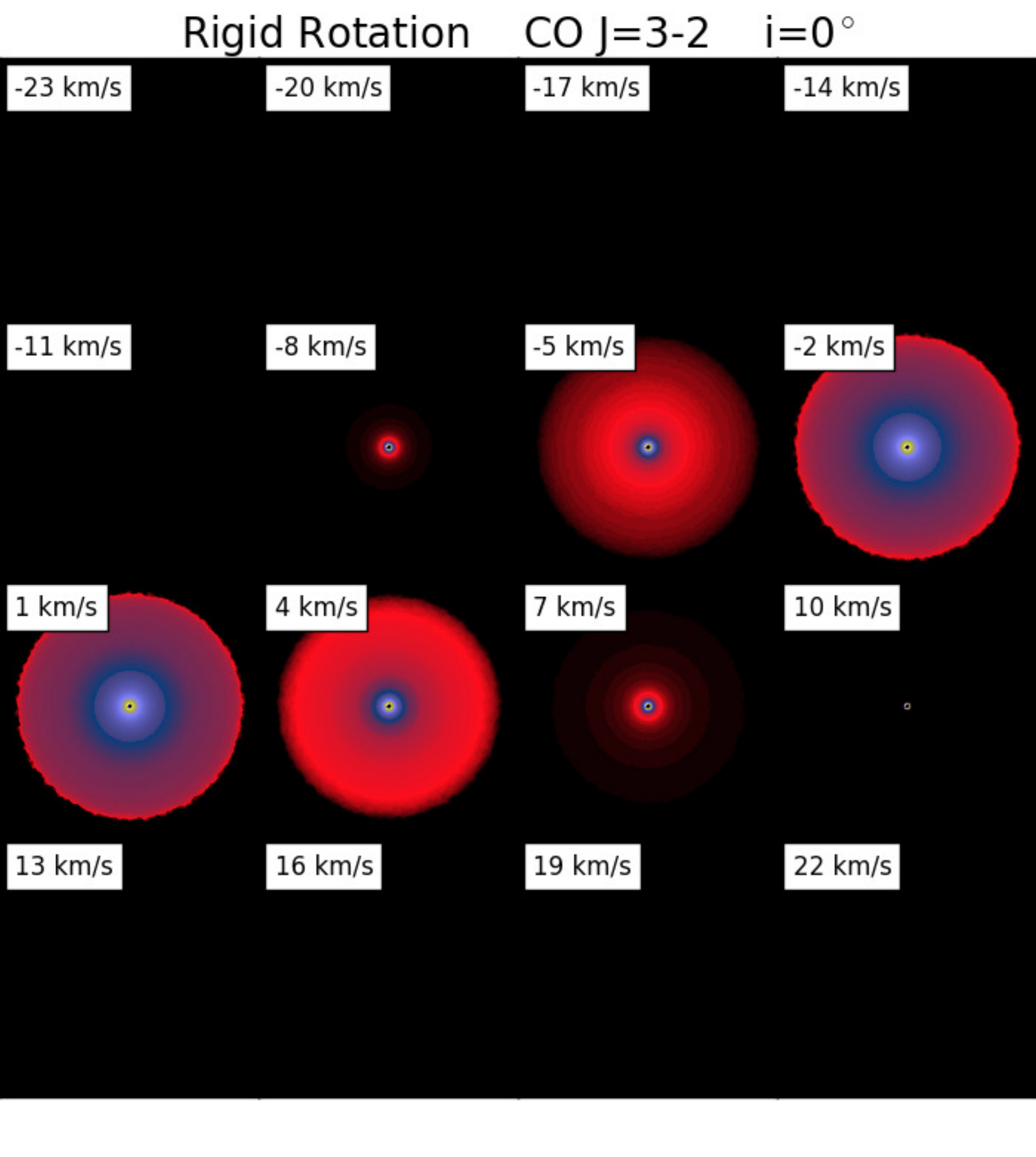}}
 \resizebox{7.5cm}{!}{\includegraphics{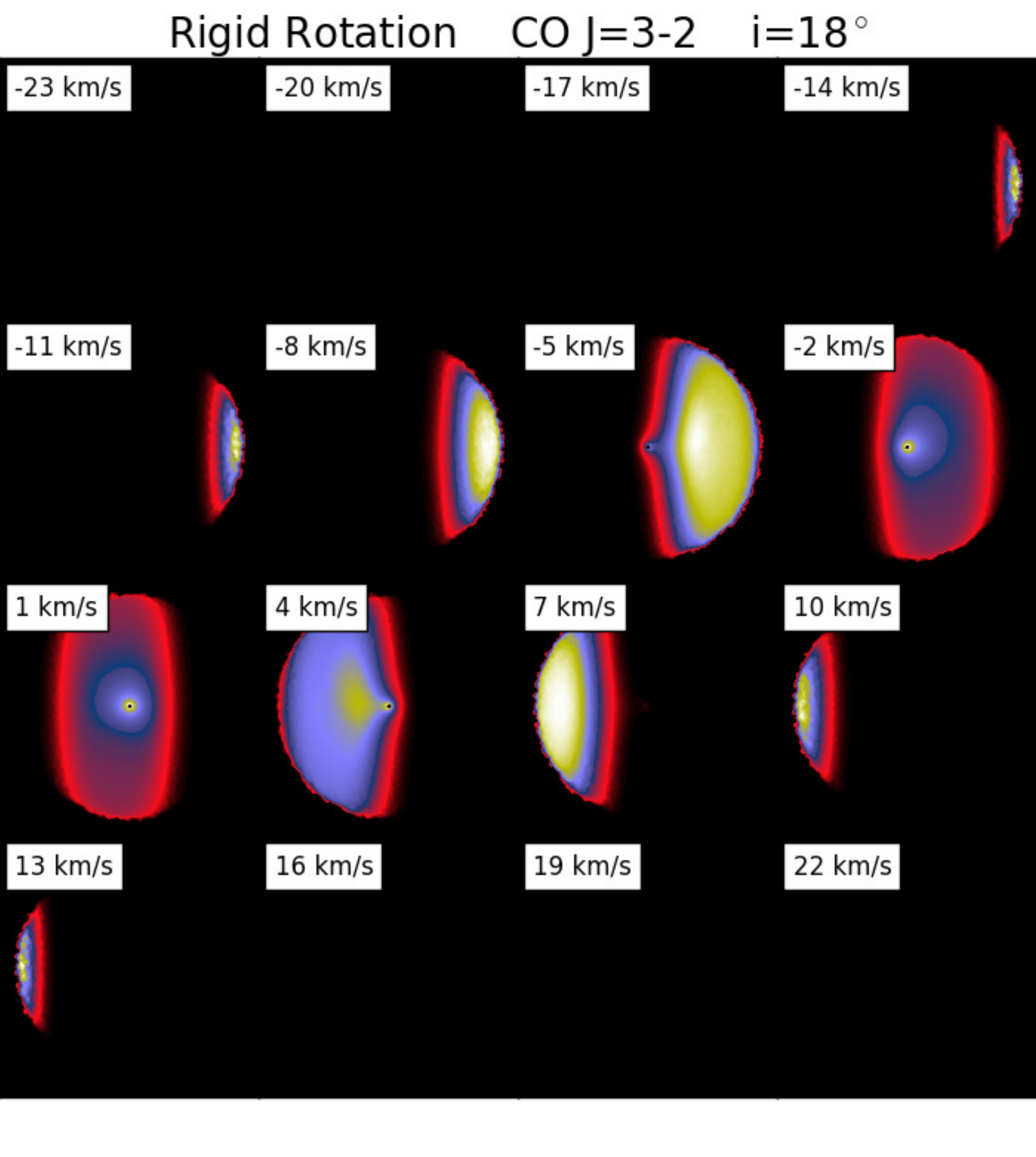}}
 \resizebox{7.5cm}{!}{\includegraphics{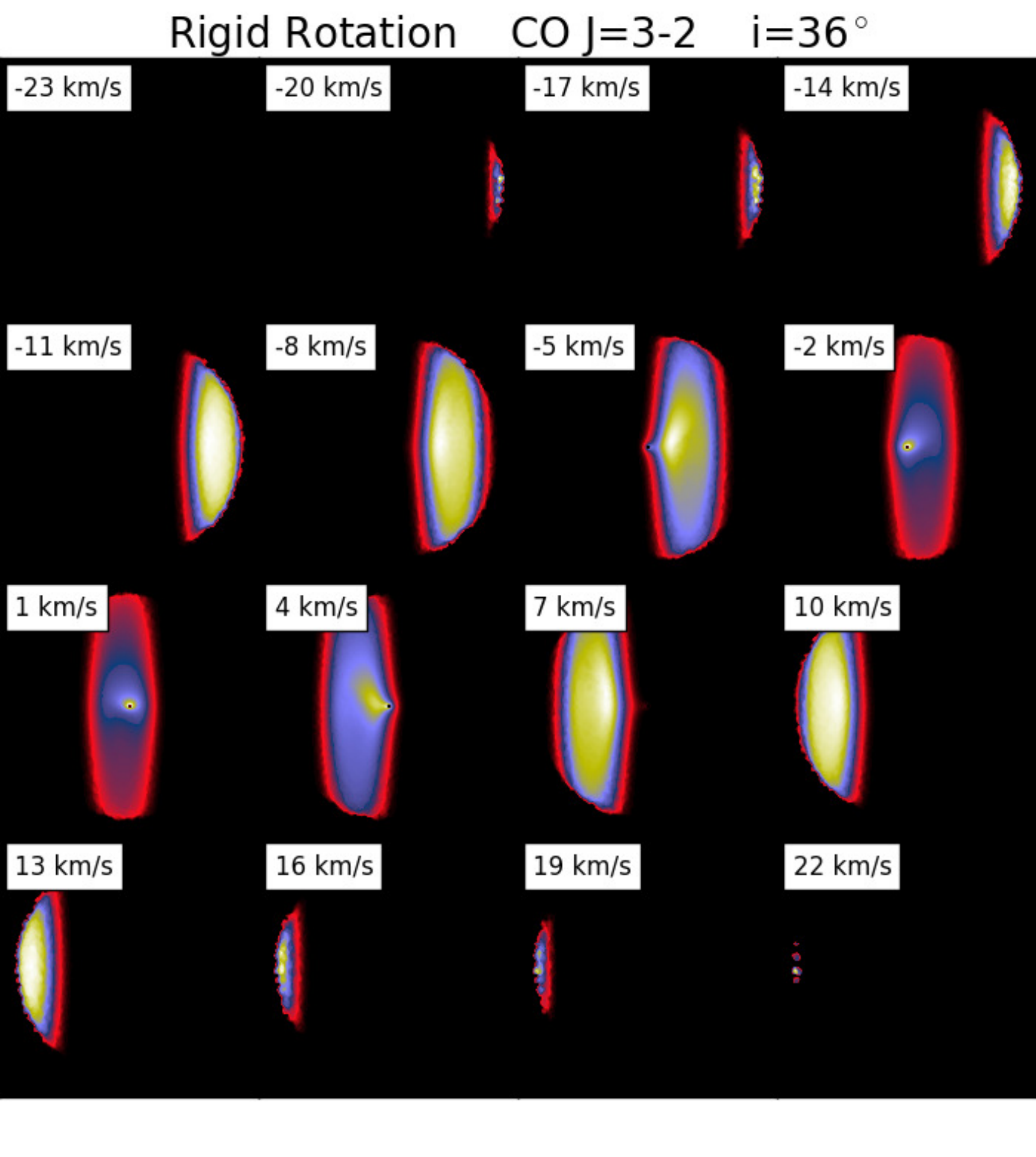}}
 \resizebox{7.5cm}{!}{\includegraphics{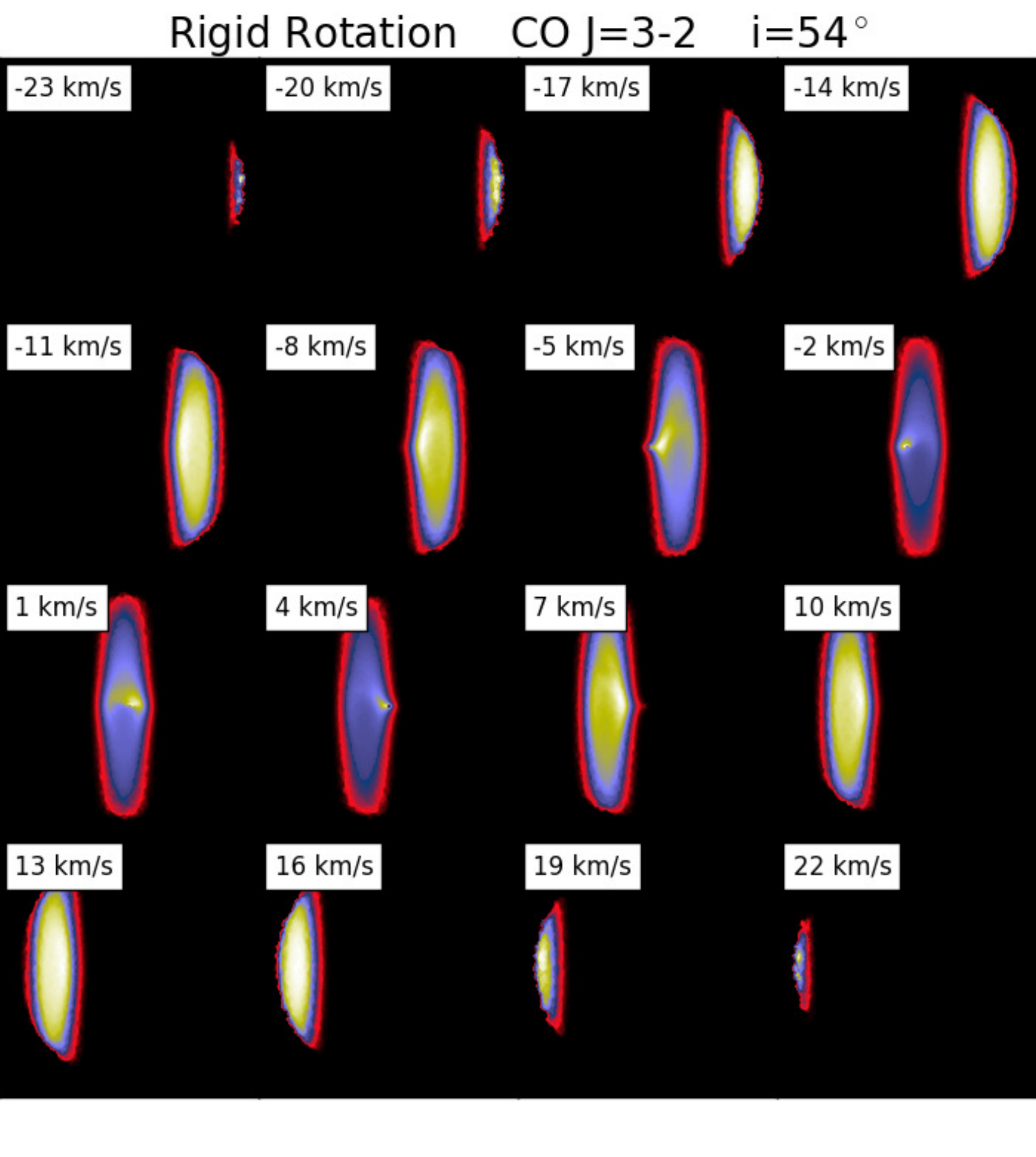}}
 \resizebox{7.5cm}{!}{\includegraphics{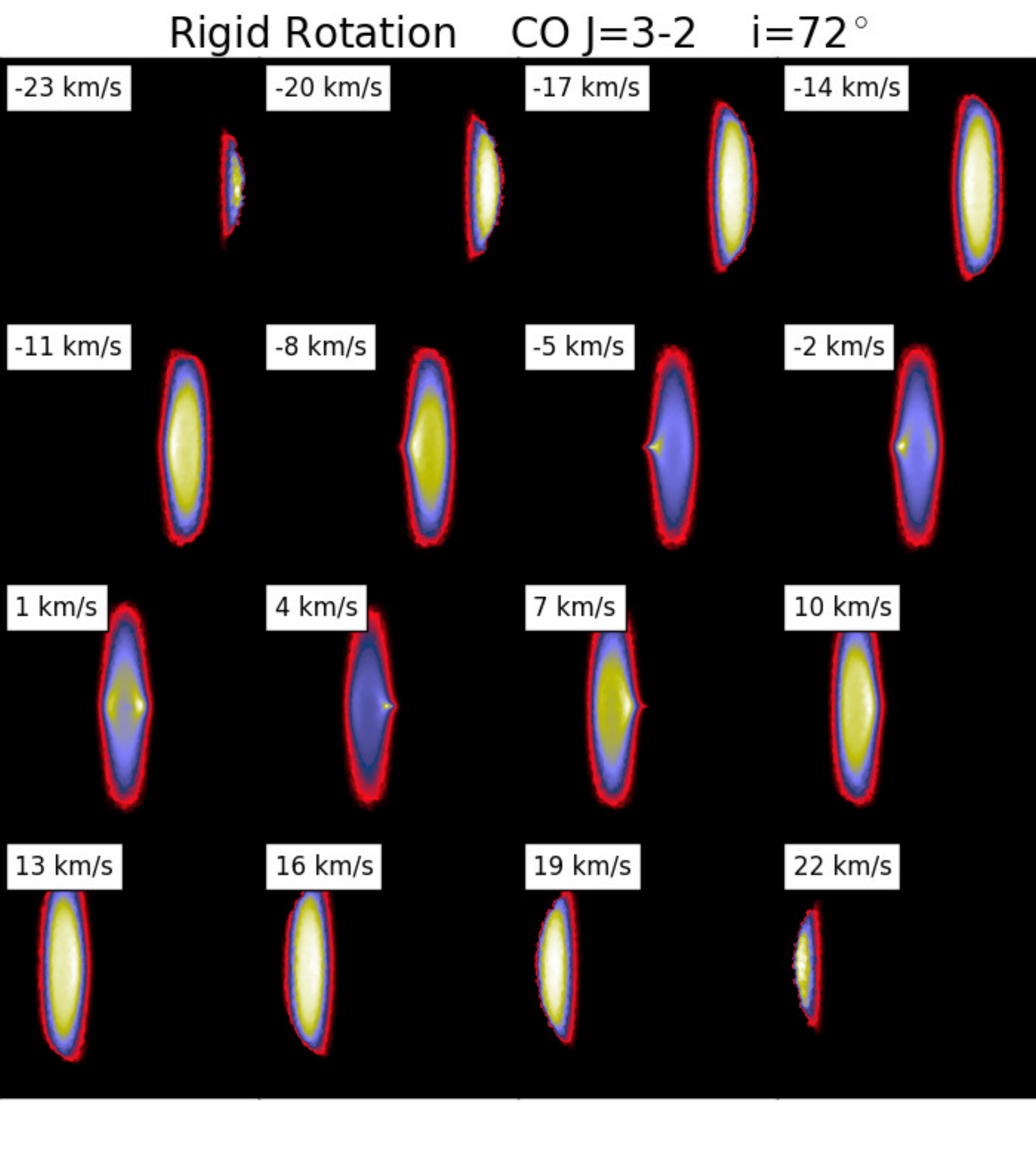}}
 \resizebox{7.5cm}{!}{\includegraphics{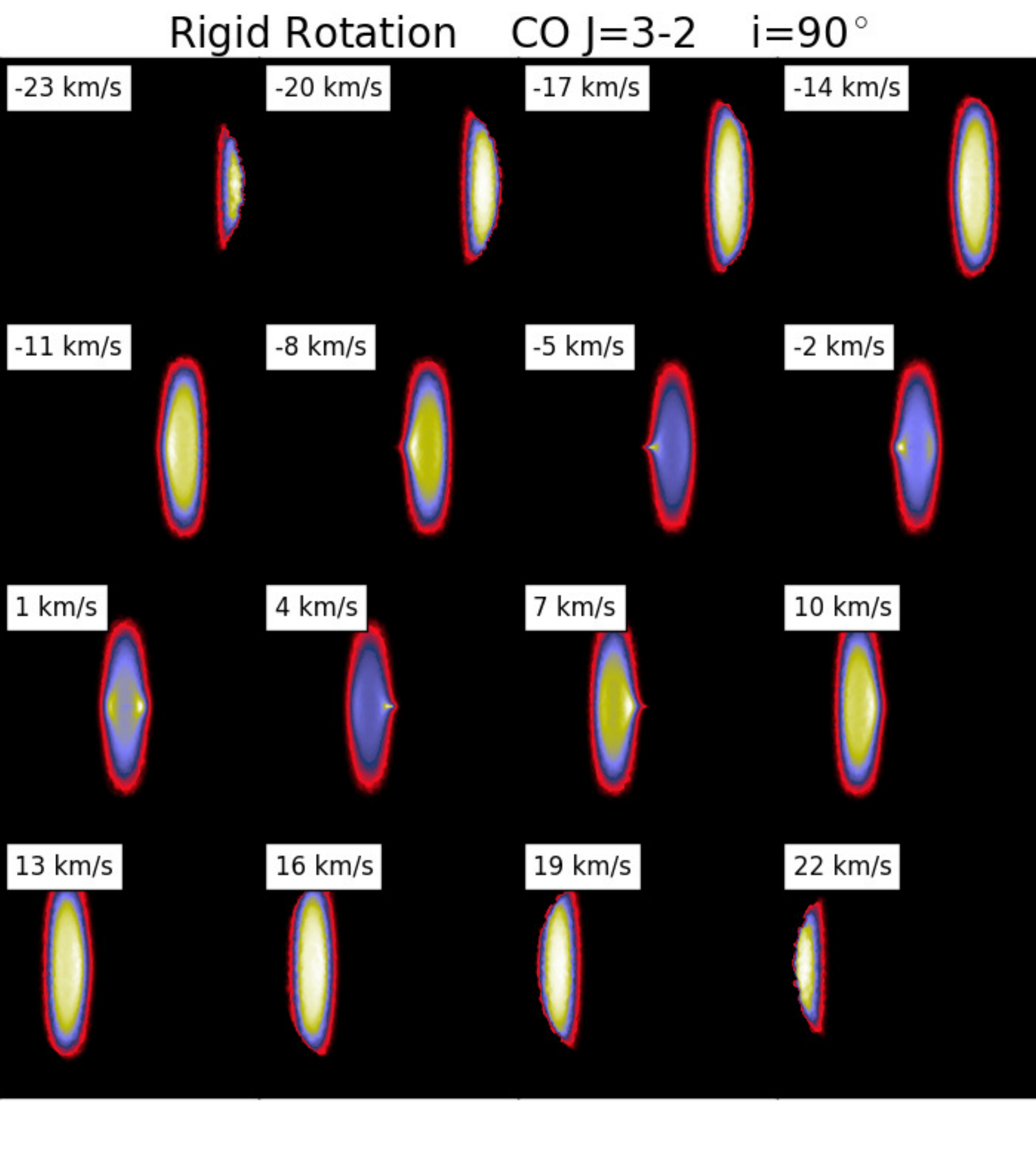}}
 \caption{Channel maps of the model described in the title of each panel plot.}
\end{figure*}

\end{appendix}

\end{document}